\documentclass[a4,draftcls,onecolumn]{IEEEtran}
\usepackage{graphicx}
\usepackage{cite}
\usepackage{array}
\usepackage{ams}
\usepackage{amsmath}
\usepackage{amssymb}
\usepackage{subfigure}
\usepackage{enumerate}
\usepackage{lscape}
\usepackage{float}

\makeatletter
\def\ifundefined{\@ifundefined}
\makeatother

\begin{document}

\title{Generalized ABBA Space-Time Block Codes}

\author{Giuseppe Thadeu Freitas de Abreu, \\
        {\tt giuseppe@ee.oulu.fi}\\
        Center for Wireless Communications, University of Oulu\\
        P.O. box 4500, 90014-Oulu, Finland}

\maketitle

\begin{abstract}
Linear space-time block codes (STBCs) of unitary rate and full
diversity, systematically constructed over arbitrary
constellations for any number of transmit antennas are introduced.
The codes are obtained by generalizing the existing ABBA STBCs,
a.k.a \emph{quasi}-orthogonal STBCs (QO-STBCs).
Furthermore, a fully orthogonal (symbol-by-symbol) decoder for the
new generalized ABBA (GABBA) codes is provided.
This remarkably low-complexity decoder relies on partition
orthogonality properties of the code structure to decompose the
received signal vector into lower-dimension tuples, each dependent
only on certain subsets of the transmitted symbols.
Orthogonal decodability results from the nested application of this
technique, with no matrix inversion or iterative signal processing
required.
The exact bit-error-rate probability of GABBA codes over generalized
fading channels with maximum likelihood (ML) decoding is evaluated
analytically and compared against simulation results obtained with
the proposed orthogonal decoder.
The comparison reveals that the proposed GABBA solution, despite its
very low complexity, achieves nearly the same performance of the
bound corresponding to the ML-decoded system, especially in systems
with large numbers of antennas.
\end{abstract}
\section{Introduction}
%
%
The design of space-time codes (STCs) as building blocks of
multiple-input-multiple-output (MIMO) systems
\cite{MyListOfPapers:Gerbert2003} has been a popular research topic
in wireless communications since the discovery of the Alamouti
scheme \cite{MyListOfPapers:Alamouti1998} and the orthogonal
space-time block-codes (OSTBC) \cite{MyListOfPapers:Tarokh1999-2,
MyListOfPapers:Tarokh1999-3}.

%
%
Although STC techniques were originally envisioned for rather modest
spatial-temporal dimensions, the last few years has seen a growing
interest on techniques suitable for large/scalable MIMO systems
\cite{MyListOfPapers:Biglieri2001, MyListOfPapers:BiglieriTCOM2005}.
Arguably, this recent owes mainly to two related developments.
The first is the advance the antenna technology enabling the design
of compact multi-mode antennas \cite{MyListOfPapers:GetuTWCOM2005,
MyListOfPapers:WaldschmidtTAP2004, MyListOfPapers:Thudor2002} which,
though small in size, can provide a large number of independent
channels by exploiting spatial, pattern and polarization diversities
\cite{MyListOfPapers:Dietrich2001}.
The second is the increasing popularity of distributed and
cooperative systems \cite{MyListOfPapers:Veen2000,
MyListOfPapers:Nicholas2003, MyListOfPapers:UysalWCNC2005}.

These recent developments provide a rationale for the feasibility of
MIMO systems of not only large but, in fact, variable dimensions.
For example, in the case of compact multi-mode antennas, MIMO setups
of different dimensions may occur because the number of channels
effectively made available by the spatial, pattern and polarization
diversities dependents on the physical scattering properties of the
environment.
Analogously, in the case of distributed architectures, variable MIMO
setups arise naturally since the cooperative transmission can only
be employed after the group of cooperating devices is assembled
\cite{MyListOfPapers:Barbarossa2004}.

MIMO techniques can significantly improve both user and rate
capacities (efficiencies) of wireless communications systems
\cite{MyListOfPapers:Cover1979, MyListOfPapers:Telatar1995}.
Although space-time transmission diversity techniques in
point-to-point links are known to be sub-optimum from a Shannon
capacity viewpoint \cite{MyListOfPapers:Sandhu2000}, these
techniques have their extremely low-complexity to their advantage.
This trade-off argument is especially compelling when considered
that the capacity achieved by MIMO systems is in practice
significantly limited by the use of digital modulation
\cite{MyListOfPapers:Wenyan2005}.
It is in distributed, cooperative diversity systems, however, that
the flexibility of scalable space-time coding techniques have the
most of their potential \cite{MyListOfPapers:Laneman-PhDThesis-2002,
MyListOfPapers:Xiaohua-SPL-2005}.

Given the above, it is fair to say that, when considering space-time
block codes as building blocks for scalable MIMO systems,
modulation-independence, generalized and systematic
constructibility, orthogonal (symbol-by-symbol) decodability and
rate-unitariness\footnote{Rate is here understood as the ratio
between the number of transmissions epochs and the number of symbols
transmitted per block. For a formal definition, please refer to
\ref{Rate}.} are the most essential features to be sought.
These requirements make a strong case for linear STBCs and since
existing techniques such as the Alamouti scheme, the OSTBCs and the
maximum-rate minimum-delay (MRMD) STBCs are either limited to a
small number of antennas, or to particular constellations, or
undermined by sub-unitary rateness \cite{MyListOfPapers:Wang2003,
MyListOfPapers:Tirkkonen-2002}, many have taken upon the challenge
of designing alternative solutions.
To cite a few examples, the \emph{universal} STCs by Gamal \emph{et
al.} \cite{MyListOfPapers:Gamal2003}, the full-rate full-diversity
(FRFD) STCs by Xiaoli \emph{et al.}
\cite{MyListOfPapers:Xiaoli2003}, the full-rate full-diversity
orthogonal STBC scheme by He \emph{et al.}
\cite{MyListOfPapers:Lei2003} and the variable-rate STBCs by Kim
\emph{et al.} \cite{MyListOfPapers:Kim2003}.

Each of the above-mentioned methods are clever techniques
representing a step forward in the quest for high-performing,
low-complexity STBCs with no or little rate-loss.
To the best of our knowledge, however, these and other similar
contributions fail to simultaneously deliver a systematic design,
scalable to any number of transmit antennas, applicable to arbitrary
constellations and allowing a linear, symbol-by-symbol decoding
method with performance comparable to that of the maximum likelihood
(ML) decoder.

In this paper, a family of space-time block codes that satisfies all
these objectives is introduced.
The codes are obtained by generalizing the ABBA (a.k.a
\emph{quasi}-orthogonal) STBCs (QO-STBCs) presented in
\cite{MyListOfPapers:Tirkkonen2000, MyListOfPapers:Jafarkhani2001,
MyListOfPapers:Papadias2003}.
A linear, orthogonal decoder for the proposed generalized ABBA
(GABBA) STBCs is also provided.
The performance of the new codes is also studied both analytically
and through computer simulations.
In particular, the exact bit error rate (BER) probability of GABBA
codes over uncorrelated memoryless fading channels with unequal
powers and arbitrary statistics is derived under the assumption of
perfect channel knowledge at the receiver and ML decoding.
Comparing those results against computer simulations, it is shown
that orthogonally decoded GABBA STBCs achieve nearly the same
performance of achievable with the ML decoder (which is
prohibitively complex for large codes and higher-order
constellations), especially in systems with a relatively large
number of antennas at the receiver.
The transmission rates theoretically achieved by the GABBA codes are
compared against the corresponding Shannon SIMO bound
\cite{MyListOfPapers:Telatar1995, MyListOfPapers:Foschini1998},
showing that GABBA codes have the potential to reach the capacity of
the SIMO channel, despite the remarkably low-complexity involved in
the technique.

\section{Systematic Construction of GABBA-STBCs}
\label{Construction}

\subsection{Preliminaries}
\label{ConstructionPreliminaries}

\setcounter{footnote}{0}
To the best of our knowledge, a simple and systematic method to
construct linear, full(unitary)-rate\footnotemark,
modulation-independent STBCs for any number of antennas is yet to be
demonstrated.
In \cite{MyListOfPapers:Tirkkonen2000}, Tirkkonen \emph{et al.}
conjectured that larger ABBA STBCs codes could ``\emph{probably be
constructed by repeatedly iterating the} ABBA \emph{scheme}.''
It will be shown here that this conjecture was almost correct,
except that the generalized construction involves two distinct ABBA
schemes.
\setcounter{footnote}{1}

%
Consider a vector $\mathbf{s}_{\!K}$ of symbols taken from a
non-ambiguous arbitrary complex constellation $\mathcal{S} \in
\mathbb{C}_{-\{0\}}^{Q}$:
\begin{equation}
\label{SymbolVector}
\mathbf{s}_{\!K} \triangleq [s_1,\cdots,s_K]^\textup{T} \in \mathcal{S}^{K \times 1},%
\end{equation}
where $^\textup{T}$ denotes transpose.

%
\newtheorem{definition}{Definition}
\begin{definition}[Linear Space-time Block Encoding Matrix]
\label{LinearSTBCs} \quad%

The term \emph{linear space-time block encoding matrix} will be used
throughout this paper in reference to matrices of the form
\begin{equation}
\mathbf{C}(\mathbf{s}_{\!K}\!) = \left[ \begin{array}{ccc}%
c_{1,1} & \cdots & c_{1,K}\\[-0.95ex]
\vdots  & \ddots & \vdots\\[-0.75ex]
c_{K,1} & \cdots & c_{K,K}
\end{array}\right] \in \mathbb{C}_{-\{0\}}^{K \times n_t},%
\end{equation}
where each element $c_{i,j}$ relates to one and only one symbol $s_k
\in \mathbf{s}_{\!K}$, such that $c_{i,j} \in
\{s_k,-s_k,s_k^*,-s_k^* \}$. \hfill \QEDopen
\end{definition}

Notice that definition \ref{LinearSTBCs} rules out constructions
based on linear combinations of entries of the symbol vector, such
as the rate $3/4$ code that appears in
\cite{MyListOfPapers:Tarokh1999-2}.

%
\begin{definition}[Raw Symbol, Raw Encoding Matrix]
\label{RawSymbol} \quad%

The symbol $s_k$ from which an entry $c_{i,j}$ of a linear
space-time block encoding matrix $\mathbf{C}(\mathbf{s}_{\!K}\!)$
derives, is referred to as the \emph{raw symbol} of $c_{i,j}$, and
is denoted by $\langle c_{i,j} \rangle$.
Likewise, the matrix $\langle \mathbf{C}(\mathbf{s}_{\!K}) \rangle$
consisting of the raw symbols of the space-time block encoding
matrix $\mathbf{C}(\mathbf{s}_{\!K})$ is referred to as the
\emph{raw encoding matrix} of $\mathbf{C}(\mathbf{s}_{\!K})$. \hfill
\QEDopen
%
%
\end{definition}

%
%
\begin{definition}[Dense STBCs]
\label{DenseSTBCs} \quad%

A linear space-time block code matrix
$\mathbf{C}(\mathbf{s}_{\!K}\!)$ is said to be \emph{dense} if and
only if (iff) $\langle c_{i,j} \rangle \neq 0 \ \forall \ (i,j)$.
\hfill \QEDopen
\end{definition}
%
%
\begin{definition}[Complete STBCs]
\label{CompleteSTBCs} \quad %

A linear space-time block code matrix
$\mathbf{C}(\mathbf{s}_{\!K}\!)$ is said to be \emph{complete} iff
it satisfies both the following conditions:
\begin{itemize}
\item[$i$)] $\begin{array}{lcl}
            \langle \mathbf{c}_{i,\downarrow}\!(\mathbf{s}_{\!K})\rangle \neq \langle \mathbf{c}_{j,\downarrow}\!(\mathbf{s}_{\!K})\rangle & \forall & i \neq j %
            \end{array}$, %
\item[$ii$)] $\begin{array}{lcl}
            \langle \mathbf{c}_{\downarrow,i}\!(\mathbf{s}_{\!K})\rangle \neq \langle \mathbf{c}_{\downarrow,j}\!(\mathbf{s}_{\!K})\rangle & \forall & i \neq j %
            \end{array}$. \hfill \QEDopen%
\end{itemize}
\end{definition}

In plain words, if the rows and columns of an encoding channel
matrix represent different transmit epochs and different transmit
spatial channels, respectively, definition \ref{DenseSTBCs} implies
that at every transmit epoch, all diversity branches are utilized.
On the other hand, according to definition \ref{CompleteSTBCs}, the
raw columns and raw rows of a complete encoding matrix
$\mathbf{C}(\mathbf{s}_{\!K}\!)$ are, simultaneously, mutually
exclusive permutations of $\mathbf{s}_{\!K}$ and
$\mathbf{s}_{\!K}^\textup{T}$, respectively.
These two properties ensure that the code fully exploits the ``raw
diversity'' of the channel, as defined in \cite[section II, $\S
1$]{MyListOfPapers:Tirkkonen2000}.
Notice that the $K$-by-$n_t$ rectangular matrices resulting from
column partitions of a dense and complete linear space-time block
encoding matrix $\mathbf{C}(\mathbf{s}_{\!K}\!)$ preserve this
property.
For simplicity, such matrices will also be, therefore, referred to
as dense and complete.

%
%
\begin{definition}[Rate]
\label{Rate} \quad %

The \emph{rate} of a linear space-time block code described by the
matrix $\mathbf{C}(\mathbf{s}_{\!K}\!)$ is defined as the ratio
between the number of independent raw symbols and the number of
linearly independent rows in $\mathbf{C}(\mathbf{s}_{\!K}\!)$. \hfill \QEDopen%
\end{definition}

Notice that, according to definition \ref{Rate}, a linear space-time
block code as defined here can, at most, achieve a unitary rate.
Therefore, the terms \emph{rate-one} and \emph{full-rate} will be
hereafter used interchangeably in reference to a linear STBCs of
rate $1$.

%
%
\begin{definition}[Real Orthogonality]
\label{DefinitionRealOrthogonality} \quad %

A pair of vectors $\mathbf{a}$ and $\mathbf{b}$ $\in \mathbb{C}^{n
\times 1}$ is said to be \emph{real orthogonal} to one another iff
\begin{equation}
\label{RealOrthogonality}%
\mathbf{a}^\textup{T}\!\cdot\!\mathbf{b} \equiv \mathbf{b}^\textup{T}\!\cdot\!\mathbf{a} = \sum\limits_{i=1}^{n}a_ib_i = 0.%
\vspace{-1em}
\end{equation}
\hfill \QEDopen
\end{definition}

Definition \ref{DefinitionRealOrthogonality} is introduced in order
make clear the distinction between the orthogonality property given
in equation \eqref{RealOrthogonality} and the notion of vector
orthogonality which is usually defined, in complex Hilbert spaces,
over the product of a vector against the transpose-conjugate of
another \cite{MyListOfPapers:Golub1996}.

%
%
\begin{definition}[Block-orthogonal STBCs]
\label{DefinitionBlockOrthogonalSTBC} \quad %

A matrix $\mathbf{C} \in \mathbb{C}^{n \times m}$ is said to be
\emph{block-orthogonal} iff there exists a permutation
$\boldsymbol{\mathcal{C}}$ of $\mathbf{C}$ such that
\begin{equation}
\boldsymbol{\mathcal{C}}\cdot\boldsymbol{\mathcal{C}}^\textup{H} =%
\left[%
\begin{array}{cc}%
\mathbf{D}_1 & \mathbf{0}\\
\mathbf{0}   & \mathbf{D}_2
\end{array}%
\right]\!,%
\end{equation}
where $^\textup{H}$ denotes Hermitian.

Furthermore, $\mathbf{C}$ is said to be \emph{real block-orthogonal}
iff a permutation $\boldsymbol{\mathcal{C}}$ exists such that
\begin{equation}
\boldsymbol{\mathcal{C}}\cdot\boldsymbol{\mathcal{C}}^\textup{T} =%
\left[%
\begin{array}{cc}%
\mathbf{D}_1 & \mathbf{0}\\
\mathbf{0}   & \mathbf{D}_2
\end{array}%
\right]\!.%
\vspace{-1.5em}
\end{equation}
\hfill \QEDopen
\end{definition}

Definitions $6$ and $7$ establish a slight generalization of the
concept of \emph{quasi}-orthogonality, in the sense presented in
\cite{MyListOfPapers:Tirkkonen2000, MyListOfPapers:Jafarkhani2001,
MyListOfPapers:Papadias2003}, with \emph{quasi}-orthogonality
resulting in the special case where the dimensions of $\mathbf{D}_1$
and $\mathbf{D}_2$ are identical.
The notion of real orthogonality will be used later in the
derivation of the GABBA orthogonal decoder, in section
\ref{OrthogonalDecoder}.

%
%
\begin{definition}[Manifold Matrices]
\label{DefinitionManifoldMatrices} \quad %

Consider the generating function defined below over the space of
complex-valued square matrices $\mathbf{A},\mathbf{B} \in
\mathbb{C}^{K \times K}$,
%
%
\begin{equation}
\label{ABBA1}%
X(\mathbf{A},\mathbf{B}) \triangleq%
\left[\!\!\!%
\begin{array}{rc}
   \mathbf{A} & \!\!\!\mathbf{B} \\
  -\mathbf{B} & \!\!\!\mathbf{A} \\
\end{array}%
\!\!\right]\!\!.%
\end{equation}

The \emph{$1^\textup{st}$ order ABBA manifold matrix} of the vector
$\mathbf{s}_2 = [s_1,s_2]$, denoted by $\mathbf{X}_{[s_1,s_2]}$, is
defined as the outcome of $X(s_1,s_2)$, with $s_i \in \mathcal{S}$.
Accordingly, the \emph{$2^\textup{nd}$ order ABBA manifold matrix}
of the symbol vector $\mathbf{s}_4\!=\![s_1,s_2,s_3,s_4]$ with
generating function $X$ is defined as the outcome of
$X(\mathbf{X}_{[s_1,s_2]},\mathbf{X}_{[s_3,s_4]})$ and denoted by
$\mathbf{X}^2_{[s_1,s_2,s_3,s_4]}$.

In general, the \emph{$n$-th order ABBA manifold matrix} of the
vector $\mathbf{s}_{\!K}$ with generating function $X$ is defined as
the outcome of
$X(\mathbf{X}^{(n-1)}_{\mathbf{s}_{\!K\!/2}},\mathbf{X}^{(n-1)}_{\tilde{\mathbf{s}}_{\!K\!/2}})$
and denoted by $\mathbf{X}^n_{\mathbf{s}_{\!K}}$, where
$\tilde{\mathbf{s}}_{\!K\!/2}$ is the \emph{complementary} symbol
given below
\begin{equation}
\label{ComplementarySymbolVector}
\tilde{\mathbf{s}}_{\!K\!/2} \triangleq [s_{\!K\!/2+1},\cdots,s_{\!K}]^\textup{T}.%
\vspace{-1em}
\end{equation} \hfill \QEDopen
\end{definition}

%
%
\newtheorem{lemma}{Lemma}
\begin{lemma}[Commutativeness of n-th Order Manifold Matrices]
\label{ManifoldCommutativeness} \quad

The $n$-th order manifold matrices of independent, equal-sized
vectors $\mathbf{a}_{\!K}$ and $\mathbf{b}_{\!K} \in \mathbb{C}^{K
\times 1}$ commute with respect to matrix product.
\end{lemma}
\begin{proof}
Let us rewrite $\mathbf{X}^n_{\mathbf{a}_{\!K}}$ and
$\mathbf{X}^n_{\mathbf{b}_{\!K}}$ in terms of their half
block-partitions:
\vspace{-0.5em}
\begin{equation}
\label{X1}
\mathbf{X}^n_{\mathbf{a}_K}\!\equiv\!\!%
\left[\!\!\!%
\begin{array}{cl}
    \ \ \ \ \!\mathbf{X}^{(n\!-\!1\!)}_{\mathbf{a}_{\!K\!/2}}   &\!\!\!\! \mathbf{X}^{(n\!-\!1\!)}_{\tilde{\mathbf{a}}_{\!K\!/2}}\\%
   -\mathbf{X}^{(n\!-\!1\!)}_{\tilde{\mathbf{a}}_{\!K\!/2}} &\!\!\!\! \mathbf{X}^{(n\!-\!1\!)}_{\mathbf{a}_{\!K\!/2}}
\end{array}%
\!\!\!\!\right]\!\!=\!\!%
\left[\!\!\!\!%
\begin{array}{rl}
    \ \mathbf{A}_1 &\!\!\!\! \mathbf{B}_1 \\
   -\mathbf{B}_1 &\!\!\!\! \mathbf{A}_1
\end{array}%
\!\!\!\!\right]\!\!,%
\vspace{-0.5em}
\end{equation}
\begin{equation}
\label{X2}
\mathbf{X}^n_{\mathbf{b}_K}\!\equiv\!\!%
\left[\!\!\!%
\begin{array}{cl}
    \ \ \ \ \!\mathbf{X}^{(n\!-\!1\!)}_{\mathbf{b}_{\!K\!/2}}   &\!\!\!\! \mathbf{X}^{(n\!-\!1\!)}_{\tilde{\mathbf{b}}_{\!K\!/2}}\\%
   -\mathbf{X}^{(n\!-\!1\!)}_{\tilde{\mathbf{b}}_{\!K\!/2}} &\!\!\!\! \mathbf{X}^{(n\!-\!1\!)}_{\mathbf{b}_{\!K\!/2}}
\end{array}%
\!\!\!\!\right]\!\!=\!\!%
\left[\!\!\!\!%
\begin{array}{rl}
    \mathbf{A}_2 &\!\!\!\! \mathbf{B}_2 \\
   -\mathbf{B}_2 &\!\!\!\! \mathbf{A}_2
\end{array}%
\!\!\!\right]\!\!.%
\end{equation}

Writing the products
$\mathbf{X}^n_{\mathbf{a}_K}\!\cdot\mathbf{X}^n_{\mathbf{b}_K}$ and
$\mathbf{X}^n_{\mathbf{b}_K}\!\cdot\mathbf{X}^n_{\mathbf{a}_K}$
explicitly and using standard block-partition properties we obtain,
\vspace{-0.5em}
\begin{equation}
\label{X1X2}
\mathbf{X}^n_{\mathbf{a}_{\!K}}\!\cdot\mathbf{X}^n_{\mathbf{b}_{\!K}}\!\!=\!%
\left[\!\!\!%
\begin{array}{cc}
    \mathbf{A}_1\!\cdot\!\mathbf{A}_2\!-\!\mathbf{B}_1\!\cdot\!\mathbf{B}_2   &\!\!\! \mathbf{A}_1\!\cdot\!\mathbf{B}_2\!+\!\mathbf{B}_1\!\cdot\!\mathbf{A}_2 \\
   -(\mathbf{A}_1\!\cdot\!\mathbf{B}_2\!+\!\mathbf{B}_1\!\cdot\!\mathbf{A}_2) &\!\!\! \mathbf{A}_1\!\cdot\!\mathbf{A}_2\!-\!\mathbf{B}_1\!\cdot\!\mathbf{B}_2
\end{array}%
\!\!\!\right]\!\!,%
\vspace{-0.1em}
\end{equation}
\begin{equation}
\label{X2X1}
\mathbf{X}^n_{\mathbf{b}_{\!K}}\!\cdot\!\mathbf{X}^n_{\mathbf{a}_{\!K}}\!\!=\!%
\left[\!\!\!%
\begin{array}{cc}
    \mathbf{A}_2\!\cdot\!\mathbf{A}_1\!-\!\mathbf{B}_2\!\cdot\!\mathbf{B}_1   &\!\!\! \mathbf{A}_2\!\cdot\!\mathbf{B}_1\!+\!\mathbf{B}_2\!\cdot\!\mathbf{A}_1 \\
   -(\mathbf{A}_2\!\cdot\!\mathbf{B}_1\!+\!\mathbf{B}_2\!\cdot\!\mathbf{A}_1) &\!\!\! \mathbf{A}_2\!\cdot\!\mathbf{A}_1\!-\!\mathbf{B}_2\!\cdot\!\mathbf{B}_1%
\end{array}%
\!\!\!\right]\!\!.%
\end{equation}

It is clear from equations \eqref{X1X2} and \eqref{X2X1} that the
condition for the $n$-th order manifold matrices
$\mathbf{X}^n_{\mathbf{a}_{\!K}}$ and
$\mathbf{X}^n_{\mathbf{b}_{\!K}}$ to commute is that their minors
$\mathbf{A}_1$, $\mathbf{B}_1$, $\mathbf{A}_2$ and $\mathbf{B}_2$
commute.
These minors are, in turn, $(n\!-\!\!1\!)$-th order manifold
matrices (see equations \eqref{X1} and \eqref{X2}) and, therefore,
commute under the same condition.
Since the minors of first-order ABBA manifold matrices are complex
scalars, the commutativeness of $\mathbf{X}^n_{\mathbf{a}_{\!K}}$
and $\mathbf{X}^n_{\mathbf{b}_{\!K}}$ with respect to matrix product
follows immediately, by induction.
\end{proof}

\subsection{GABBA Encoding Matrices}
\label{MotherCodeConstruction}

We can now introduce the major result of this section.
%
%
\newtheorem{theorem}{Theorem}
\begin{theorem}[Construction of GABBA Matrices]
\label{GABBAConstruction} \quad%

For any given number of transmit antennas $n_t \in \mathbb{N}^+$,
there exists a dense, complete and block-orthogonal GABBA matrix for
rate-one space-time block encoded transmission of symbols taken from
any complex-valued arbitrary constellation $\mathcal{S}$.
\end{theorem}

\begin{proof}
The proof is constructive.
For any given $n_t$, let $n\!=\!\lceil \log_2 n_t\!\rceil$ and $K =
2^n$, where $\lceil x \rceil$ denotes the smallest integer greater
or equal to $x$.
Next, construct the $(n\!\!-\!\!1\!)$-th order ABBA manifold matrix
$\mathbf{X}^{(n\!-\!1\!)}_{\mathbf{s}_{\!K\!/2}}$ of the symbol
vector $\mathbf{s}_{\!K\!/2}$, and the $(n\!\!-\!\!1\!)$-th ABBA
order manifold matrix
$\mathbf{X}^{(n\!-\!1\!)}_{\tilde{\mathbf{s}}_{\!K\!/2}}$ of the
complimentary symbol vector $\tilde{\mathbf{s}}_{\!K\!/2}$, both
with the generating function $X$ given in equation \eqref{ABBA1}.

Next, construct the dense and complete GABBA \emph{mother encoding
matrix} $\mathbf{C}(\mathbf{s}_{\!K}\!)$ by computing
\begin{equation}
\label{Equation}%
\mathbf{C}(\mathbf{s}_{\!K}\!) = \mathbf{Y}^n_{\!\mathbf{s}_{\!K}} = Y(\mathbf{X}^{(n\!-\!1\!)}_{\mathbf{s}_{\!K\!/2}},\mathbf{X}^{(n\!-\!1\!)}_{\tilde{\mathbf{s}}_{\!K\!/2}}), %
\end{equation}
with $Y(\mathbf{A},\mathbf{B})$ defined as
\begin{equation}
\label{ABBA2}%
Y(\mathbf{A},\mathbf{B}) \triangleq%
\left[\!\!\!%
\begin{array}{cl}
   \ \mathbf{A}            & \!\!\!\mathbf{B} \\
  -\mathbf{B}^\textup{H} & \!\!\!\mathbf{A}^\textup{H} \\
\end{array}%
\!\!\right]\!\!.%
\end{equation}

For obvious reasons, $Y(\mathbf{A},\mathbf{B})$ is referred to as
the \emph{wrapping} function.

Notice that the manifold matrices $\mathbf{X}^{(n\!-\!i\!)}$ are
constructed over $2^{(n\!-\!i\!)}$ vectors each carrying a mutually
exclusive partition of $\mathbf{s_K}$ tuples of symbols and,
furthermore, that at each iteration of $X$, these arguments are in
opposite quadrants of the resulting matrix.
The same occurs with the final ``wrapping'' operation, $i.e.$, the
application of equation \eqref{ABBA2}.
It is therefore clear that the GABBA mother encoding matrices are
dense and complete.

It is also evident that $\mathbf{Y}^n_{\!\mathbf{s}_{\!K}}$ are
square matrices of size $K = 2^n$.
Therefore, the capital scalar $K$ will, hereafter, exclusively
denote integer powers of two.
Rectangular ($K$--by--$n_t$) GABBA encoding matrix are obtained by
simply selecting any $n_t$ columns of
$\mathbf{C}(\mathbf{s}_{\!K}\!)$.
Since columnwise partition does not affect the density and
completeness of $\mathbf{C}(\mathbf{s}_{\!K}\!)$, GABBA encoding
matrices $\mathbf{C}_{\!n_t}$ are also dense and complete, which
concludes the proof.
\end{proof}

Hereafter we shall simplify the notation slightly and write
$\mathbf{C}_{\!K}$, instead of $\mathbf{C}(\mathbf{s}_{\!K}\!)$, for
the $K$-by-$K$ GABBA mother encoding matrix.
Accordingly, the matrix obtained by selecting $n_t$ columns of
$\mathbf{C}_{\!K}$ will be denoted by $\mathbf{C}_{\!n_t}$.

The following property is an immediate consequence of Theorem
\ref{GABBAConstruction}.
%
%
\newtheorem{corollary}{Corollary}
\begin{corollary}[Block-orthogonality of GABBA Encoding Matrices]
\label{GABBAQuasiOrthogonality} \quad%

GABBA mother encoding matrices are quasi-orthogonal and GABBA
encoding matrices are block-orthogonal.
\end{corollary}
\begin{proof}
The $K$--by--$K$ GABBA mother encoding matrix is given by
\begin{equation}
\label{GABBAMother}
\mathbf{C}(\mathbf{s}_{\!K}\!)\!\triangleq\!\!%
\left[\!\!\!%
\begin{array}{cc}
    \ \ \ \!\mathbf{X}^{(n\!-\!1\!)}_{\mathbf{s}_{\!K\!/2}}      &\!\!\!\!\! \mathbf{X}^{(n\!-\!1\!)}_{\tilde{\mathbf{s}}_{\!K\!/2}}\\%
    -\!\left(\!\mathbf{X}^{(n\!-\!1\!)}_{\tilde{\mathbf{s}}_{\!K\!/2}}\!\right)^\textup{\!\!H}%
                                                                 &\!\!\!\! \left(\mathbf{X}^{(n\!-\!1\!)}_{\mathbf{s}_{\!K\!/2}}\right)^\textup{\!\!H}
\end{array}%
\!\!\!\!\right]\!\!=\!\!%
\left[\!\!\!\!%
\begin{array}{cl}
    \ \mathbf{A} &\!\!\!\! \mathbf{B} \\
   -\mathbf{B}^\textup{H} &\!\!\!\! \mathbf{A}^\textup{\!H}
\end{array}%
\!\!\!\right]\!\!.%
\end{equation}

Writing the product
$\mathbf{C}(\mathbf{s}_{\!K}\!)\!\cdot\!\mathbf{C}(\mathbf{s}_{\!K}\!)^\textup{H}$
explicitly, and after trivial algebraic manipulations, we obtain
\begin{equation}
\label{CkCkH}
\mathbf{C}(\mathbf{s}_{\!K}\!)\!\cdot\!\mathbf{C}(\mathbf{s}_{\!K}\!)^\textup{H}\!=\!%
\left[\!\!\!%
\begin{array}{cc}
   \mathbf{A}\!\cdot\!\mathbf{A}^\textup{\!H} + \mathbf{B}\!\cdot\!\mathbf{B}^\textup{\!H}  & \!\!\! \mathbf{B}\!\cdot\!\mathbf{A} -  \mathbf{A}\!\cdot\!\mathbf{B}\\
   (\mathbf{B}\!\cdot\!\mathbf{A} -  \mathbf{A}\!\cdot\!\mathbf{B})^\textup{\!H}            & \!\! \mathbf{A}\!\cdot\!\mathbf{A}^\textup{\!H} + \mathbf{B}\!\cdot\!\mathbf{B}^\textup{\!H}
\end{array}%
\!\!\!\right]\!\!.%
\end{equation}

Equation \eqref{CkCkH} indicates that the \emph{quasi}-orthogonality
of GABBA mother matrices is conditioned on the commutativeness of
its minors $(\mathbf{A}, \mathbf{B})$ and, therefore, follows
directly from Lemma $1$.

Rectangular ($K$--by--$n_t$) GABBA matrices $\mathbf{C}_{\!n_t}$ are
obtained by arbitrarily selecting $n_t$ columns of
$\mathbf{C}(\mathbf{s}_{\!K}\!)$, that is, by selection a total of
$n_t$ column vectors out of the sets
$\{\mathbf{c}_{\downarrow,1}\!(\mathbf{s}_{\!K}), \cdots,
\mathbf{c}_{\downarrow,K\!/2}\!(\mathbf{s}_{\!K})\}$ and
$\{\mathbf{c}_{\downarrow,{1+K\!/2}}\!(\mathbf{s}_{\!K}), \cdots,
\mathbf{c}_{\downarrow,K}\!(\mathbf{s}_{\!K})\}$.
Since this columnwise partition does not affect the orthogonality
between these two sets, the corresponding mother encoding matrix,
$\mathbf{C}_{\!n_t}$ are obviously \emph{block}-orthogonal, which
concludes the proof.
\end{proof}

An example of a GABBA mother encoding matrix with $K=32$ and a
Matlab implementation of the encoder described in this section are
given in Appendix \ref{Appendix_GABBAEncodingMatrix}.
The reader is encouraged to inspect the example provided and to
utilize the program in order to generate GABBA matrices of various
sizes and verify their block-orthogonality\footnote{Instructions on
how to generate a GABBA encoding matrix of symbolic variables are
given as comments in the Matlab code provided. Note that the
Symbolics Tool Box is required. To encode a numerical symbol vector
{\tt s}, simply call the function as usual giving {\tt s} as input.
To inspect the block orthogonality of the GABBA encoding matrices,
first generate a symbolic GABBA mother encoding matrix $<${\tt C}$>$
and then type $<${\tt C'*C}$>$.}.

\section{Quasi-Orthogonal Decoding GABBA Codes}
\label{QausiOrthogonalDecoder}%

\subsection{Preliminaries}%
\label{DecoderPreliminaries}%

Consider the problem of decoding the received signal vector
corresponding to the transmission of a GABBA matrix
$\mathbf{C}_{n_t}$ through a memoryless block-fading channel
$\mathbf{h}_{\!n_t}$ in the presence of an additive noise vector
$\mathbf{n}_{\!K}$.
At a given receive antenna, such a received signal vector can be
modeled as
\begin{equation}
\label{ReceivedSignalVector}%
\mathbf{r} = \mathbf{C}_{n_t} \!\cdot \mathbf{h}_{\!n_t} + \mathbf{n}_{\!K} = \mathbf{H}_{\!K} \cdot \bar{\mathbf{s}}_{\!K} + \mathbf{n}_{\!K},%
\end{equation}
where $\bar{\mathbf{s}}_{\!K}$ is the \emph{augmented symbol
vector}, defined as
\begin{equation}
\label{AugmentedSymbolVector}
\bar{\mathbf{s}}_{\!K} \triangleq [s_1,\cdots,s_K,s_1^{*},\cdots,s_K^{*}]^\textup{T},%
\end{equation}
where $^{*}$ denotes complex conjugate.

Examples of the use of alternative representations of the linear
STBC receive vector, similar to that given in equation
\eqref{ReceivedSignalVector}, can be found in the literature.
For instance, in \cite{MyListOfPapers:Li2001}, the idea was employed
in order derive a simpler decoder for OSTBCs, and in
\cite{MyListOfPapers:Papadias2003}, the transformation  was used to
construct the \emph{quasi}-orthogonal decoder for the $4$-by-$4$
code there proposed.

We have shown, however, that this technique can be expanded well
beyond the relatively straightforward applications suggested in
\cite{MyListOfPapers:Li2001} and \cite{MyListOfPapers:Papadias2003}.
In particular, we have shown in
\cite{MyListOfPapers:Abreu-ISIT-2003,
MyListOfPapers:Giuseppe-IEE2004} that a more elaborate utilization
of these transformation can be exploited to construct a novel
decoder for OSTBCs which enables its use in time-selective fading
channels.
In this section, we build on the concepts laid in
\cite{MyListOfPapers:Abreu-ISIT-2003,
MyListOfPapers:Giuseppe-IEE2004} in order to show that GABBA codes
are indeed orthogonally decodable.

\subsection{Structure of GABBA Encoded Channel Matrices}
\label{StructureOfGABBAEncodedChannelMatrices}%

Equation \eqref{ReceivedSignalVector} indicates that
$\mathbf{H}_{\!K}$ are constructed by simply re-arranging the
corresponding channel entries appropriately.
These matrices are, therefore, hereafter referred to as the GABBA
\emph{encoded channel matrices}.

Begin by noticing that when the transmitter has a number $n_t\!<\!K$
of antennas, an $n_t$ column-partition of the corresponding
$K$-by-$K$ GABBA mother encoding matrix is utilized.
Mathematically, this is equivalent to the case when $K-n_t$ columns
of $\mathbf{C}_{\!K}$ and rows of $\mathbf{h}$ (corresponding to the
``unused'' antennas) assume zero values\footnote{This is the reason
why we maintain the notation $\mathbf{H}_{\!K}$ in equation
\eqref{ReceivedSignalVector}}, leading to a sparse
$\mathbf{H}_{\!K}$, but leaving unaffect its real
\emph{quasi}-orthogonality properties.
It is, therefore, sufficient to consider mother matrices
$\mathbf{H}_{\!K}$ when studying the properties of GABBA encoded
channel matrices $\mathbf{H}_{\!n_t}$.

Before we proceed, let us introduce some additional terminology .
%
%
\begin{definition}[Notation for Matrix Partitions]
\label{DefinitionPartitions} \quad%

The symbol $[\mathbf{A}]^{\textup{P}:p}$ denotes the \emph{left}
(P=L), \emph{right} (P=R), \emph{upper} (P=U) or \emph{lower} (P=D)
partition of the matrix $\mathbf{A}$, containing its $p$
\emph{leftmost}, \emph{rightmost},\emph{uppermost} and
\emph{lowermost} columns/rows, respectively.
\end{definition}
%
%
\begin{definition}[Notation for Matrix Reorientation]
\label{DefinitionFlip} \quad%

The symbols $\mathbf{A}^{\!\dagger}$ and $\mathbf{A}^{\!\ddagger}$
denote, respectively, the \emph{bottom-up} and the
\emph{left-to-right} reorientations of a matrix $\mathbf{A}$ (a.k.a.
\emph{flip upside-down} and \emph{flip left-right} operations).
Trivial properties of these operations are:
\begin{subequations}
\begin{eqnarray}
\label{FlipUDProperty}%
&\left(\mathbf{A}^{\!\dagger}\right)^{\!\dagger} = \mathbf{A},&\\[-0.5ex] %
\label{FlipUDProductProperty}%
&\mathbf{A}^{\!\dagger}\!\cdot\!\mathbf{b} =
\left(\mathbf{A}\!\cdot\!\mathbf{b}\right)^{\dagger} =
\mathbf{A}^{\!\dagger\ddagger}\!\cdot\!\mathbf{b}^{\dagger} \!\!.&
\end{eqnarray}
where $\mathbf{A} \in \mathbb{C}^{n \times m}$ and $\mathbf{b} \in \mathbb{C}^{m \times 1}$.%
\end{subequations}
\end{definition}

Next, observe that due to the structures of $\mathbf{C}_{\!K}$ and
$\bar{\mathbf{s}}_{\!K}$, $\mathbf{H}_{\!K}$ can be written as
\begin{equation}
\label{EncodedChannelStructure}%
\mathbf{H}_{\!K} =
\left[%
\begin{array}{cc}
  \mathbf{H}_{\!K_1} \!\!  & \!\!   \mathbf{0}   \!\!\\%
  \mathbf{0}   \!\!  & \!\!   \mathbf{H}_{\!K_2} \!\!\\%
\end{array}%
\right]\!\!,%
\end{equation}
where $\mathbf{H}_{\!K_1}$ and $\mathbf{H}_{\!K_2}$ are both
$K\!/2$--by--$K$ matrices.

The following result can then be stated and proved.

\begin{lemma}[Construction of GABBA Encoded Channel Matrices]
\label{ConstructionOfGABBAEncodedChannelMatrices} \quad%

The non-zero minors $\mathbf{H}_{\!K_1}$ and $\mathbf{H}_{\!K_2}$ of
the GABBA encoded channel matrix $\mathbf{H}_{\!K}$ are given by
\begin{eqnarray}
&\mathbf{H}_{\!K_1} = [\mathbf{H}^{n}_{\mathbf{h}_{\!K}^{+}}]^{\textup{U}:K\!/2}\!,&\\ %
&\mathbf{H}_{\!K_2} = [\tilde{\mathbf{H}}^{n}_{\mathbf{h}_{\!K}^{\circlearrowright}}]^{\textup{U}:K\!/2}\!,& %
\end{eqnarray}
where $\mathbf{H}^{n}_{\mathbf{h}_{\!K}^{+}}$ and
$\tilde{\mathbf{H}}^{n}_{\mathbf{h}_{\!K}^{\circlearrowright}}$ are
$n$-th order manifold matrices of the \emph{extended channel vector}
$\mathbf{h}_{\!K}^{+}$ and the \emph{modified channel vector}
$\mathbf{h}^\circlearrowright_{\!K}$, respectively,
\begin{eqnarray}
\label{AugmentedChannelVector} %
&\mathbf{h}_{\!K}^{+} \triangleq [\underbrace{h_1,\cdots,0,\cdots,h_{k},\cdots,0,\cdots,h_{n_t},\cdots,0}_{\mathbf{h}_{n_t} \textup{ extended with } K-n_t \textup{ zero-entries}}]^\textup{T}\!,& \\ %
\label{ModifiedChannelVector}
&\mathbf{h}^\circlearrowright_{\!K} \triangleq [[\mathbf{h}_{\!K}^{+}]^{\textup{D}:K\!/2} \ [\mathbf{h}_{\!K}^{+}]^{\textup{U}:K\!/2}] %
                                    = [\underbrace{h_{\!K\!/2+1},\cdots,0,\cdots,h_{n_t},\cdots,0}_{\textup{last } K\!/2 \textup{ entries of } \mathbf{h}_{\!K}^{+}}%
                                       \underbrace{,h_1,\cdots,h_k,\cdots,0,\cdots,h_{\!K\!/2}}_{\textup{first } K\!/2 \textup{ entries of } \mathbf{h}_{\!K}^{+}}]^\textup{T}\!,& %
\end{eqnarray}
with the corresponding generating functions $H$ and $\tilde{H}$
given by
\begin{eqnarray}
\label{ABBA3}%
&H(\mathbf{A},\mathbf{B}) \triangleq%
\left[\!\!\!%
\begin{array}{cr}
   \mathbf{A} & \!\!\!\mathbf{B} \\
   \mathbf{B} & \!\!\!-\mathbf{A} \\
\end{array}%
\!\!\right]\!\!,&\\ %
\label{ABBA4}%
&\tilde{H}(\mathbf{A},\mathbf{B}) \triangleq%
\left[\!\!\!%
\begin{array}{cr}
   \mathbf{A} & \!\!\!-\mathbf{B} \\
   \mathbf{B} & \!\!\! \mathbf{A} \\
\end{array}%
\!\!\right]\!\!.%
\end{eqnarray}
\end{lemma}

\begin{proof}
The proof is constructive.
The noise-independent terms of the first $K/2$ entries of
$\mathbf{r}$ can be written as
%
%
\begin{eqnarray}
\label{H1Structure}
%
%
&[\mathbf{C}_{\!n_t}]^{\textup{U}:\tfrac{K}{2}} \!\cdot \mathbf{h}\!\!=\!\!\!%
\left[
\begin{tabular}{cc}
$\begin{array}{rl}
\!\!\!\!\!\!  s_1 & \!\!\!s_2\\%
\!\!\!\!\!\! -s_2 & \!\!\!s_1
\end{array}$ \!\!\!\!\vline \!\!\!\!&\!\!\!\!  $\left[[\mathbf{C}_{\!K}]^{\textup{U}:2}\right]^{\!\textup{R}:(\!K-\!2)}$\!\!\!\!\\%
\hline\\[-4ex]%
\multicolumn{2}{c}{$\big[[\mathbf{C}_{\!K}]^{\textup{U}:\tfrac{K}{2}}\big]^{\textup{D}:2}$}\!\!\!\!\\
\end{tabular}%
\right]\!\!\cdot\!\!%
\left[%
\begin{array}{c}
\!\!\!h_1\!\!\!\!\!\\
\!\!\!h_2\!\!\!\!\!\\
\hline\\[-4ex]%
\!\!\![\mathbf{h}]^{\textup{D}:(\!K-\!2)}\!\!\!\!\!\\%
\end{array}
\right]\!\!\!=\!\!%
%
%
\left[
\begin{tabular}{cc}
$\begin{array}{lr}
\!\!\!\!\!\!\!  h_1 & \!\!\!\! h_2\\%
\!\!\!\!\!\!\!  h_2 & \!\!\!\!-h_1
\end{array}$ \!\!\!\!\vline \!\!\!\!&\!\!\!\!  $\left[[\mathbf{H}_{\!K_1}]^{\textup{U}:2}\right]^{\textup{R}:(\!K-\!2)}$\!\!\!\!\\%
\hline\\[-4ex]%
\multicolumn{2}{c}{$[\mathbf{H}_{\!K_1}]^{\textup{D}:(\!K\!/2-\!2)}$}\!\!\!\!\\
\end{tabular}%
\right]\!\!\cdot\!\!%
\left[\!%
\begin{array}{c}
\!\!s_1\!\!\!\!\\
\!\!s_2\!\!\!\!\\
\hline\\[-4ex]%
\!\![\mathbf{s}]^{\textup{D}:(\!K-\!2)}\!\!\!\!\\%
\end{array}
\!\right]\!\!\!=\!\mathbf{H}_{\!K_1}\!\cdot\mathbf{s}_{\!K}.&\nonumber\\[-3.5ex]
\end{eqnarray}

Equation \eqref{H1Structure} derives directly from the structure of
$\mathbf{C}_{\!K}$ (see section \ref{MotherCodeConstruction}), and
reveals that $\mathbf{H}_{\!K_1}$ is the upper half-partition of the
$n$-th order manifold matrix of $\mathbf{h}_{\!K}^{+}$, with the
generating function given by equation \eqref{ABBA3}.

Analogously, the noise-independent terms of the last $K/2$ entries
of $\mathbf{r}$ (flipped upside-down), can be written as
%
%
\begin{eqnarray}
\label{H2Structure}
%
%
&&\hspace{-2em}\big[[\mathbf{C}_{\!n_t}]^{\textup{D}:\tfrac{K}{2}}\big]^{\!\dagger}
\!\cdot \mathbf{h}\! = \nonumber\\%
&&\hspace{-1.5em}=\!\!\left[\!
\begin{tabular}{cccc}
$\begin{array}{ll}
\!\!\!\!\!\! -\!s_{\!K}^*     & \!\!\!\!\!\!\!\!-\!s_{\!K-\!1}^*\\%
\!\!\!\!\!\! -\!s_{\!K-\!1}^* & \!\!\!\!\!         s_{\!K}^*
\end{array}$ \!\!\!\!\!\vline \!\!\!\!&\!\!\!\!
$[\mathbf{C}_{\!K}^\dagger]
{%
    \!\!\!\tiny
    \begin{array}{l}
    \textup{U}\!:\!2\\
    \textup{R}\!:\!\!K\!-\!2\\
    \textup{L}\!:\!\!\frac{K}{2}\!\!-\!2
    \end{array}
}$ \!\!\!\!\!\! \vline \!\!\!&\!\!\!\!\!%
$\begin{array}{lr}
\! s_{\!\tfrac{K}{2}\!}^*   & \!\!\!\!\!\!   s_{\!\tfrac{K}{2}-\!1}^*\\%
\! s_{\!\tfrac{K}{2}-\!1}^* & \!\!\!\!\!\!-\!s_{\!\tfrac{K}{2}\!}^*
\ \ \ \
\end{array}$ \!\!\!\!\!\!\! \vline \!\!\!&\!\!\!\!\!
$[\mathbf{C}_{\!K}^\dagger]
{%
    \!\!\!\tiny
    \begin{array}{l}
    \textup{U}\!:\!2\\
    \textup{R}\!:\!\!\tfrac{K}{2}\!\!-\!2
    \end{array}
}$\!\!\!\!\!\!\!\!\\[4ex]%
\hline\\[-4ex]%
\multicolumn{4}{c}{$[\mathbf{C}_{\!K}^{\!\dagger}]
{%
    \!\!\!\tiny
    \begin{array}{l}
    \textup{U}\!:\!\!K/2\\
    \textup{D}\!:\!\!K/2\!\!-\!2
    \end{array}
}$}\\
\end{tabular}%
\right]\!\!\cdot\!%
\mathbf{h}_{\!K}\!\!=\!\!%
%
%
\left[
\begin{tabular}{clcl}
$\begin{array}{rr}
\!\!\!\!\!\!\!\!-\!h_1 & \!\!\!\!\! -\!h_2\\%
\!\!\!\!\!\!\!\!   h_2 & \!\!\!\!\! -\!h_1
\end{array}$ \!\!\!\!\! \vline \!\!\!\!&\!\!\!\!
$[\mathbf{H}_{\!K_2}^{\dagger\ddagger}]
{%
    \!\!\!\tiny
    \begin{array}{l}
    \textup{U}\!:\!2\\
    \textup{R}\!:\!\!K\!-\!2\\
    \textup{L}\!:\!\!\tfrac{K}{2}\!\!-\!\!2
    \end{array}
}$ \!\!\!\!\!\! \vline \!\!\!\!\! &\!\! %
$\begin{array}{rl}
\!\!\!\!   h_{\!\tfrac{K}{2}+1} & \!\!\!\!\! h_{\!\tfrac{K}{2}+2}\\%
\!\!\!\!-\!h_{\!\tfrac{K}{2}+2} & \!\!\!\!\! h_{\!\tfrac{K}{2}+1}
\end{array}$ \!\!\!\!\!\! \vline \!\!\!\!&\!\!\!\!
$[\mathbf{H}_{\!K_2}^{\dagger\ddagger}]
{%
    \!\!\!\tiny
    \begin{array}{l}
    \textup{U}\!:\!2\\
    \textup{R}\!:\!\!\tfrac{K}{2}\!\!-\!\!2
    \end{array}
}$ \!\!\!\!\!\!\!\!\!\!\\[4ex]%
\hline\\[-4ex]%
\multicolumn{4}{c}{$[\mathbf{H}_{\!K_2}^{\dagger\ddagger}]^{{\scriptscriptstyle \textup{D}:\!K\!/2\!-\!2}}$}\!\!\!\!\\
\end{tabular}%
\right]\!\!\cdot\!%
\mathbf{s}^{*\!\dagger}
\!\!=\nonumber\\
&&\hspace{4em} =
\mathbf{H}_{\!K_2}^{\dagger\ddagger}\!\cdot\!\mathbf{s}^{*\dagger} =%
\left(\mathbf{H}_{\!K_2}\!\cdot\!\mathbf{s}^*\right)^\dagger =%
\mathbf{H}_{\!K_2}^{\dagger}\!\cdot\!\mathbf{s}^{*}.%
\end{eqnarray}

From equation \eqref{H2Structure} it is found that
$\mathbf{H}_{\!K_2}$ has the following structure:
\begin{eqnarray}
\label{H2StructureSimple}
%
%
\mathbf{H}_{\!K_2} = \left[
\begin{tabular}{clcl}
$\begin{array}{rr}
\!\!\!\! h_{\!K\!/2+1} & \!\!\!\!\!-\!h_{\!K\!/2+2}\\%
\!\!\!\! h_{\!K\!/2+2} & \!\!\!\!\!   h_{\!K\!/2+1}
\end{array}$ \!\!\!\!\! \vline \!\!\!\!&\!\!\!\!
$[\mathbf{H}_{\!K_2}]
{%
    \!\!\!\tiny
    \begin{array}{l}
    \textup{U}\!:\!2\\
    \textup{R}\!:\!\!K\!-\!2\\
    \textup{L}\!:\!\!K\!/2\!\!-\!\!2
    \end{array}
}$ \!\!\!\!\!\! \vline \!\!\! &\!\! %
$\begin{array}{rr}
\!\!\!\!\!\!-\!h_1 & \!\!\!    h_2\\%
\!\!\!\!\!\!-\!h_2 & \!\!\! -\!h_1
\end{array}$ \!\!\!\!\!\! \vline \!\!\!\!&\!\!\!\!
$[\mathbf{H}_{\!K_2}]
{%
    \!\!\!\tiny
    \begin{array}{l}
    \textup{U}\!:\!2\\
    \textup{R}\!:\!\!K\!/2\!\!-\!\!2
    \end{array}
}$ \!\!\!\!\!\!\!\!\!\!\\[2ex]%
\hline\\[-4ex]%
\multicolumn{4}{c}{$[\mathbf{H}_{\!K_2}]^{{\scriptscriptstyle \textup{D}:\!K\!/2\!-\!2}}$}\!\!\!\!\\
\end{tabular}%
\right]\!\!.
\end{eqnarray}

Equation \eqref{H2StructureSimple} finally reveals that
$\mathbf{H}_{\!K_2}$ is the upper half-partition of the $n$-th order
manifold matrix of $\mathbf{h}^\circlearrowright_{\!K}$, with the
generating function given in equation \eqref{ABBA4}, which concludes
the proof.
\end{proof}

Lemma \ref{ConstructionOfGABBAEncodedChannelMatrices} gives an
algorithm to construct GABBA encoded channel matrices of any size,
and reveals the similarity in the construction of the GABBA mother
encoding, and encoded channel matrices.
As an example, the minors $\mathbf{H}_{\!K_1}$ and
$\mathbf{H}_{\!K_2}$ of the encoded channel matrix corresponding to
$\mathbf{C}_{32}$ are shown in Appendix
\ref{Appendix_GABBAEncodedChannelMatrix}.
A Matlab implementation of the algorithm is also provided.
The reader is again encouraged to inspect the example, as well as to
utilize the program to generate GABBA encoded channel matrices of
various sizes and verify the veracity of equation
\eqref{ReceivedSignalVector}.

\subsection{Properties of GABBA Encoded Channel Matrices}
\label{PropertiesOfGABBAEncodedChannelMatrices}

%

Let us now introduce the following property of GABBA encoded channel
matrices.
\begin{lemma}[Quasi-orthogonality of GABBA Encoded Channel Matrices]
\label{QuasiOrthogonalityOfGABBAEncodedChannelMatrices} \quad

The following property holds and is referred to as the
\emph{quasi-orthogonality} of GABBA Encoded Channel
Matrices\footnote{This represents a slight abuse of the terminology
introduced earlier, given that the property is defined over a sum of
matrix products. We believe, however, that this does not contribute
to any confusion and simplifies that language somewhat. The reason
why such a product sum of $\mathbf{H}_{\!\mathbf{h}_{\!K}^{+}}^n$
and $\mathbf{H}_{\!\mathbf{h}^\circlearrowright_{\!K}}^n$ is
considered is to became clear soon, when the procedure for
orthogonal decoding is introduced.}:
\begin{equation}
\label{H_Product}%
\mathbf{H}_{\!\mathbf{h}_{\!K}^{+}}^{n \textup{H}}\!\cdot\!\mathbf{H}_{\!\mathbf{h}_{\!K}^{+}}^n + \mathbf{H}_{\!\mathbf{h}^\circlearrowright_{\!K}}^{n \textup{T}}\!\cdot\!\mathbf{H}_{\!\mathbf{h}^\circlearrowright_{\!K}}^{n *} \triangleq \mathbf{P} = %
\left[\!\!\!%
\begin{array}{cc}
   \hat{\mathbf{H}}_{1} & \!\!\!\mathbf{0} \\
   \mathbf{0}           & \!\!\! \hat{\mathbf{H}}_{2} \\
\end{array}%
\!\!\right]\!\!,%
\end{equation}
where $\hat{\mathbf{H}}_{1}$ and $\hat{\mathbf{H}}_{2}$ are
$K\!/2$--by--$K\!/2$ matrices.
\end{lemma}
\begin{proof}
Let $\mathbf{A}_{i}$ and $\mathbf{B}_{i}$ be manifold matrices of
the sub-vectors $[\mathbf{h}_{\!K}^{+}]^\textup{U:K/2}$ and
$[\mathbf{h}^\circlearrowright_{\!K}]^\textup{U:K/2}$, such that:
\begin{enumerate}[$a$)]
\item $\mathbf{A}_{1}$ and $\mathbf{B}_{1}$ are structured as described by the generating function given in equation \eqref{ABBA3};\\[-3.5ex] %
\item $\mathbf{A}_{2}$ and $\mathbf{B}_{2}$ are structured as described by the generating function given in equation \eqref{ABBA4};\\[-3.5ex] %
\item $\mathbf{A}_{i}$ depend on the $K\!/2$-tuple $\mathbf{h}_a\!\triangleq\![h_1,\cdots,h_{\!K\!/2}]$, while $\mathbf{B}_{i}$ depend on the $K\!/2$-tuple %
      $\mathbf{h}_b\!\triangleq\![h_{\!K\!/2+1},\cdots,h_{\!K}]$.%
\end{enumerate}

From equations \eqref{ABBA3} and \eqref{ABBA4} we then have
\begin{equation}
\label{StructureOfP}%
\mathbf{P}\!=\!\!%
\left[\!\!\!%
\begin{array}{ll}
   \hat{\mathbf{H}}_{1}        & \hat{\mathbf{P}} \\
   \hat{\mathbf{P}}^\textup{H} & \hat{\mathbf{H}}_{2} \\
\end{array}%
\!\!\right]\!\!,%
\end{equation}
where
\begin{eqnarray}
\label{StructureOfH1_Hat}%
&\hat{\mathbf{H}}_1 =|\mathbf{A}_{1}\! |^2\!\!+|\mathbf{B}_{1}\!|^2\!+%
              \!\mathbf{A}_{2}^\textup{\!T}\!\cdot\!\mathbf{A}_{2}^*\!\!+\!\mathbf{B}_{2}^\textup{T}\!\cdot\!\mathbf{B}_{2}^{*},& \\%
\label{StructureOfH2_Hat}%
&\hat{\mathbf{H}}_2 =|\mathbf{A}_{\!2}\! |^2\!\!+|\mathbf{B}_{2}|^2\!+%
              \!\mathbf{A}_{1}^\textup{\!T}\!\cdot\!\mathbf{A}_{1}^*\!\!+\!\mathbf{B}_{1}^\textup{T}\!\cdot\!\mathbf{B}_{1}^{*},& \\%
\label{StructureOfP_Hat}%
&\hat{\mathbf{P}}= \hat{\mathbf{P}}_{1} - \hat{\mathbf{P}}_{1}^\textup{\!H},&\\ %
\label{StructureOfP_Hat1}%
&\hat{\mathbf{P}}_{1} \triangleq \mathbf{A}_{1}^\textup{\!H}\!\cdot\!\mathbf{B}_{1}\!\!+\!\mathbf{A}_{2}^\textup{T}\!\cdot\!\mathbf{B}_{2}^{*},& \\[-4.5ex]\nonumber %
\end{eqnarray}

Form equation \eqref{StructureOfP_Hat} it is seen hat the
\emph{quasi}-orthogonality of the GABBA encoded channel matrices, as
defined in equation \eqref{H_Product}, can result from two
independent, and both sufficient, conditions:
\begin{itemize}
\item[$i$)]  $\hat{\mathbf{P}}_{1} = \hat{\mathbf{P}}_{1}^\textup{H}$ (\textup{``weak'' condition})
\item[$ii$)] $\hat{\mathbf{P}}_{1} = \mathbf{0}$             (\textup{``strong'' condition})
\end{itemize}

The commutativeness of manifold matrices established by Lemma
\ref{ManifoldCommutativeness} implies\footnote{Although the
generating function considered in Lemma
\ref{ManifoldCommutativeness} differs from those given by equations
\eqref{ABBA3} and \eqref{ABBA4}, the extension of the proof to these
generating functions is trivial.} that $\hat{\mathbf{P}}_{1} =
\hat{\mathbf{P}}_{1}^\textup{T}$.
It is therefore evident that the weak condition is only satisfied if
either the matrices $\mathbf{A}_{i}$ and $\mathbf{B}_{i}$ or,
equivalently, $\mathbf{C}_{\!K}$ are/is real-valued\footnote{In the
later case because then the linear combination of the received
signal vector can be performed independently over the in-phase and
quadrature components of the channel vector $\mathbf{h}_{n_t}$,
splitting the corresponding encoded channel matrix
$\mathbf{H}_{\!K}$ into real and imaginary parts.}.

The strong condition $ii$, on the other hand, is generally
satisfied, as shown below.

Let $\mathbf{A}_{1}\!=\!H(\mathbf{A}_{1,1},\mathbf{B}_{1,1}),
\mathbf{B}_{1}\!=\!H(\mathbf{A}_{1,2},\mathbf{B}_{1,2}),
\mathbf{A}_{2}\!=\!\tilde{H}(\mathbf{A}_{2,1},\mathbf{B}_{2,1}),
\mathbf{B}_{2}\!=\!\tilde{H}(\mathbf{A}_{2,2},\mathbf{B}_{2,2})$,
where $\mathbf{A}_{1,j}$ and $\mathbf{B}_{1,j}$ are build over the
the first and second half partitions of $\mathbf{h}_a$,
respectively, while $\mathbf{A}_{2,j}$ and $\mathbf{B}_{2,j}$ are
build over the first and second half partitions of $\mathbf{h}_b$,
respectively.
Then, $\hat{\mathbf{P}}_{1}$ can be written as:
\begin{equation}
\label{P_HatMoreExplicitly}%
\hat{\mathbf{P}}_{1} = %
\overbrace{%
\left[\!\!\!%
\begin{array}{cr}
   \mathbf{A}_{1,1}^\textup{H} & \!\!\! \mathbf{B}_{1,1}^\textup{H} \\
   \mathbf{B}_{1,1}^\textup{H} & \!\!\!-\mathbf{A}_{1,1}^\textup{H} \\
\end{array}%
\!\!\!\right]}^{\scriptstyle \textup{dependent on} \atop [h_1,\cdots,h_{\!K\!/\!4}]} %
\!\!\cdot\!\!%
\underbrace{%
\left[\!\!\!%
\begin{array}{cr}
   \mathbf{A}_{1,2} & \!\!\! \mathbf{B}_{1,2} \\
   \mathbf{B}_{1,2} & \!\!\!-\mathbf{A}_{1,2} \\
\end{array}%
\!\!\!\right]}_{\scriptstyle \textup{dependent on} \atop [h_{\!K\!/\!4\!+1},\cdots,h_{\!K\!/\!2}]} %
\!+%
\overbrace{%
\left[\!\!\!%
\begin{array}{rc}
    \mathbf{B}_{2,1}^\textup{T} & \!\!\! \mathbf{A}_{2,1}^\textup{T} \\
   -\mathbf{A}_{2,1}^\textup{T} & \!\!\! \mathbf{B}_{2,1}^\textup{T} \\
\end{array}%
\!\!\!\right]}^{\scriptstyle \textup{dependent on} \atop [h_{\!K\!/\!2\!+1},\cdots,h_{\!3K\!/\!4}]} %
\!\!\cdot\!\!%
\underbrace{%
\left[\!\!\!%
\begin{array}{cr}
   \mathbf{B}_{2,2}^* & \!\!\!-\mathbf{A}_{2,2}^* \\
   \mathbf{A}_{2,2}^* & \!\!\! \mathbf{B}_{2,2}^* \\
\end{array}%
\!\!\!\right]}_{\scriptstyle \textup{dependent on} \atop [h_{3\!K\!/\!4\!+1},\cdots,h_{\!K}]}\!=%
\left[\!\!\!%
\begin{array}{cr}
   \hat{\mathbf{P}}_{1,1} & \!\!\! \hat{\mathbf{P}}_{1,2} \\
   \hat{\mathbf{P}}_{2,1} & \!\!\! \hat{\mathbf{P}}_{2,2} \\
\end{array}%
\!\!\!\right]\!\!. %
\end{equation}

From equations \eqref{H_Product}, \eqref{StructureOfP},
\eqref{StructureOfP_Hat1} and \eqref{P_HatMoreExplicitly}, and due
to the properties $a$, $b$ and $c$ listed above, it is evident that
the sum-product structure of $\mathbf{P}$ propagates to
$\hat{\mathbf{P}}$, to $\hat{\mathbf{P}}_{1}$, to its blocks
$\hat{\mathbf{P}}_{i,j}$, and so fourth.
For example $\hat{\mathbf{P}}_{1,1}$ is given by
\begin{eqnarray}
\hat{\mathbf{P}}_{1,1}&\!\!\!\!=\!\!\!\!&(\mathbf{A}_{1,1}^\textup{\!H}\!\cdot\!\mathbf{A}_{1,2}\!+\!\mathbf{B}_{1,1}^\textup{H}\!\cdot\!\mathbf{B}_{1,2}\!+\!%
\mathbf{B}_{2,1}^\textup{T}\!\cdot\!\mathbf{B}_{2,2}^*\!-\!\mathbf{A}_{2,1}^\textup{\!T}\!\cdot\!\mathbf{A}_{2,2}^*)\!-\!%
(\mathbf{A}_{1,1}^\textup{\!H}\!\cdot\!\mathbf{A}_{1,2}\!+\!\mathbf{B}_{1,1}^\textup{H}\!\cdot\!\mathbf{B}_{1,2}\!+\!%
\mathbf{B}_{2,1}^\textup{T}\!\cdot\!\mathbf{B}_{2,2}^*\!-\!\mathbf{A}_{2,1}^\textup{\!T}\!\cdot\!\mathbf{A}_{2,2}^*)^\textup{H}\!=\nonumber\\%
&\!\!\!\!=\!\!\!\!&\!\overbrace{\!\!\!\!\!%
\underbrace{\mathbf{A}_{1,1}^\textup{\!H}}_{\scriptstyle \textup{dependent on} \atop [h_1,\cdots,h_{\!K\!/\!4}]}\!\!\!\!\!\!\!\!\cdot\mathbf{A}_{1,2}\!%
-\!\mathbf{A}_{1,1}\!\cdot\!\!\!\!\!\!\!\!\!\!\!\!%
\underbrace{\mathbf{A}_{1,2}^\textup{\!H}}_{\scriptstyle \textup{dependent on} \atop [h_{\!K\!/\!4\!+1},\cdots,h_{\!K\!/\!2}]}\!\!\!\!\!\!\!\!\!\! %
}^{\textup{same generating function}}\!+ \cdots %
=\underbrace{\tilde{\mathbf{A}}_{1}^\textup{\!H}\!\cdot\!\tilde{\mathbf{B}}_{1}\!-\!\tilde{\mathbf{B}}_{1}\!\cdot\!\tilde{\mathbf{A}}_{1}^\textup{\!H}%
}_{\scriptstyle \textup{same structure as } \hat{\mathbf{P}}_1}\!+ \cdots%
\end{eqnarray}

Ultimately, this structure propagates to each of the scalar elements
$\hat{p}_{_{\scriptstyle 1(n,m)}}$ of $\hat{\mathbf{P}}_{1}$,
yielding
\begin{equation}
\label{StructureOfPElements}%
\hat{p}_{_{\scriptstyle 1(n,m)}} =\!\sum\limits_{i=1}^{K/2} (h_{\! a_i}^*h_{\! b_i}\!\!+\!h_{\! a_i}h_{\! b_i}^*)\!-\!(h_{\! a_i}^*h_{\! b_i}\!\!+\!h_{\! a_i}h_{\! b_i}^*)^*\!\!= 0, %
\end{equation}
where the set of pairs $\{(h_{\! a_1},h_{\! b_1}),\cdots,(h_{\!
a_{\!K/2}},h_{\! b_{\!K/2}})\}$ are different for each $(n,m)$-th
element of the matrix.

Equation \eqref{StructureOfPElements}, together with the fact that
$\hat{\mathbf{H}}_1$ and $\hat{\mathbf{H}}_2$ are matrices of the
same size, implies the \emph{quasi}-diagonality of
$\hat{\mathbf{P}}$, which concludes the proof.
\end{proof}

The reader is invited to utilize the Matlab code provided in
Appendix \ref{ExampleGABBAEncodedChannelMatrix} to generate GABBA
channel encoded matrices of various sizes and verify its
quasi-orthogonality property algebraically\footnote{This is
accomplished as follows. First, give the variable $K$ an appropriate
(power-of-two) value. For example, type $<${\tt K=32;}$>$ in the
command line (the brackets $< >$ not included) to make $K=32$. Next,
generate the corresponding symbolic GABBA encoded channel matrix
$\mathbf{H}_{\!K}$ and its non-zero minors $\mathbf{H}_{\!K_1}$ and
$\mathbf{H}_{\!K_2}$ with the command $<${\tt
[H,H1,H2]=GABBAEncodedChannelMatrix([],K);}$>$. If desired, omit the
semicolon or type $<${\tt H1}$>$ or $<${\tt H2}$>$ to verify that
the matrices generated are not numeric, but symbolic (algebraic)
entities. Compute the algebraic outcome of the sum-product
$\mathbf{P}$ by typing $<${\tt P=H1'*H1+H2.'*conj(H2);}$>$. Verify
the algebraic values of the minors of $\mathbf{P}$ corresponding to
$\hat{\mathbf{P}}$ and $\hat{\mathbf{P}}^\textup{H}$ by typing {\tt
P(1:K/2,K/2+1:end)} and {\tt P(K/2+1:end,1:K/2)}, respectively.
Notice also that there are additional pairs of orthogonal vectors in
the GABBA encoded matrices. This property will be exploited in
section \ref{OrthogonalDecoder} to devise a fully orthogonal decoder
for the GABBA codes. In order to inspect the
\emph{quasi}-orthogonality property over real matrices, compute the
the sum-product $\mathbf{P}$ with $<${\tt P=H1.'*H1+H2.'*H2;}$>$.
Notice the interesting difference in the results. For example, for
$K=8$, the sum-product $\mathbf{P}$ is in fact diagonal!}.

\subsection{Quasi-orthogonal Decodability}
\label{QausiOrthogonalDecodability}%

The \emph{quasi}-orthogonality of GABBA encoded channel matrices has
the following consequence.

\begin{theorem}[GABBA \emph{Quasi}-Orthogonal Decodability]
\label{FirstOrderReduction}%
\quad

The receive vector $\mathbf{r}_{\!K}$ corresponding to a GABBA STBC
$K$-by-$n_t$ transmitted through a memoryless block-fading channel
with additive noise can be linearly combined so as to produce the
following pair of vectors $\hat{\mathbf{r}}_{\!K_1}$ and
$\hat{\mathbf{r}}_{\!K_2}$, respectively dependent on the mutually
exclusive $K\!/2$-tuple of symbols, as shown below
\begin{eqnarray}
\label{LinearSystem1Final}%
&\hat{\mathbf{r}}_{\!K_1} \!\!=\!\! \mathbf{H}_{\!K\!/2}\!\cdot\!\mathbf{s}_{\!K\!/2} + \mathbf{n}_{\!K\!/2_1},& \\%
\label{LinearSystem2Final}%
&\hat{\mathbf{r}}_{\!K_2} \!\!=\!\! \mathbf{H}_{\!K\!/2}\!\cdot\!\tilde{\mathbf{s}}_{\!K\!/2} + \mathbf{n}_{\!K\!/2_2}.&%
\end{eqnarray}
\end{theorem}
\begin{proof}
Consider the following linear combination:
\begin{equation}
\label{LinearSystem}%
\hat{\mathbf{r}}_{\!K} = [\mathbf{H}_{\!K}^\textup{H}]^{\textup{U}:K}\!\cdot\!\mathbf{r}_{K} %
                            +  \mathbf{r}_{K}^*\!\cdot\![\mathbf{H}_{\!K}^\textup{T}]^{\textup{R}:K}.
\end{equation}

Due to the block structure of $\mathbf{H}_{\!K}$, it is clear that
the noise independent-terms in equation \eqref{LinearSystem} can be
split into the following independent components:
\begin{eqnarray}
\label{LinearSystem1}%
&&\hspace{-2em}\hat{\mathbf{r}}_{\!K_1} \!\!=\!\! \big[[\mathbf{H}_{\!K_1}^\textup{H}]^{\textup{U}:\tfrac{K}{2}} \ \mathbf{0}\big] \!\!\cdot\!\!%
\left[\!\!\begin{array}{l}
         \mathbf{H}_{\!K_1}\!\!\cdot\!\mathbf{s}_{\!K}\\
         \mathbf{H}_{\!K_2}\!\!\cdot\!\mathbf{s}_{\!K}^*
       \end{array}
\!\!\!\right]\!\!+\!%
\big[\mathbf{0} \ \ [\mathbf{H}_{\!K_2}^\textup{T}]^{\textup{U}:\tfrac{K}{2}}\big] \!\!\cdot\!\!%
\left[\!\!\begin{array}{l}
         \mathbf{H}_{\!K_1}^*\!\!\cdot\!\mathbf{s}_{\!K}^*\\
         \mathbf{H}_{\!K_2}^*\!\!\cdot\!\mathbf{s}_{\!K}
       \end{array}
\!\!\!\right]
\!\!\!=\!\! \left([\mathbf{H}_{\!K_1}^\textup{H}]^{\textup{U}:\tfrac{K}{2}}\!\cdot\!\mathbf{H}_{\!K_1} %
\!\!+\![\mathbf{H}_{\!K_2}^\textup{T}]^{\textup{U}:\tfrac{K}{2}}\!\cdot\!\mathbf{H}_{\!K_2}^*\right)\!\!\cdot\!\mathbf{s}_{\!K}\!=%
\hat{\mathbf{H}}_{\!K_1}\!\!\cdot\!\mathbf{s}_{\!K\!/2},\nonumber \\[-2ex] \\%
\label{LinearSystem2}%
&&\hspace{-2em}\hat{\mathbf{r}}_{\!K_2} \!\!\!=\!\! \big[[\mathbf{H}_{\!K_1}^\textup{H}]^{\textup{D}:\tfrac{K}{2}} \ \mathbf{0}\big] \!\!\cdot\!\!%
\left[\!\!\begin{array}{l}
         \mathbf{H}_{\!K_1}\!\!\cdot\!\mathbf{s}_{\!K}\\
         \mathbf{H}_{\!K_2}\!\!\cdot\!\mathbf{s}_{\!K}^*
       \end{array}
\!\!\!\right]\!\!+\!%
\big[\mathbf{0} \ \ [\mathbf{H}_{\!K_2}^\textup{T}]^{\textup{D}:\tfrac{K}{2}}\big] \!\!\cdot\!\!%
\left[\!\!\begin{array}{l}
         \mathbf{H}_{\!K_1}^*\!\!\cdot\!\mathbf{s}_{\!K}^*\\
         \mathbf{H}_{\!K_2}^*\!\!\cdot\!\mathbf{s}_{\!K}
       \end{array}
\!\!\!\right]
\!\!\!=\!\! \left([\mathbf{H}_{\!K_1}^\textup{H}]^{\textup{D}:\tfrac{K}{2}}\!\cdot\!\mathbf{H}_{\!K_1} %
\!\!+\![\mathbf{H}_{\!K_2}^\textup{T}]^{\textup{D}:\tfrac{K}{2}}\!\cdot\!\mathbf{H}_{\!K_2}^*\right)\!\!\cdot\!\mathbf{s}_{\!K}\!=%
\hat{\mathbf{H}}_{\!K_2}\!\!\cdot\!\tilde{\mathbf{s}}_{\!K\!/2}.\nonumber \\[-2ex]%
\end{eqnarray}

It is left to prove that the matrices $\hat{\mathbf{H}}_{\!K_1}$ and
$\hat{\mathbf{H}}_{\!K_2}$, which are referred to as
\emph{equivalent encoded channel matrices}, are identical.
%
%
From equations \eqref{ABBA3} and \eqref{H1Structure}; and from
equations \eqref{ABBA4} and \eqref{H1Structure}, we have
\begin{eqnarray}
\label{HK1}%
&\mathbf{H}_{\!K_1} =%
\left[\!\!\!%
\begin{array}{cr}
   \mathbf{A}_1 & \!\!\!\mathbf{B}_1 \\
   \mathbf{B}_1 & \!\!\!-\mathbf{A}_1 \\
\end{array}%
\!\!\right]\!\!,&\\%
\label{HK1}%
&\mathbf{H}_{\!K_2} =%
\left[\!\!\!%
\begin{array}{cr}
   \mathbf{B}_2 & \!\!\!-\mathbf{A}_2 \\
   \mathbf{A}_2 & \!\!\! \mathbf{B}_2 \\
\end{array}%
\!\!\right]\!\!.&%
\end{eqnarray}

Substituting these into equations \eqref{LinearSystem1} and
\eqref{LinearSystem2}, we have, straightforwardly:
\begin{equation}
\label{StructureOfHhatK1}%
\hat{\mathbf{H}}_{\!K_1} = %
\left[
[\! %
\begin{array}{lr} %
\mathbf{A}_{1}^\textup{\!H} & \mathbf{B}_{1}^\textup{H}
\end{array} %
\!]\!\cdot\!\!%
\left[\!\!\!%
\begin{array}{cr}
   \mathbf{A}_1 & \!\!\!\mathbf{B}_1 \\
   \mathbf{B}_1 & \!\!\!-\mathbf{A}_1 \\
\end{array}%
\!\!\!\!\right]\!+\! %
[\! %
\begin{array}{lr} %
\mathbf{B}_{2}^\textup{T} & \mathbf{A}_{2}^\textup{\!T}
\end{array} %
\!]\!\cdot\!\!%
\left[\!\!\!%
\begin{array}{cr}
   \mathbf{B}_2^* & \!\!\!-\mathbf{A}_2^* \\
   \mathbf{A}_2^* & \!\!\! \mathbf{B}_2^* \\
\end{array}%
\!\!\right]\right]^{\textup{U}:\tfrac{K}{2}} \!\!\!\!\!\!\!\!\!\!\!\!= %
|\mathbf{A}_{1}\!|^2\!+\!|\mathbf{B}_{1}\!|^2\!+\!\mathbf{B}_{2}^\textup{T}\!\cdot\!\mathbf{B}_{2}^{*}\!+\!\mathbf{A}_{2}^\textup{\!T}\!\cdot\!\mathbf{A}_{2}^*%
\triangleq\mathbf{H}_{\!K\!/2},%
\end{equation}
\begin{equation}
\label{StructureOfHhatK2}%
\hat{\mathbf{H}}_{\!K_2} = %
\left[
[\! %
\begin{array}{lr} %
\mathbf{B}_{1}^\textup{H} & -\mathbf{A}_{1}^\textup{\!H}
\end{array} %
\!]\!\cdot\!\!%
\left[\!\!\!%
\begin{array}{cr}
   \mathbf{A}_1 & \!\!\!\mathbf{B}_1 \\
   \mathbf{B}_1 & \!\!\!-\mathbf{A}_1 \\
\end{array}%
\!\!\!\!\right]\!+\! %
[\! %
\begin{array}{ll} %
-\mathbf{A}_{2}^\textup{\!T} & \mathbf{B}_{2}^\textup{T}
\end{array} %
\!]\!\cdot\!\!%
\left[\!\!\!%
\begin{array}{cr}
   \mathbf{B}_2^* & \!\!\!-\mathbf{A}_2^* \\
   \mathbf{A}_2^* & \!\!\! \mathbf{B}_2^* \\
\end{array}%
\!\!\right]\right]^{\textup{D}:\tfrac{K}{2}} \!\!\!\!\!\!\!\!\!\!\!\!= %
|\mathbf{A}_{1}\!|^2\!+\!|\mathbf{B}_{1}\!|^2\!+\!\mathbf{B}_{2}^\textup{T}\!\cdot\!\mathbf{B}_{2}^{*}\!+\!\mathbf{A}_{2}^\textup{\!T}\!\cdot\!\mathbf{A}_{2}^*%
\triangleq\mathbf{H}_{\!K\!/2}, %
\end{equation}
which concludes the proof.
\end{proof}

Referring to equations \eqref{LinearSystem1} through \eqref{HK1}, it
is clear that equations \eqref{StructureOfHhatK1} and
\eqref{StructureOfHhatK2} can be combined into
\begin{equation}
\label{StructureOfGABBAReducedEncodedChannelMatrix}%
\hat{\mathbf{H}}_{\!K_1} =\hat{\mathbf{H}}_{\!K_2} =\mathbf{H}_{\!K\!/2} = %
\dfrac{1}{2}\left(\mathbf{H}_{\!K_1}^\textup{H}\cdot\mathbf{H}_{\!K_1} + \mathbf{H}_{\!K_2}^\textup{T}\cdot\mathbf{H}_{\!K_2}^*\right), %
\end{equation}
which removes the need for partitioning the minors
$\mathbf{H}_{\!K_1}$ and $\mathbf{H}_{\!K_2}$ when computing the
GABBA reduced encoded channel matrix.

The reader is invited to inspect the \emph{quasi}-orthogonal
decodability of GABBA codes using the Matlab programs provided in
Appendices \ref{Appendix_GABBAEncodingMatrix} and
\ref{Appendix_GABBAEncodedChannelMatrix}\footnote{
To work with the examples of Appendices
\ref{Appendix_GABBAEncodingMatrix} and
\ref{Appendix_GABBAEncodedChannelMatrix}, type $<${\tt K=32;}$>$.
Generate the GABBA encoded channel matrix with the command $<${\tt
[H,H1,H2]=GABBAEncodedChannelMatrix([],K);}$>$.
Next, type $<${\tt for k=1:K/2,eval(['syms s' num2str(k)]),end;}$>$
to define $K\!/2$ symbolic variables.
Construct the symbol vector with its last $K\!/2$ entries reduced to
zeros with $<${\tt s=[];for k=1:K/2,eval(['s=[s;s' num2str(k)
'];']),end}$>$ and $<${\tt szeros=[s;zeros(K/2,1)];}$>$.
Apply the GABBA encoder with $<${\tt C=GABBAEncoder(szeros);}$>$.
Type $<${\tt r=H*[szeros;conj(szeros)];}$>$ to obtain the receive
vector and, finally, type $<${\tt
rhat2=H(:,K/2+1:K)'*r+(r'*H(:,3*K/2+1:2*K)).';}$>$ and $<${\tt
expand(rhat2)}$>$, obtaining a vector of algebraic zeros which
confirms that the outcome of the linear combination given by
equation \eqref{LinearSystem2} is independent of the first $K\!/2$
symbols in $\mathbf{s}_{\!K}$.
Similarly, type $<${\tt szeros=[zeros(K/2,1);s];}$>$ and $<${\tt
C=GABBAEncoder(szeros);}$>$ to generate and encode a symbol vector
with its first $K\!/2$ entries padded with zeros.
Compute the received signal vector with $<${\tt
r=H*[szeros;conj(szeros)];}$>$ and the linear combination given by
equation \eqref{LinearSystem1} with $<${\tt
rhat1=H(:,1:K/2)'*r+(r'*H(:,K+1:3*K/2)).';}$>$.
Finally, inspect the result with $<${\tt expand(rhat1)}$>$.}.

In light of the similarities between equations
\eqref{LinearSystem1Final} and \eqref{LinearSystem2Final} and the
equivalent system description given in equation
\eqref{ReceivedSignalVector}, the matrix $\mathbf{H}_{\!K\!/2}$ will
be hereafter referred to as the GABBA \emph{reduced encoded channel
matrix}.

Theorem \ref{FirstOrderReduction} establishes that the GABBA codes
admit a \emph{quasi}-orthogonal decoder similar to that described in
\cite{MyListOfPapers:Tirkkonen2000, MyListOfPapers:Jafarkhani2001,
MyListOfPapers:Papadias2003} and, therefore, can be seen as a
complete generalization of the ABBA scheme previously known for
limited number of antennas only.
In the next section it will be shown, however, that a fully
orthogonal decoder for the proposed GABBA codes also exists.

\section{Orthogonal Decoding of GABBA Codes}
\label{OrthogonalDecoder}%

\subsection{Properties of GABBA Reduced Encoded Channel Matrices}
\label{PropertiesOfGABBAReducedEncodedChannelMatrices}

The essence behind Theorem \ref{FirstOrderReduction} is that the
$n$-th order manifold structure of the GABBA encoded channel
matrices, which ensures their \emph{quasi}-orthogonality, ultimately
enables the decomposition of the received vector into two vectors
each dependent of mutually exclusively $K/2$-tuples of transmitted
symbols.
Similarly, it will be shown that the symbol-by-symbol decodability
of GABBA codes derives from additional orthogonalities in the
reduced encoded channel matrix, beyond those proved in Theorem
\ref{FirstOrderReduction}.

To this end, let us first introduce the following result.
\begin{lemma}[Structure of GABBA Reduced Encoded Channel Matrices]
\label{StructureOfGABBAReducedEncodedChannelMatrices} \quad%

GABBA reduced encoded channel matrices $\mathbf{H}_{\!K\!/2}$ are
$(n-1)$-th order manifold matrices with generating function as given
in equation \eqref{ABBA4}.
\end{lemma}

\begin{proof}
In similarity to Lemma
\ref{QuasiOrthogonalityOfGABBAEncodedChannelMatrices}, let us write
the minors $\mathbf{A}_i$ and $\mathbf{B}_i$ in equations
\eqref{StructureOfHhatK1} and \eqref{StructureOfHhatK2} one step
more explicitly, yielding
$\mathbf{A}_{1}\!=\!H(\mathbf{A}_{1,1},\mathbf{B}_{1,1})$,
$\mathbf{B}_{1}\!=\!H(\mathbf{A}_{1,2},\mathbf{B}_{1,2})$,
$\mathbf{A}_{2}\!=\!\tilde{H}(\mathbf{A}_{2,1},\mathbf{B}_{2,1})$
and
$\mathbf{B}_{2}\!=\!\tilde{H}(\mathbf{A}_{2,2},\mathbf{B}_{2,2})$.
Then, the equivalent encoded channel matrix $\mathbf{H}_{\!K\!/2}$
can be written as
\begin{equation}
\label{StructureOfH_K_over_2}%
\mathbf{H}_{\!\tfrac{K}{2}}\!\!=\!\!%
\sum\limits_{i=1}^{2}{\!\! %
\left[\!\!\!\begin{array}{cc}
         |\mathbf{A}_{1,i}|^2\!\!+\!|\mathbf{B}_{1,i}|^2\!\!+\!\mathbf{B}_{2,i}^\textup{T}\!\cdot\!\mathbf{B}_{2,i}^{*}\!\!+\!\mathbf{A}_{2,i}^\textup{\!T}\!\cdot\!\mathbf{A}_{2,i}^* \!\!\!&\!\!\!%
         \mathbf{A}_{1,i}^\textup{\!H}\!\cdot\!\mathbf{B}_{1,i}\!-\!\mathbf{B}_{1,i}^\textup{H}\!\cdot\!\mathbf{A}_{1,i}\!-\!\mathbf{B}_{2,i}^\textup{T}\!\cdot\!\mathbf{A}_{2,i}^{\!*}\!\!+\!\mathbf{A}_{2,i}^\textup{\!T}\!\cdot\!\mathbf{B}_{2,i}^{\!*} \\[1ex] %
         \mathbf{B}_{1,i}^\textup{H}\!\cdot\!\mathbf{A}_{1,i}\!-\mathbf{A}_{1,i}^\textup{\!H}\!\cdot\!\mathbf{B}_{1,i}\!+\!\mathbf{B}_{2,i}^\textup{T}\!\cdot\!\mathbf{A}_{2,i}^{\!*}\!\!-\!\mathbf{A}_{2,i}^\textup{\!T}\!\cdot\!\mathbf{B}_{2,i}^{\!*} \!\!\!&\!\!\! %
         |\mathbf{A}_{1,i}|^2\!\!+\!|\mathbf{B}_{1,i}|^2\!\!+\!\mathbf{B}_{2,i}^\textup{T}\!\cdot\!\mathbf{B}_{2,i}^{*}\!\!+\!\mathbf{A}_{2,i}^\textup{\!T}\!\cdot\!\mathbf{A}_{2,i}^*%
       \end{array}
\!\!\!\right]}\!\!=\!\! %
\left[\!\!\!\!%
\begin{array}{cr}
    \mathbf{A} & \!\!\!-\mathbf{B} \\[-0.5ex]
    \mathbf{B} & \!\!\! \mathbf{A}
\end{array}%
\!\!\!\!\right]\!\!.%
\end{equation}

It is evident from equation \eqref{StructureOfH_K_over_2} that
$\mathbf{H}_{\!K\!/2}$ has the structure described by equation
\eqref{ABBA4}.
Proceeding in a similar manner it is straightforward to show that,
furthermore, the minors $\mathbf{A}$ and $\mathbf{B}$ also have the
same structure.
In other words, $\mathbf{H}_{\!K\!/2}$ can be constructed by
iterating the generating function given by equation \eqref{ABBA4}
$n\!-\!1$ times and, therefore, is a $(n\!-\!1)$-th order manifold
matrix.
\end{proof}

The fact that GABBA reduced encoded channel matrices
$\mathbf{H}_{\!K\!/2}$ have a manifold structure similar to that
$\mathbf{H}_{\!K}$ prompted us to investigate its orthogonality
properties.
One difficulty in dealing with the reduced matrices in algebraic
form, however, is that its elements are sums of products of the
channel estimates.
A solution to this problem is to relabel the entries of
$\mathbf{H}_{\!K\!/2}$.
A Matlab program that can be used to systematically relabel the
entries of GABBA reduced channel matrices is provided in Appendix
\ref{ProgramRelabelGABBAReducedChannelMatrix}\footnote{Before
running the program, type in a value for the variable $K$, $e.g.$,
$<${\tt K=32;}$>$ and generate the corresponding GABBA encoded
channel matrix and its minors with the command $<${\tt
[H,H1,H2]=GABBAEncodedChannelMatrix([],K);}$>$.
Finally, type $<${\tt [Hhat1,Hhat2]=RelabelH(H1,H2,1)}$>$ to
generate and display the GABBA reduced encoded channel matrices.
Notice that the outcomes $<${\tt Hhat1}$>$ and $<${\tt Hhat2}$>$ are
generated independently and yield the same results, as proven in
Theorem \ref{FirstOrderReduction}.
It is also evident from the results that GABBA reduced encoded
channel matrices are $(n-1)$-th order manifold matrices with
generating function $\tilde{H}(\mathbf{A},\mathbf{B}) =
\left[{{\mathbf{A} \ \! -\mathbf{B}} \atop {\mathbf{B} \! \ \ \
\mathbf{A}}}\right]$, as proved in Lemma
\ref{StructureOfGABBAReducedEncodedChannelMatrices}.}.
Another solution is to simply rely on the proof of Lemma
\ref{StructureOfGABBAReducedEncodedChannelMatrices}, and construct
the relabeled GABBA reduced encoded channel as the manifold matrix
of a channel vector of length with the .
The corresponding Matlab code is provided in Appendix
\ref{ProgramGenerateGABBAReducedChannelMatrix}\footnote{To use the
program, assuming that the variable $K$ has been assigned a
power-of-two value, simply type $<${\tt
H=RelabeledGABBAReducedEncodedChannelMatrix(16)}$>$. Notice that, in
order to remain coherent to the notation used throughout the paper,
the relabeled matrix outputted is of size $K\!/2$.}.
Finally, one could also use the program given in Appendix
\ref{ProgramGABBAEncodedChannelMatrix} and generate the
$(n\!-\!1)$-th manifold matrix as the second minor of the GABBA
encoded channel matrix ($i.e.$, the matrix $<${\tt H2}$>$), using a
modified channel vector as input\footnote{To generate the modified
channel vector, type $<${\tt h=[];for k=1:K/2, eval(['h=[h;h'
num2str(k) '];']),end}$>$ and $<${\tt h=[h(K/4+1:K/2);h(1:K/4)]}$>$
into the command line, after giving the variable $<${\tt K}$>$ an
appropriate (power of $2$) value.}.

Empowered by the ability to algebraically generate GABBA reduced
channel matrices of any size, one can easy inspect their properties.
Such an inspection quickly reveals that GABBA reduced channel
matrices are real \emph{quasi}-orthogonal (see definitions
\ref{DefinitionRealOrthogonality} and
\ref{DefinitionBlockOrthogonalSTBC}), $i.e.$, there is a permutation
that can be applied to a reduced GABBA encoded matrices such that
the first half columns(rows) of the resulting permuted matrix are
real-orthogonal to the remaining half columns(rows).

For example, it is found that first and fourth columns(rows) of the
$4$-by-$4$ GABBA reduced encoded channel matrix are real-orthogonal
to its second and third columns(rows).
Likewise, the $\{1, 4, 6, 7\}$-th columns(rows) of the $8$-by-$8$
GABBA reduced encoded channel matrix are real-orthogonal to its $\{
2, 3, 5, 8\}$-th columns(rows)..
Analogously, the $\{1, 4, 6, 7, 10, 11, 13, 16\}$-th and $\{ 2, 3,
5, 8, 9, 12, 14, 15\}$-th columns and rows of the $16$-by-$16$ GABBA
reduced encoded channel matrix are mutually real-orthogonal, and so
fourth.
Though rather hard to prove in a concise and generalized fashion,
these results are easy to verify algebraically (for example, using
the Matlab programs given in Appendix
\ref{Appendix_RelabelGABBAReducedChannelMatrix}), and indicate a
clear pattern of real-orthogonality between the vectors of the GABBA
reduced encoded channel matrices, which is conjectured to behave as
follows.

\newtheorem{conjecture}{Conjecture}
\begin{conjecture}[Real Quasi-Orthogonality Pattern of GABBA Reduced Encoded Channel Matrices]
\label{RealQuasiOrthogonalityOfGABBAReducedChannelMatrices} \quad

Consider the function
\begin{equation}
\label{OrthogonalityFunction}%
p_N(x) = \!\!\!\!\!\!\mod\!\!\!\!\left(x\!\cdot\!\left\lfloor\log_2\!N\right\rfloor - x + \!\!\!\!\!\sum\limits_{n=1}^{\left\lfloor\log_2{\!N}\right\rfloor\!-\!1}{\!\!\left\lfloor\dfrac{x}{2^n}\right\rfloor},2\right), %
\end{equation}
where $\lfloor x \rfloor$ denotes the largest integer no greater
than $x$ and $\!\!\!\!\mod\!\!(x,2)$ denotes the rest of division by
$2$.

Let the sets $\boldsymbol{p}_{0:N}$ and $\boldsymbol{p}_{1:N}$ be
respectively given by
\begin{eqnarray}
\label{OrthogonalityIndexVector1} %
&\boldsymbol{p}_{0:N} = \{ 1, \cdots, i, \cdots \}, \textup{ such that } p_N(i) = 0, (i \leq N),& \\[-0.5ex] %
\label{OrthogonalityIndexVector2} %
&\boldsymbol{p}_{1:N} = \{ 2, \cdots, j, \cdots \}, \textup{ such that } p_N(j) = 1, (j \leq N).& %
\end{eqnarray}

If $\mathbf{H}_{\!K\!/2}$ denotes the $K\!/2$-by-$K\!/2$ GABBA
reduced encoded channel matrix, then the permuted matrix given below
is real \emph{quasi}-orthogonal
\begin{equation}
\label{PermutedReducedGABBAEncodedChannelMatrix} %
\boldsymbol{\mathcal{H}}_{\!K\!/2} = \big[[\mathbf{H}_{\!K\!/2}]^{\boldsymbol{p}_{0:K\!/2}} [\mathbf{H}_{\!K\!/2}]^{\boldsymbol{p}_{1:K\!/2}} \big]. %
\end{equation}

That is, the $\boldsymbol{p}_{0:N}$-th columns(rows) of the
$N$-by-$N$ GABBA reduced encoded channel matrix are real-orthogonal
to its $\boldsymbol{p}_{1:N}$-th columns(rows). \hfill \QEDopen
\end{conjecture}

A Matlab program to generate the permutation indexes
$\boldsymbol{p}_{0:N}$ and $\boldsymbol{p}_{1:N}$ is provided in
Appendix \ref{Appendix_GeneratePermutationIndexes}.

Conjecture \ref{RealQuasiOrthogonalityOfGABBAReducedChannelMatrices}
is algebraically verifiable for matrices of very large sizes and,
therefore, is nearly as strong as theorem for most practical
purposes.
Indeed, using the programs provided in Appendices
\ref{Appendix_RelabelGABBAReducedChannelMatrix} and
\ref{Appendix_GeneratePermutationIndexes}, it is found that
conjecture \ref{RealQuasiOrthogonalityOfGABBAReducedChannelMatrices}
is hold true for matrices as large as Matlab can handle in symbolic
form\footnote{This is dependent on the amount of memory available in
platform used. We have verified the conjecture to hold over GABBA
encoded channel matrices of sizes up to $K=1024$.}, which is beyond
the dimension of virtually any foreseeable MIMO application!
The reader is invited to test our claims\footnote{After generating
an algebraic and relabeled GABBA reduced encoded channel matrix,
construct the vectors $\boldsymbol{p}_{0:N}$ and
$\boldsymbol{p}_{1:N}$ with the command $<${\tt
[p0,p1]=PermutationIndexes(1:K/2)}$>$ and type $<${\tt
H(:,p0).'*H(:,p1)}$>$, $<${\tt H(:,p1).'*H(:,p0)}$>$, $<${\tt
H(p0,:)*H(p1,:).'}$>$ or $<${\tt H(p1,:)*H(p0,:).'}$>$ to verify the
real-orthogonalities of the column- and row-wise partitions of
$<${\tt H}$>$.}.

\subsection{Orthogonal Decodability of the Original ABBA Code}
\label{OrthogonalDecodabilityOfOriginalABBACode}

The real \emph{quasi}-orthogonality of GABBA reduced encoded
matrices implies that the vectors $\hat{\mathbf{r}}_{\!K_1}$ and
$\hat{\mathbf{r}}_{\!K_2}$, of equations \eqref{LinearSystem1} and
\eqref{LinearSystem2}, can be further decomposed into a total of $4$
vectors, each dependent of mutually exclusively $K\!/\!4$-tuples of
transmitted symbols as follows.

First, notice that the real \emph{quasi}-orthogonality property of
$\mathbf{H}_{\!K\!/2}$ ensures that,
\begin{equation}
\label{H_K_over_2_Product}%
\boldsymbol{\mathcal{H}}_{\!\tfrac{K}{2}}^{\!\textup{T}}\!\cdot\!\boldsymbol{\mathcal{H}}_{\!\tfrac{K}{2}} = %
\left[\!\!\!%
\begin{array}{cc}
   \boldsymbol{\mathcal{H}}_{1} & \!\!\!\mathbf{0} \\
   \mathbf{0}           & \!\!\! \boldsymbol{\mathcal{H}}_{2} \\
\end{array}%
\!\!\right]\!\!.%
\end{equation}

Let $\hat{\boldsymbol{r}}_{\!K_1}$, $\hat{\boldsymbol{r}}_{\!K_2}$,
$\boldsymbol{s}_{\!K\!/2}$ and $\tilde{\boldsymbol{s}}_{\!K\!/2}$ be
the permuted versions of the vectors $\hat{\mathbf{r}}_{\!K_1}$,
$\hat{\mathbf{r}}_{\!K_2}$, $\mathbf{s}_{\!K\!/2}$ and
$\tilde{\mathbf{s}}_{\!K\!/2}$, respectively.
Then, equations \eqref{LinearSystem1} and \eqref{LinearSystem2} can
be rewritten as (see also equations \eqref{StructureOfHhatK1} and
\eqref{StructureOfHhatK2}),
\begin{eqnarray}
\label{PermutedVector_r_hatK1}
\hat{\boldsymbol{r}}_{\!K_1} = \boldsymbol{\mathcal{H}}_{\!K\!/2}\cdot\boldsymbol{s}_{\!K\!/2},\\ %
\label{PermutedVector_r_hatK2}
\hat{\boldsymbol{r}}_{\!K_2} = \boldsymbol{\mathcal{H}}_{\!K\!/2}\cdot\tilde{\boldsymbol{s}}_{\!K\!/2}.%
\end{eqnarray}

Substituting equation \eqref{H_K_over_2_Product} into these
equations we obtain,
\begin{eqnarray}
\label{LinearSystem11}
\hat{\boldsymbol{r}}_{\!K_{1,1}} = [\boldsymbol{\mathcal{H}}_{\!K\!/2}^{\textup{T}}\cdot\hat{\boldsymbol{r}}_{\!K_1}]^{\textup{U:K/4}} = \boldsymbol{\mathcal{H}}_{1}\cdot[\boldsymbol{s}_{\!K\!/2}]^{\textup{U:K/4}} = \boldsymbol{\mathcal{H}}_{1}\cdot[\mathbf{s}_{\!K}]^{[\boldsymbol{p}_{0:K\!/2}]^\textup{U:K/4}},\\ %
\label{LinearSystem12}
\hat{\boldsymbol{r}}_{\!K_{1,2}} = [\boldsymbol{\mathcal{H}}_{\!K\!/2}^{\textup{T}}\cdot\hat{\boldsymbol{r}}_{\!K_1}]^{\textup{D:K/4}} = \boldsymbol{\mathcal{H}}_{2}\cdot[\boldsymbol{s}_{\!K\!/2}]^{\textup{D:K/4}} = \boldsymbol{\mathcal{H}}_{2}\cdot[\mathbf{s}_{\!K}]^{[\boldsymbol{p}_{1:K\!/2}]^\textup{U:K/4}},\\ %
\label{LinearSystem21}
\hat{\boldsymbol{r}}_{\!K_{2,1}} = [\boldsymbol{\mathcal{H}}_{\!K\!/2}^{\textup{T}}\cdot\hat{\boldsymbol{r}}_{\!K_2}]^{\textup{U:K/4}} = \boldsymbol{\mathcal{H}}_{1}\cdot[\tilde{\boldsymbol{s}}_{\!K\!/2}]^{\textup{U:K/4}} = \boldsymbol{\mathcal{H}}_{1}\cdot[\mathbf{s}_{\!K}]^{[\boldsymbol{p}_{0:K\!/2}]^\textup{D:K/4}},\\ %
\label{LinearSystem22}
\hat{\boldsymbol{r}}_{\!K_{2,2}} = [\boldsymbol{\mathcal{H}}_{\!K\!/2}^{\textup{T}}\cdot\hat{\boldsymbol{r}}_{\!K_2}]^{\textup{D:K/4}} = \boldsymbol{\mathcal{H}}_{2}\cdot[\tilde{\boldsymbol{s}}_{\!K\!/2}]^{\textup{D:K/4}} = \boldsymbol{\mathcal{H}}_{2}\cdot[\mathbf{s}_{\!K}]^{[\boldsymbol{p}_{1:K\!/2}]^\textup{D:K/4}}. %
\end{eqnarray}

The last terms in equations \eqref{LinearSystem11} through
\eqref{LinearSystem22} are obtained by recalling that the product
$\mathbf{H}_{\!K\!/2}^{\textup{T}}\cdot\mathbf{H}_{\!K\!/2}$ is a
sparse matrix with the elements of $\boldsymbol{\mathcal{H}}_{1}$
distributed across its $\boldsymbol{p}_{0:K\!/2}$ columns and rows,
and with the elements of $\boldsymbol{\mathcal{H}}_{2}$ distributed
across its $\boldsymbol{p}_{1:K\!/2}$ columns and rows, such that
the half-partitions of $\boldsymbol{p}_{0:K\!/2}$ are, respectively,
the indexes of the symbols in the permuted vectors
$[\boldsymbol{s}_{\!K\!/2}]^{\textup{U:K/4}}$ and
$[\tilde{\boldsymbol{s}}_{\!K\!/2}]^{\textup{U:K/4}}$ and, likewise,
$\boldsymbol{p}_{1:K\!/2}$ is the set of indexes of the symbols in
$[\boldsymbol{s}_{\!K\!/2}]^{\textup{D:K/4}}$ and
$[\tilde{\boldsymbol{s}}_{\!K\!/2}]^{\textup{D:K/4}}$, respectively.

In particular, for the case of $K=4$, where the GABBA code reduces
to the a variation of the original ABBA code, equations
\eqref{LinearSystem11} through \eqref{LinearSystem22} ensure the
full decoupling of all four symbols encoded.
In that case, it is trivial to show that
$\boldsymbol{\mathcal{H}}_{1}$ and $\boldsymbol{\mathcal{H}}_{2}$
reduce to identical scalar numbers, such that equations
\eqref{LinearSystem11} through \eqref{LinearSystem22} need only be
normalized accordingly to yield soft estimates of the encoded
symbols.

For the convenience of the reader, explicit calculations for the
orthogonal decoding of the $4$-by-$4$ GABBA code are put together in
a Matlab script file provided in Appendix
\ref{ProgramOrthogonalDecodingOfOriginalABBACode}.
One needs only to study and execute the file to verify the veracity
of our claims.

\subsection{Orthogonal Decodability of GABBA Codes}
\label{OrthogonalDecodabilityOfGABBACodes}

It was shown, so far, that the \emph{quasi}-orthogonality of GABBA
encoded channel matrices and the real \emph{quasi}-orthogonality of
GABBA reduced encoded channel matrices enable the orthogonal
decodability of the $4$-by-$4$ GABBA code, which is a variation of
the original ABBA scheme proposed in
\cite{MyListOfPapers:Tirkkonen2000, MyListOfPapers:Jafarkhani2001,
MyListOfPapers:Papadias2003}.

In this subsection, it will be shown that the orthogonal
decodability is a general property of the GABBA codes.
To this end, it is sufficient to show that the real
\emph{quasi}-orthogonality of GABBA reduced encoded channel matrices
propagates to the minors $\boldsymbol{\mathcal{H}}_{1}$ and
$\boldsymbol{\mathcal{H}}_{2}$, and so forth.

\begin{conjecture}[Real Quasi-Orthogonality of Second-Order GABBA Reduced Encoded Channel Matrices]
\label{RealQuasiOrthogonalityOfSecondOrderGABBAReducedChannelMatrices} \quad %

Let $\boldsymbol{\mathcal{H}}_{1}$ and
$\boldsymbol{\mathcal{H}}_{2}$ denote the minors of the real-product
of two $K\!/2$-by-$K\!/2$ permuted GABBA reduced encoded channel
matrices, as given in equation \eqref{H_K_over_2_Product}.
Then, the \emph{second-order} GABBA reduced encoded channel matrix,
defined as
\begin{equation}
\label{SecondOrderGABBAReducedEncodedChannelMatrix} %
\mathbf{H}^2_{\!K\!/4} \triangleq \boldsymbol{\mathcal{H}}_{1}^\textup{T}\cdot\boldsymbol{\mathcal{H}}_{2} = \boldsymbol{\mathcal{H}}_{2}^\textup{T}\cdot\boldsymbol{\mathcal{H}}_{1}, %
\end{equation}
is a real \emph{quasi}-orthogonal matrix.

Furthermore, let the $\boldsymbol{\mathcal{H}}^2_{\!K\!/4}$ denote
the permuted version of $\mathbf{H}^2_{\!K\!/4}$ such that
\begin{equation}
\label{HH_K_over_4_Product}%
\boldsymbol{\mathcal{H}}^2_{\!\tfrac{K}{4}}\!^{\!\textup{T}}\!\cdot\!\boldsymbol{\mathcal{H}}^2_{\!\tfrac{K}{4}} = %
\left[\!\!\!%
\begin{array}{cc}
   \boldsymbol{\mathcal{H}}^2_{1} & \!\!\!\mathbf{0} \\
   \mathbf{0}                     & \!\!\! \boldsymbol{\mathcal{H}}^2_{2} \\
\end{array}%
\!\!\right]\!\!.%
\end{equation}

Such a permutated matrix is given by
\begin{equation}
\label{SecondOrderPermutedGABBAReducedEncodedChannelMatrix} %
\boldsymbol{\mathcal{H}}^2_{\!K\!/4} = \big[[\mathbf{H}^2_{\!K\!/4}]^{\boldsymbol{p}_{0:K\!/4}} [\mathbf{H}^2_{\!K\!/2}]^{\boldsymbol{p}_{1:K\!/4}} \big], %
\end{equation}
where $\boldsymbol{p}_{0:K\!/4}$ and $\boldsymbol{p}_{1:K\!/4}$ are
as in equations \eqref{OrthogonalityIndexVector1} and
\eqref{OrthogonalityIndexVector2}. \hfill \QEDopen
\end{conjecture}

Again, the veracity of conjecture
\ref{RealQuasiOrthogonalityOfSecondOrderGABBAReducedChannelMatrices}
can be easily verified for values of $K$ as large as Matlab is able
to handle\footnote{For instance, input a value for the variable $K$,
$e.g.$, $<${\tt K=32;}$>$, and type in the command $<${\tt
H=RelabeledGABBAReducedEncodedChannelMatrix(K);}$>$ to generate a
relabeled, algebraic GABBA reduced encoded channel matrix of desired
size. Next, compute the second-order GABBA reduced encoded channel
matrix algebraically through the real-product $<${\tt HH=H.'*H;}$>$,
as in equation \eqref{SecondOrderGABBAReducedEncodedChannelMatrix}.
Verify the commutativeness of the product as implied in that
equation with the command $<${\tt HH==H*H.'}$>$, which should return
a matrix of logical ``ones''. Then, compute the permutation indexes
with $<${\tt [p1,p2]=PermutationIndexes(1:K/2);}$>$ and, finally,
type in $<${\tt HH(p1,p2)}$>$ and $<${\tt HH(p2,p1)}$>$ to verify
that the off-diagonal minors of the permuted second-order GABBA
reduced encoded channel matrix are indeed null-matrices. If desired,
type in $<${\tt HH(p1,p1)}$>$ and $<${\tt HH(p2,p2)}$>$ to inspect
the diagonal minors corresponding to
$\boldsymbol{\mathcal{H}}^2_{1}$ and
$\boldsymbol{\mathcal{H}}^2_{2}$ in equation
\eqref{HH_K_over_4_Product}.} and, therefore, has the strength of a
theorem form most practical purposes.

The similarity between Conjectures
\ref{RealQuasiOrthogonalityOfGABBAReducedChannelMatrices} and
\ref{RealQuasiOrthogonalityOfSecondOrderGABBAReducedChannelMatrices}
is evident.
In fact, these results can be can be combined and generalized as
follows.

\begin{conjecture}[Real Quasi-Orthogonality of $n$-th Order GABBA Reduced Encoded Channel Matrices]
\label{RealQuasiOrthogonalityOfNthOrderGABBAReducedChannelMatrices} \quad %

Let the $n$-th GABBA reduced encoded channel matrix
defined\footnote{For convenience, equation
\eqref{StructureOfGABBAReducedEncodedChannelMatrix} is repeated in
equation \eqref{FirstOrderGABBAReducedEncodedChannelMatrix}.} by
\begin{eqnarray}
\label{FirstOrderGABBAReducedEncodedChannelMatrix} %
\hspace{8.5em}&\mathbf{H}_{\!K\!/2} \triangleq %
    \dfrac{1}{2}\left(\mathbf{H}_{\!K_1}^\textup{H}\cdot\mathbf{H}_{\!K_1} + \mathbf{H}_{\!K_2}^\textup{T}\cdot\mathbf{H}_{\!K_2}^*\right),& \quad \qquad (n = 1) \\ %
\label{NthOrderGABBAReducedEncodedChannelMatrix} %
&\mathbf{H}^n_{\!K\!/{2^n}} \triangleq (\boldsymbol{\mathcal{H}}^{n\!-\!1}_{1})^\textup{T}\cdot\boldsymbol{\mathcal{H}}^{n\!-\!1}_{2}= %
                                      (\boldsymbol{\mathcal{H}}^{n\!-\!1}_{2})^\textup{T}\cdot\boldsymbol{\mathcal{H}}^{n\!-\!1}_{1}& \quad (1 < n \leq \log_2{\!K}). %
\end{eqnarray}

Then, for all $1 \leq n \leq \log_2{\!K}$, the permutation
\begin{equation}
\label{NthOrderPermutedGABBAReducedEncodedChannelMatrix} %
\boldsymbol{\mathcal{H}}^n_{\!K\!/{2^n}} = %
            \big[[\mathbf{H}^n_{\!K\!/{2^n}}]^{\boldsymbol{p}_{0:K\!/{2^n}}} [\mathbf{H}^n_{\!K\!/{2^n}}]^{\boldsymbol{p}_{1:K\!/{2^n}}} \big], %
\end{equation}
is such that
\begin{equation}
\label{H_K_over_2n_Product}%
\boldsymbol{\mathcal{H}}^n_{\!\tfrac{K}{2^n}}\!^{\!\textup{T}}\!\cdot\!\boldsymbol{\mathcal{H}}^n _{\!\tfrac{K}{2^n}} = %
\left[\!\!\!%
\begin{array}{cc}
   \boldsymbol{\mathcal{H}}^n_{1} & \!\!\!\mathbf{0} \\
   \mathbf{0}                     & \!\!\! \boldsymbol{\mathcal{H}}^n_{2} \\
\end{array}%
\!\!\right]\!\!,%
\end{equation}
with $\boldsymbol{\mathcal{H}}^n_{1}$ and
$\boldsymbol{\mathcal{H}}^n_{2}$ reducing to a complex scalar at $n
= \log_2{\!K}$.

In other words, GABBA reduced encoded channel matrices of $n$-th
order are real \emph{quasi}-orthogonal. \hfill \QEDopen
\end{conjecture}

The claims stated in conjecture
\ref{RealQuasiOrthogonalityOfNthOrderGABBAReducedChannelMatrices}
are, once more, verifiable to matrices of very large sizes and,
therefore, are a solid results that can be used with the strength of
a theorem for all cases of practical interest.
A Matlab script file is provided in Appendix
\ref{ProgramOrthogonalDecodabilityOfGABBACodes} to facilitate this
verification.

The orthogonal decodability of GABBA codes is an immediate and
obvious consequence of the real \emph{quasi}-orthogonality of GABBA
reduced encoded channel matrices.
The following algorithm to obtain orthogonal estimates of all
symbols encoded in GABBA encoding matrix can be inferred directly
from conjecture
\ref{RealQuasiOrthogonalityOfNthOrderGABBAReducedChannelMatrices}.

\newtheorem{algorithm}{Algorithm}
\begin{algorithm}[Orthogonal Decoding of GABBA Codes with a Multi-antenna Receiver]
\label{GABBAOrthogonalDecodingAlgorithm} \quad %

Consider a $K$-by-$n_r$ matrix constructed from the collection of
independent receive vectors
$\mathbf{r}_{{\!K_1}:1},\cdots,\mathbf{r}_{{\!K_1}:n_r}$, each
obtained at a different receive antenna, corresponding to the
transmission of a $K$-by-$n_t$ GABBA encoding matrix through a MIMO
channel consisting of $n_t$ transmit and $n_r$ receive antennas.
Assume that the $n_t$-by-$1$ channel vectors
$\mathbf{h}_{n_t:1},\cdots,\mathbf{h}_{n_t:nr}$ at all receive
antennas are perfectly known.
Then, orthogonal estimates of all symbols encoded in GABBA encoding
matrix are obtained as follows:
\begin{enumerate}[1)]
\item For every $i$-th receive antenna, construct the corresponding $K$-by-$2K$ GABBA encoded channel matrix as described in Lemma \ref{ConstructionOfGABBAEncodedChannelMatrices};%
\item For every $i$-th receive antenna, combine the received vector as shown in equations \eqref{LinearSystem1} and \eqref{LinearSystem2}, obtaining two independent reduced %
      receive vectors $\hat{\mathbf{r}}_{\!K_1:i}$ and $\hat{\mathbf{r}}_{\!K_2:i}$; %
\item For every $i$-th receive antenna, use equation \eqref{FirstOrderGABBAReducedEncodedChannelMatrix} to construct the $1^\textup{st}$ order GABBA reduced encoded channel matrix $\mathbf{H}_{\!K\!/2:i}$; %
\item Add the first reduced receive vectors $\hat{\mathbf{r}}_{\!K_1:i}$ across the receive antennas, obtaining $\hat{\mathbf{r}}_{\!K_1:\Sigma}$;
\item Add the second reduced receive vector $\hat{\mathbf{r}}_{\!K_2:i}$ across the receive antennas, obtaining $\hat{\mathbf{r}}_{\!K_2:\Sigma}$;
\item Add the $1^\textup{st}$ order GABBA reduced encoded channel matrix across the receive antennas, obtaining $\mathbf{H}_{\!K\!/{2^n}:\Sigma}$; %
\item Set $n = 1$; %
\item Multiply the aggregate reduced receive vectors $\hat{\mathbf{r}}_{\!K_j:\Sigma}$, $(j = 1,\cdots,2^n)$ by the transpose of the aggregate %
      $n$-th order GABBA reduced encoded channel matrix $\mathbf{H}_{\!K\!/{2^n}:\Sigma}$, obtaining $2^{n+1}$ new aggregate reduced receive vectors, %
      each dependent on mutually exclusive $K\!/{2^{(n+1)}}$-tuples of transmitted symbols (see equations \eqref{LinearSystem11} through \eqref{LinearSystem22}); %
\item Compute the permutation indexes $\boldsymbol{p}_{0:K/{2^n}}$ and $\boldsymbol{p}_{1:K/{2^n}}$; %
\item Obtain a higher-order GABBA reduced encoded channel matrices by partitioning $\mathbf{H}_{\!K\!/{2^n}:\Sigma}$, as in equation \eqref{NthOrderPermutedGABBAReducedEncodedChannelMatrix}; %
\item If $n < \log_2{\!K} - 1$, increment $n$ by $1$ and return to step $8$;
\item Otherwise, the symbols have been fully decoupled. Normalize the combined signals\footnote{The notation is slightly abused here for the sake of simplicity.} %
      $\hat{\mathbf{r}}_{\!K_j:\Sigma}$, $(j = 1,\cdots,2^n)$ by the GABBA reduced encoded channel matrix (which at this point has been reduced to a scalar) to obtain %
      the soft estimate of the corresponding symbol. \hfill \QEDopen
\end{enumerate}
\end{algorithm}

Notice that right after step $6$ of algorithm
\ref{GABBAOrthogonalDecodingAlgorithm}, we have:
\begin{eqnarray}
&\hat{\mathbf{r}}_{\!K_1:\Sigma} = \sum\limits_{i=1}^{nr}{(\hat{\mathbf{r}}_{\!K_1:i} + \mathbf{n}_{\!K_1:i})} = %
     \sum\limits_{i=1}^{nr}{(\mathbf{H}_{\!K\!/2:i}\cdot\mathbf{s}_{\!K\!/2} + \mathbf{n}_{\!K_1:i})} = %
     \mathbf{H}_{\!K\!/2:\Sigma} \cdot \mathbf{s}_{\!K\!/2} + \mathbf{n}_{\!K_1:\Sigma},& \\ %
&\hat{\mathbf{r}}_{\!K_2:\Sigma} = \sum\limits_{i=1}^{nr}{(\hat{\mathbf{r}}_{\!K_2:i} + \mathbf{n}_{\!K_2:i})} = %
     \sum\limits_{i=1}^{nr}{(\mathbf{H}_{\!K\!/2:i}\cdot\tilde{\mathbf{s}}_{\!K\!/2} + \mathbf{n}_{\!K_2:i})} = %
     \mathbf{H}_{\!K\!/2:\Sigma} \cdot \mathbf{s}_{\!K\!/2} + \mathbf{n}_{\!K_2:\Sigma},& %
\end{eqnarray}
where $\mathbf{n}_{\!K_1:i}$, $\mathbf{n}_{\!K_2:i}$,
$\mathbf{n}_{\!K_1:\Sigma}$ and $\mathbf{n}_{\!K_2:\Sigma}$ are
noise vectors.

It is clear from equation \eqref{StructureOfH_K_over_2} that the
diagonal elements of each $1^\textup{st}$ order GABBA reduced
encoded channel matrix $\mathbf{H}_{\!K\!/2:i}$ are real scalars
given by $\sum_{j=1}^{n_t}{|\mathbf{h}_{j:i}|^2}$, while all
remaining elements are independent complex numbers.
Consequently, the summation of the vectors
$\hat{\mathbf{r}}_{\!K_1:\Sigma}$ and
$\hat{\mathbf{r}}_{\!K_2:\Sigma}$ across the receive antennas --
which incurs in no loss of information, nor in any decrease of
signal to noise ratio -- significantly reduces the complexity the
orthogonal soft-decoding procedure in a multi-antenna receiver, by
eliminating the need for the parallel processing of the signal
streams at all receive antennas independently.

Notice also that the soft estimates obtained at step $12$ of
\ref{GABBAOrthogonalDecodingAlgorithm} are not in a trivial
ascending ordered.
In fact, although the symbols are split into two $K\!/2$-tuples with
indexes arranged in ascending order up to the first linear
combination, at the following steps each of the tuples are split
into two tuples arranged according to the sequences
$\boldsymbol{p}_{0:K/{2^n}}$ and $\boldsymbol{p}_{1:K/{2^n}}$.
Therefore, the soft symbol estimates obtained by the GABBA
orthogonal decoder are ordered in accordance with the sequence
computed as follows.

\begin{algorithm}[Order of GABBA Orthogonal Soft Estimates]
\label{OrthogonalEstimateReordering} \quad %

Start with the matrix
\begin{equation}
\boldsymbol{P}_1 = \left[%
\begin{array}{cccc}
1         & 2         & \cdots & K\!/2\\[-1ex]
K\!/2 + 1 & K\!/2 + 2 & \cdots & K
\end{array}
\right].
\end{equation}

Next, for $n = \{2,\cdots,\log_2{\!K}\}$, partition the matrix
$\boldsymbol{P}_n$ in half (columnwise) and stack the two
partitions, obtaining
\begin{equation}
\boldsymbol{P}_{n} = \left[%
\begin{array}{c}
\\[-5ex]
[\boldsymbol{P}_{n-1}]^{\boldsymbol{p}_{0:K/{2^n}}}\\[-1ex]
[\boldsymbol{p}_{n-1}]^{\boldsymbol{p}_{1:K/{2^n}}}
\end{array}
\right],
\end{equation}
where $\boldsymbol{p}_{0:N}$ and $\boldsymbol{p}_{1:N}$ are computed
as described in conjecture
\ref{RealQuasiOrthogonalityOfGABBAReducedChannelMatrices}.

Then, the $j$-th entry of the resulting column vector
$\boldsymbol{P}_{\log_2{\!K}}$ is the index of the symbol
corresponding to the $j$-th orthogonal soft estimate output by
algorithm \ref{GABBAOrthogonalDecodingAlgorithm}. \hfill \QEDopen
\end{algorithm}

A Matlab implementation of algorithms
\ref{GABBAOrthogonalDecodingAlgorithm} and
\ref{OrthogonalEstimateReordering}, combined into a fully orthogonal
decoder for the GABBA codes, is provided in Appendix
\ref{Appendix_GABBAOrthogonalDecoder}.
The reader is encouraged to test the generality, speed and numeric
efficient of the technique\footnote{For instance, after giving the
variables $K$ and $n_t$ desired values, type $<${\tt
s=(randn(K,1)+j*randn(K,1));}$>$ to generate a symbol vector taking
from a complex Gaussian constellation and and GABBA-encode it with
$<${\tt C=GABBAEncoder(s);}$>$. Then, generate a Rayleigh
block-fading channel matrix with the command $<${\tt
hh=(randn(K,nr)+j*randn(K,nr));}$>$ and compute the corresponding
receive signal matrix with $<${\tt rr=C*hh;}$>$. Employ the GABBA
decoder to obtain the soft orthogonal estimates of the transmitted
symbols calling the function $<${\tt sx=GABBADecoder(rr,hh);}$>$.
Finally, verify the results by listing the original and the
orthogonal estimate symbol vector next to one another with $<${\tt
[s sx]}$>$, or compute the norm of their difference with $<${\tt
norm(s-sx)}$>$.}.
Try large values of $K$ and $n_r$ over numeric symbol taken for
different constellations to experience the remarkable low-complexity
and flexibility of the GABBA decoder\footnote{Note that Matlab's
floating point relative accuracy is of approximately
$2\cdot10^{-16}$. Extremely large values of $K$ may cause Matlab to
overflow. This problem can be avoided by normalizing the
intermediate quantities.}.
Emulate the case of $n_t < K$ by simply puncturing arbitrary columns
of the GABBA mother encoding matrix and padding the rows of the
corresponding encoded channel matrix with zeros.

As a final remark for this section, we emphasize that the decoding
procedure described in algorithm
\ref{GABBAOrthogonalDecodingAlgorithm} is not to be confused with a
zero forcing or any iterative technique such as the minimum mean
square error (MMSE) decoder.
Indeed, in the proposed decoder, all transmitted symbols are
decoupled through linear combinations of the received signal vector,
without any assumptions on the modulation scheme or any
knowledge/estimation of the noise statistics, and without resorting
to matrix inversions.
Instead, as shown in the preceding sections, the orthogonality of
the soft symbol estimated is ensured by the
\emph{quasi}-orthogonality of the encoded channel matrix and the
real \emph{quasi}-orthogonality of the reduced encoded channel
matrix of different orders.

In fact, it is evident that the nested calculation described in
algorithm \ref{GABBAOrthogonalDecodingAlgorithm} can be reduced to a
single-step implementation, where the linear combination weights
$\mathbf{F}_{\!k:1}$ and $\mathbf{F}_{\!k:2}$ required to obtain the
orthogonal estimate of each symbol $s_k$ are first computed from the
corresponding products of GABBA reduced encoded channel matrices,
and then applied directly over the original received signal vector,
as shown in equation \eqref{VectorOrthogonalSymbolEstimate}
\begin{equation}
\label{VectorOrthogonalSymbolEstimate}%
\hat{\mathbf{s}}_k = \mathbf{F}_{\!k:1}^\textup{H}\!\cdot\!\mathbf{r}_K + \mathbf{r}_K^*\!\cdot\!\mathbf{F}_{\!k:2}^\textup{T} = \alpha_k\cdot\mathbf{s}_k + \tilde{\mathbf{n}}_k,%
\end{equation}
where $\alpha_k$ is a real scalar proportional to
$\sum\limits_{(i,j)=(1,1)}^{(n_r,n_t)}{|\mathbf{h}_{i,j}|^2}$, while
$\tilde{\mathbf{n}}_k$ is the linear combination of the original
noise terms $\mathbf{n}_k$ (see equation
\eqref{ReceivedSignalVector}) with independent, complex
coefficients.

In other words, it can be said that the GABBA orthogonal decoding
``algorithm'' is a true extension of the Alamouti decoder, currently
available for the case of $n_t = 2$.
The flexibility in handling with GABBA codes of different sizes
(scalability), and the relatively easy expressions involved are,
nevertheless, advantages of the nested implementation here
described.

\section{Performance of GABBA Codes}%
\label{PerformanceResults}

In this section, performance of GABBA codes in uncorrelated fading
channels of generalized statistics is studied both analytically and
through simulations.

Given the difficulty in computing explicitly the coefficients
$\alpha_k$, as well as the corresponding coefficients multiplying
the noise terms in $\tilde{\mathbf{n}}_k$, the exact BER probability
of orthogonally decoded GABBA codes is hard to determine in a
generalized and fashion.
Therefore, we shall take the approach of deriving analytical
expressions (bounds) corresponding to the ML decoder (whose
complexity is prohibitive for large systems and high-order
constellations), and compare the curves with simulation results
obtained using the proposed GABBA orthogonal decoder.

\subsection{Performance of GABBA Codes with ML Decoding: Analysis of Exact BER Probability}
\label{GABBA_BER Analytical}

The BER probability of digitally modulated signals in a fading
channel can be computed by averaging the corresponding error
probability in the additive white Gaussian noise (AWGN) channel over
the statistics of the fading process
\cite{MyListOfPapers:Alouini2000}.
In a setting with $n_t$ transmit and $n_r$ receive antennas, a
ML-decoded rate-one STBC with full diversity is equivalent to a
$(n_t \times n_r)$-branch maximum ratio combined receive diversity
system with receive power equal to $n_r$, $i.e.$, where the unitary
transmit power is shared amongst all the $n_t$ transmit channels
\cite{MyListOfPapers:Proakis2000}.

Consider an $n_t$-by-$n_r$ MIMO set-up in which the
constellation-energy-to-noise-power ratio (Es/No) of the $n$-th pair
of transmit-to-receive antennas is denoted $\gamma_n$, and where the
envelope fading process corresponding to that diversity branch is
described by a continuous probability density function (PDF)
$p_{_{\Gamma_n}}\!\!(\gamma_n)$.
Let $P_b(\gamma_n)$ denote the BER probability of the digital
modulation system over the $n$-th branch under AWGN conditions.
Following the arguments in the previous paragraph, the BER
probability of the full-rate full-diversity GABBA codes over this
MIMO structure is given by
\begin{equation}
\label{BER_MultipleIntegralFormula}%
\bar{P}_b(\gamma) = \underbrace{\int\limits_{0}^{\infty}\cdots\int\limits_{0}^{\infty}\hspace{-0.5em}}_{(n_t \times n_r)-\textup{fold}} %
                    {\!\prod\limits_{n=1}^{n_t \times n_r}{\!\!\!P_b(\gamma_n)\cdot\!p_{_{\Gamma_n}}\!\!(\gamma_n)}} \; d_{\gamma_1}\cdots d_{\gamma_{n_t n_r}}. %
\end{equation}

\subsubsection{$M$-ary PSK Modulation}
\label{BER_DerivationPSK} \quad %

The BER of $M$-ary PSK modulated signals over the AWGN channel were
(believed to be) calculated exactly by Lee in
\cite{MyListOfPapers:LeeCOM1986}.
A minor error (occurring at higher-order constellations) was
recently discovered by Lassing \emph{et al.} and corrected in
\cite{MyListOfPapers:LassingISIT2003,
MyListOfPapers:LassingCOM2003}.
In particular, from \cite[eqs. (2) and
(3)]{MyListOfPapers:LassingISIT2003}, we have
\begin{equation}
\label{BER_AWGN_PSK}%
P_{b:\textup{PSK}}(\gamma|M) = \frac{1}{\log_2{\!M}}\sum\limits_{k=1}^{M-1}{\bar{d}_{k:\textup{PSK}} \cdot P_{k:\textup{PSK}}(\gamma|M)}, %
\end{equation}
with
\begin{equation}
\label{dK}%
\bar{d}_{k:\textup{PSK}} = 2 \left| \frac{k}{M} + \left\lfloor \frac{k}{M} \right\rceil \right| + %
            2 \sum\limits_{i=2}^{\log_2{\!M>2}}{\left| \frac{k}{2^i} + \left\lfloor \frac{k}{2^i} \right\rceil \right|}, %
\end{equation}
where $\lfloor x \rceil$ rounds $x$ to its nearest integer, and
$P_{k:\textup{PSK}}(\gamma|M)$ is the probability that a received
$M$-ary PSK symbol falls $k$ sectors away from the sector it belongs
to due to the effect of AWGN, under the signal-to-noise ratio
$\gamma$.

A convenient formula for $P_{k:\textup{PSK}}(\gamma|M)$ can be
derived\footnote{To this end, notice that in our context
\cite[$\Delta\Phi=0$]{MyListOfPapers:PawulaCOM2001}. From \cite[eq.
(18b)]{MyListOfPapers:PawulaCOM2001}, this gives \cite[$A =
-\Phi$]{MyListOfPapers:PawulaCOM2001}. Finally, let
\cite[$A$]{MyListOfPapers:PawulaCOM2001} be given by the left and
right angles defining each PSK sector. The difference between the
formulas at adjacent delimiting angles gives equation
\eqref{Pk_PSK}.} from \cite[eq. (24)]{MyListOfPapers:PawulaCOM2001},
yielding
\begin{equation}
\label{Pk_PSK}%
P_{k:\textup{PSK}}(\gamma|M) = \frac{1}{2\pi}\left[ %
              \int\limits_{0}^{\pi(1-\delta_k^{-})}{\!\!\!\!\!e^{-\gamma\cdot\tfrac{\sin^2(\pi\delta_k^{-})}{\sin^2(\theta)}}}d\theta \;\;-\!\!\! %
              \int\limits_{0}^{\pi(1-\delta_k^{+})}{\!\!\!\!\!e^{-\gamma\cdot\tfrac{\sin^2(\pi\delta_k^{+})}{\sin^2(\theta)}}}d\theta \right], %
\end{equation}
where, for uniform $M$-ary PSK constellations, we have
\begin{eqnarray}
\label{DeltaKminus} %
&\delta_k^{-} = \dfrac{2k-1}{M},&\\
\label{DeltaKplus} %
&\delta_k^{+} = \dfrac{2k+1}{M}.&
\end{eqnarray}

Equation \eqref{Pk_PSK} also appears (with a minor error, however)
in \cite[pp. 201, eq. (129)]{MyListOfPapers:Alouini2000}.

Substituting equation \eqref{Pk_PSK} into equation
\eqref{BER_AWGN_PSK}, and the consequent result into equation
\eqref{BER_MultipleIntegralFormula}, one realizes that the moment
generating function (MGF) approach (see
\cite{MyListOfPapers:Alouini2000}) can be easily applied in the
computation of the exact average BER probability of $M$-ary PSK over
fading channels.
In fact, equation \eqref{BER_MultipleIntegralFormula} can be
rewritten as
\begin{equation}
\label{BER_Fading_PSK_MGF}%
\bar{P}_{b:\textup{PSK}}(\gamma|M) = \frac{1}{2\log_2{\!M}}\sum\limits_{k=1}^{M-1}{\bar{d}_k \cdot \left( I(\delta_k^{-},g_{_\textup{PSK}}(\delta_k^{-}),n_t \cdot n_r) - I(\delta_k^{+},g_{_\textup{PSK}}(\delta_k^{+}),n_t \cdot n_r)\right)}, %
\end{equation}
with the functions $g_{_\textup{PSK}}(\delta)$ and $I(\delta,g,N)$
respectively given by
\begin{eqnarray}
\label{gPSK}%
&g_{_\textup{PSK}}(\delta) = \sin^2(\pi\delta),&\\ %
\label{ElementaryMGF_Integral}%
&I(\delta,g,N) = \dfrac{1}{\pi}\cdot\!\!\!\!\!\int\limits_{0}^{\pi(1-\delta)}{\prod\limits_{n=1}^{N}{\mu_{\gamma_n}\!\!\left(-\dfrac{g}{\sin^2(\theta)}\right)}}d\theta,& %
\end{eqnarray}
where $\mu_{\gamma_n}\!(x)$ is the MGF of the PDF
$p_{_{\Gamma_n}}\!\!(\gamma_n)$.

The exact BER integral formula given in equation
\eqref{BER_Fading_PSK_MGF} can be easily extended to the case of a
STBC with rate $\rho$ and transmit diversity order\footnote{Transmit
diversity order is understood as the ratio between the diversity
gain achieved by the STBC, to the number of transmit antennas.}
$\eta$.
To this end, one needs only to realize that in a linear STBC of rate
$\rho$ and diversity $\eta$, each symbol is transmitted $1/\rho$
times by each transmit antenna (so that the energy of the transmit
constellation increases proportionally); and that the diversity
order $\eta < n_t$ can be translated as a uniform loss in the
diversity contributions of all branches in the systems.
These observations lead to
\begin{equation}
\label{BER_Fading_PSK_MGF_GeneralRateGain}%
\bar{P}_{b:\textup{PSK}}(\gamma|M,\rho,\eta) = \frac{1}{2\log_2{\!M}}\sum\limits_{k=1}^{M-1}{\bar{d}_k \cdot \left( \tilde{I}(\delta_k^{-},g_{_\textup{PSK}}(\delta_k^{-}),n_t \cdot n_r|\rho,\eta) - \tilde{I}(\delta_k^{+},g_{_\textup{PSK}}(\delta_k^{+}),n_t \cdot n_r|\rho,\eta)\right)}, %
\end{equation}
where
\begin{equation}
\label{ElementaryMGF_IntegralRhoEta}%
\tilde{I}(\delta,g,N|\rho,\eta) = \dfrac{1}{\pi}\cdot\!\!\!\!\!\int\limits_{0}^{\pi(1-\delta)}{\left(\prod\limits_{n=1}^{N}{\mu_{\gamma_n}\!\!\left(-\dfrac{g}{\rho\cdot\sin^2(\theta)}\right)}\right)^{\!\!\!\eta}}d\theta. %
\end{equation}

Equation \eqref{BER_Fading_PSK_MGF_GeneralRateGain} is a general,
simple and exact formula for the BER probability of linear STBCs of
rate $\rho$ and diversity order $\eta$ employing uniform $M$-ary PSK
modulation over an ensemble of uncorrelated fading channels with
arbitrary and unequal statistics.

\subsubsection{$M$-ary QAM Modulation}
\label{BER_DerivationQAM} \quad %

In similarity to the previous subsection, an exact formula for the
performance of linear STBCs over $M$-ary QAM constellation can be
derived as follows.
The exact BER probability for QAM modulation in the AWGN channel was
derived by Cho \emph{et al.} \cite{MyListOfPapers:ChoCOM2002}.
In particular, for regular $M$-ary QAM constellations we have, from
\cite[eq. (16)]{MyListOfPapers:ChoCOM2002},
\begin{equation}
\label{BER_AWGN_QAM}%
P_{b:\textup{QAM}}(\gamma|M) = \frac{1}{\log_2{\!\sqrt{M}}}\sum\limits_{k=1}^{\log_2{\!\sqrt{M}}}{P_{k:\textup{QAM}}(\gamma|M)}. %
\end{equation}

In equation \eqref{BER_AWGN_QAM}, $P_{k:\textup{QAM}}(\gamma|M)$ is
the probability that the $k$-th bit in the $M$-ary symbol is in
error, which can be put from \cite[eq.
(14)]{MyListOfPapers:ChoCOM2002} into the more convenient form shown
below
\begin{equation}
\label{Pk_QAM}%
P_{k:\textup{QAM}}(\gamma|M)  = \frac{2}{\sqrt{M}}%
                     \sum\limits_{i=0}^{(1-2^{-k})\sqrt{M}-1}{\!\!\!\!\!d_{i:\textup{QAM}} \cdot %
                     Q\left((2i+1)\sqrt{(\tfrac{2\gamma}{M-1})}\right)},
\end{equation}
where
\begin{equation}
\label{d_i_QAM} %
d_{i:\textup{QAM}} = (-1)^{\left\lfloor\tfrac{i \cdot 2^{k-1}}{\sqrt{M}}\right\rfloor} %
                     \left(2^{k-1} - \left\lfloor\tfrac{i \cdot 2^{k-1}}{\sqrt{M}}+\tfrac{1}{2}\right\rfloor \right), %
\end{equation}
and $Q(x)$ is the Gaussian $Q$-function, given by
\begin{equation}
\label{Q_Function} %
Q(x) = \dfrac{1}{\pi}\cdot\int\limits_{0}^{\tfrac{\pi}{2}} {e^{\tfrac{-x^2}{\sin^2(\theta)}}} d\theta. %
\end{equation}

The similarities between equations \eqref{BER_AWGN_PSK} and
\eqref{BER_AWGN_QAM}, and between equations \eqref{Pk_PSK} and
\eqref{Pk_QAM} are evident.
Substituting equations \eqref{BER_AWGN_QAM} through
\eqref{Q_Function} into equation
\eqref{BER_MultipleIntegralFormula}, and invoking the MGF method, we
obtain
\begin{equation}
\label{BER_Fading_QAM_MGF}%
\bar{P}_{b:\textup{QAM}}(\gamma|M) = \frac{2}{\sqrt{M}\log_2{\!\sqrt{\!\!M}}}\sum\limits_{k=1}^{\log_2{\!\sqrt{M}}} %
                     \left[\sum\limits_{i=0}^{(1-2^{-k})\sqrt{M}-1}{\!\!\!\!\!d_{i:\textup{QAM}} \cdot I\left(\dfrac{1}{2},g_{_\textup{QAM}}(i),n_t \cdot n_r\right)}\right], %
\end{equation}
where
\begin{equation}
\label{gQAM}%
g_{_\textup{QAM}}(i) = \dfrac{3}{2}\cdot\dfrac{(2i+1)^2}{(M-1)}. %
\end{equation}

Again, equation \eqref{BER_Fading_QAM_MGF} can be generalized to
STBCs of rate $\rho$ and diversity order $\eta$ by replacing
equation \eqref{ElementaryMGF_Integral} with equation
\eqref{ElementaryMGF_IntegralRhoEta}, which gives
\begin{equation}
\label{BER_Fading_QAM_MGF_GeneralRateGain}%
\bar{P}_{b:\textup{QAM}}(\gamma|M,\rho,\eta) = \frac{2}{\sqrt{M}\log_2{\!\sqrt{\!\!M}}}\sum\limits_{k=1}^{\log_2{\!\sqrt{M}}} %
                     \left[\sum\limits_{i=0}^{(1-2^{-k})\sqrt{M}-1}{\!\!\!\!\!d_{i:\textup{QAM}} \cdot \tilde{I}\left(\dfrac{1}{2},g_{_\textup{QAM}}(i),n_t \cdot n_r|\rho,\eta\right)}\right]. %
\end{equation}

For convenience, a Matlab implementation of the above-derived
analytical BER probability formulas over uncorrelated and unequal
fading channels of Hoyt, Rice, Rayleigh and Nakagami statistics
\cite{MyListOfPapers:Hoyt1947, MyListOfPapers:Rice1948,
MyListOfPapers:Rayleigh1889, MyListOfPapers:Nakagami1960} is
provided in Appendix \ref{Program_BER_STBCGeneral}.

Closed form formulas corresponding to equations
\eqref{BER_Fading_PSK_MGF_GeneralRateGain} and
\eqref{BER_Fading_QAM_MGF_GeneralRateGain} can most likely be
derived (at least under certain conditions) using ideas found, for
instance, in \cite{MyListOfPapers:Alouini2000}.
Given the simplicity and conciseness of the formulas provided, as
well as the accurate results obtained with these equations, however,
it is unlikely that the resulting expressions (which oftentimes
involve terms dependent on factorials) would be of use.

\subsection{Performance of Orthogonally Decoded GABBA Codes: Simulations and Comparisons}
\label{GABBA_BER_Simulated}

In this section, the performance of orthogonally decoded GABBA codes
are compared to the exact formulas (bounds) derived above.

\subsubsection{Equipower Rayleigh Block-fading Channels}
\label{PerformanceRayleighBlock-fadingChannels} \quad %

\begin{figure}[h!]
\includegraphics[width=\columnwidth]{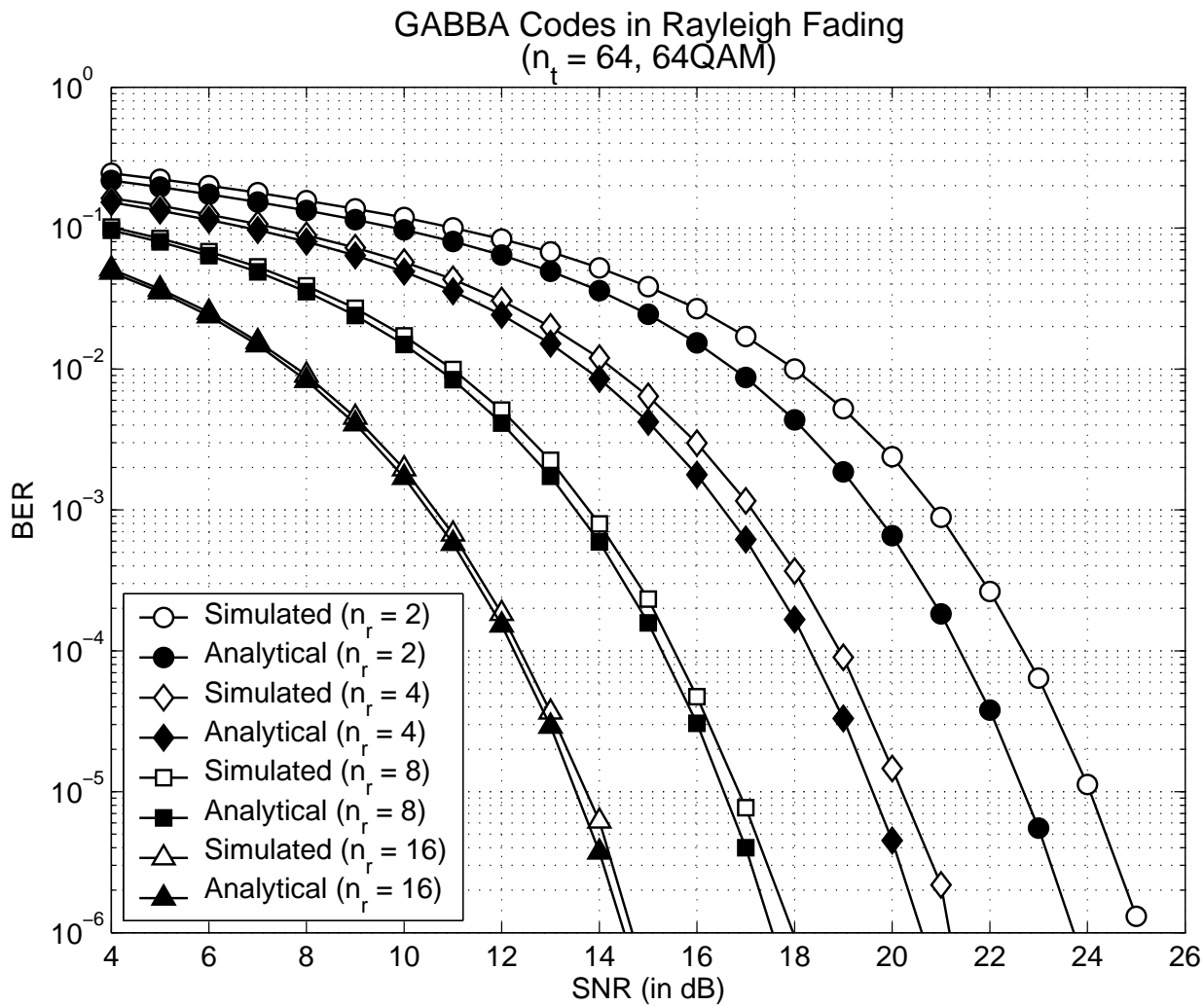}
\caption{Performance of the GABBA code $\mathbf{C}_{64}$ over
$64$-QAM modulation in Rayleigh block-fading channels with various
numbers of receive antennas. Simulated results (white markers) were
obtained using the orthogonal GABBA decoder described in section
\ref{OrthogonalDecodabilityOfGABBACodes}, while analytical results
(black markers) were obtained with equation
\eqref{BER_Fading_QAM_MGF}, with $\rho=1$ and $\eta=1$ (full-rate
full-diversity).}%
\label{GABBANtx64_NrxVAR_64QAM_m1}%
\end{figure}

First, in figure \ref{GABBANtx64_NrxVAR_64QAM_m1}, the BER
performance of a large GABBA code ($n_t = 64$) simulated over
independent and identically distributed (i.i.d.) Rayleigh fading
channels is compared against the corresponding analytical BER
probability bound.
The simulated results (white markers) were obtained with the
orthogonal, symbol-by-symbol decoder described in subsection
\ref{OrthogonalDecodabilityOfGABBACodes}, while the analytical
results (black markers) were computed using equation
\eqref{BER_Fading_QAM_MGF}, with $\rho=1$ and $\eta=1$.

A number of important conclusions can be inferred from the
comparison.
Firstly, the remarkably low complexity of the scheme is emphatically
illustrated.
Indeed, in order to decode a single block, the ML-decoder, whose
performance is associated to the analytical (analytical) curves
shown in the figure, would require the minimization of a decision
metric over all possible $64$-tuples composed of $64$-QAM
constellation symbols, resulting in the prohibitive decoding
complexity of $\mathcal{O}(64^{64})$.
Likewise, a \emph{quasi}-orthogonal decoding procedure would require
the minimization of two independent decision metrics over all
possible $32$-tuples of $64$-QAM symbols, yielding an equally
prohibitive decoding complexity of $\mathcal{O}(2\cdot64^{32})$.
In contrast, the proposed orthogonal decoder requires merely the
comparison of each orthogonal symbol estimate against the $64$
possible symbols in the constellation, the same complexity of an
uncoded system.

Secondly, hard evidence is provided to the fact that the GABBA
orthogonal decoder is not a mere zero-forcing technique.
This is true because the performance loss of the symbol-by-symbol
decoding procedure relative to its bound decreases with the number
of receive antennas.
In other words, no systematic penalty in signal-to-noise ratio is
payed by the use of the GABBA orthogonal decoder.
It is difficult to determine exactly the reason why the simulated
results obtained with the orthogonal decoder do not match exactly
the bounds.
The difference between the analytical and simulated curves shown in
figure \ref{GABBANtx64_NrxVAR_64QAM_m1} could, in principle, be
caused by coloration of noise samples in the decoupled symbol
estimates.
Since the noise terms at the output of the decoder are linear
combinations of the original noise vector, with coefficients given
by sum-products of the uncorrelated channel, noise coloration is
rather unlikely, especially with large codes and more so when
multiple receive antennas are used.
This author is led to conclude therefore, that the problem is (even
if partly, but substantially) caused by rounding errors inserted by
limitations of the simulation platform utilized (in this case
Matlab).
Anyhow, given the extreme contrast in the complexities involved by
each of the alternatives compared, the results shown in figure
\ref{GABBANtx64_NrxVAR_64QAM_m1} are nothing but remarkable.

\begin{figure}[h!]
\includegraphics[width=\columnwidth]{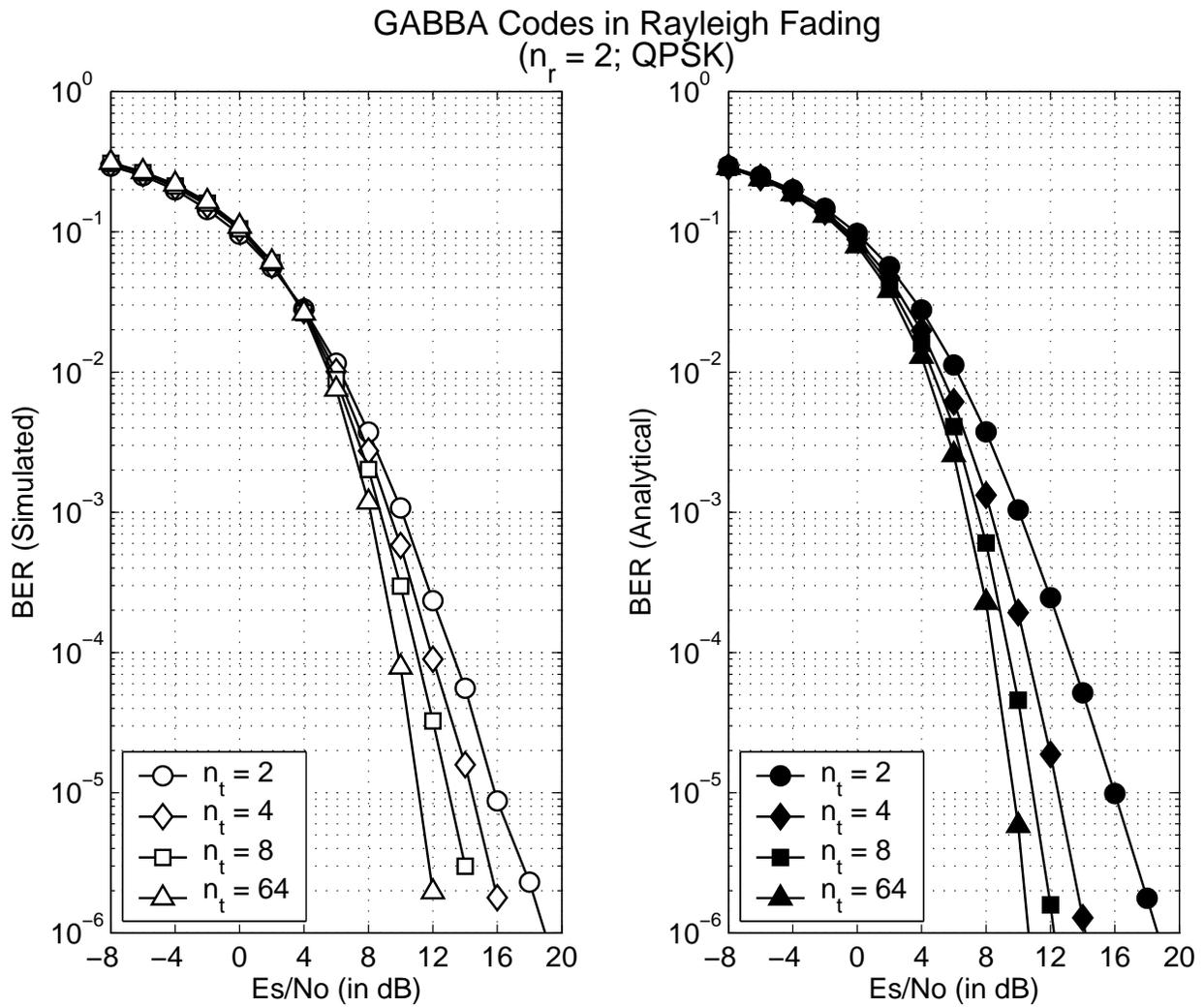}
\caption{Performance of GABBA codes of different sizes over QPSK
symbols in Rayleigh block-fading channels with $2$ receive antennas.
Simulated results (white markers) were obtained using the orthogonal
GABBA decoder described in section
\ref{OrthogonalDecodabilityOfGABBACodes}, while analytical results
(black markers) were computed using equation
\eqref{BER_Fading_QAM_MGF}, with $\rho=1$ and $\eta=1$ (full-rate
full-diversity).}%
\label{GABBANtxVar_Nrx2_QPSK_m1}%
\end{figure}

Finally, it is clearly shown that the diversity order attained by
the GABBA STBC with its orthogonal decoder is the same as the one
theoretically achievable with the ML decoder.
This claim is consistent with the discussion of section
\ref{OrthogonalDecoder}, and is further supported by the results
shown in figure \ref{GABBANtxVar_Nrx2_QPSK_m1}.
In this figure, the simulated and analytical BER probabilities of
GABBA codes of various sizes over QPSK modulated signals in the
Rayleigh fading channel are compared.
It is found that even with a small number of receive antennas
($n_r=2$), the performance of orthogonally decoded GABBA codes is
close to the theoretical bound, and that the difference between does
not increase with the number of transmit antennas $n_t$, $i.e.$, as
larger codes are used\footnote{Except when going from $n_t = 2$ to
$n_t=4$. This is because at $n_t = 2$ the GABBA code reduces to the
Alamouti scheme, which admits a linear orthogonal decoder identical
to the ML decoder.}.

\subsubsection{Unequal Block-fading Channels of Same Statistics}
\label{PerformanceUnequalBlock-fadingChannelsSameStatistics} \quad %

The point of figures \ref{GABBANtx64_NrxVAR_64QAM_m1} and
\ref{GABBANtxVar_Nrx2_QPSK_m1} was to illustrate how the GABBA codes
and the corresponding orthogonal decoder provide a realistic,
high-performing solution for a large, flexible, linear STBC
achieving full-rate and full-diversity over i.i.d. fading channels.

Next, we turn our attention to the effects of unequal diversity
branches and on the performance of GABBA codes or, in other words,
to the effect of utilizing large GABBA codes in the presence of
non-i.i.d. fading channels.
Our objectives are two-fold, first, to verify that the impact of
unequal power and/or unequal statistics of the space-time channels
is less significant as the number of transmit antennas increases
(larger codes); and, second, to illustrate the flexibility and
accuracy of the BER expression given in equation
\eqref{BER_Fading_QAM_MGF}.

First, consider the case of channels with the same fading statistics
but different powers
\cite{MyListOfPapers:AndersenComMag1995,MyListOfPapers:Vega1997}.
This case can be associated to a distributed scenario
\cite{MyListOfPapers:Nicholas2003} with cooperative GABBA space-time
encoded transmission, where the multiple cooperating transmitters
are in the same environment (same channel statistics), but at
different locations (different signal strengths at the receiver).
The cooperation is assumed to be perfect, $i.e.$, cooperating
transmitters are perfectly synchronized and coordinated, such that
their transmit signals arrive at the receiver at the same time, at
every transmission epoch\footnote{A practical example would be an
indoor OFDM multi-cast system, where several base-stations are
reliably connected to one another ($e.g.$, via cable) and transmit
the same signal ($e.g.$ a video steam) cooperatively to a wireless
user. The OFDM set-up not only provides for memoryless channels but
also ensures that a synchronous transmission (synchronized
acess-points) results in synchronous reception for all users in the
premises.}.

It is also assumed that the average power of each transmit channel
is an independent random uniformly distributed in the interval
$[0,P_\textup{max}]$.
Let the average power of the $k$-th transmit channel be denoted by
$\bar{p}_k$.
We start by asking ourselves what is the power distribution profile
(across the different transmit channels) that best describes the
average scenario faced by the system.
Notice that if the statistics of all channels are the same, all
diversity branches can, without loss of generality, be relabeled
into ascending order.
Mathematically we have:
\begin{equation}
\label{PowerKtuple} %
\wp_1 \leq \cdots \leq \wp_k \leq \cdots \leq \wp_K, %
\end{equation}

The random $\wp_k$ is referred to as the $k$-th order statistic of
the ensemble $\{\bar{p}_1,\cdots,\bar{p}_K\}$.
For $P_\textup{max}=1$, the expected value of the $k$-th order
statistic $\wp_k$ is given by \cite[pp. 665, eq.
(7.9.20)]{MyListOfPapers:Zwillinger1996}
\begin{equation}
\label{MeanOrderStatistic} %
\bar{\wp}_k = \frac{K!}{(k-1)!(K-k)!} \cdot \int\limits_{0}^{1}{x^k \cdot (1-x)^{K-k}dx}. %
\end{equation}

The integral in equation \eqref{MeanOrderStatistic} is a variation
of the well-known Beta function \cite[pp. 544, eq.
(6.12.1)]{MyListOfPapers:Zwillinger1996} (a.k.a. Euler's integral of
the first kind), which has no general solution.
Fortunately, for $\{k,K\} \in \mathbb{N}$, it is easy to show (see
Appendix \ref{DerivationEulerIntegralSolution}) that
\begin{equation}
\label{EulerIntegralSolution} %
\int\limits_{0}^{1}{x^k \cdot (1-x)^{K-k}dx} = \frac{(K-k)!k!}{(K+1)!}. %
\end{equation}

From equations \eqref{MeanOrderStatistic} and
\eqref{EulerIntegralSolution}, we finally obtain, for a general
value of $P_\textup{max}$,
\begin{equation}
\label{MeanOrderStatisticSolution} %
\bar{\wp}_k = \frac{k}{K+1}P_\textup{max}.
\end{equation}

\begin{figure}[h!]
\includegraphics[width=\columnwidth]{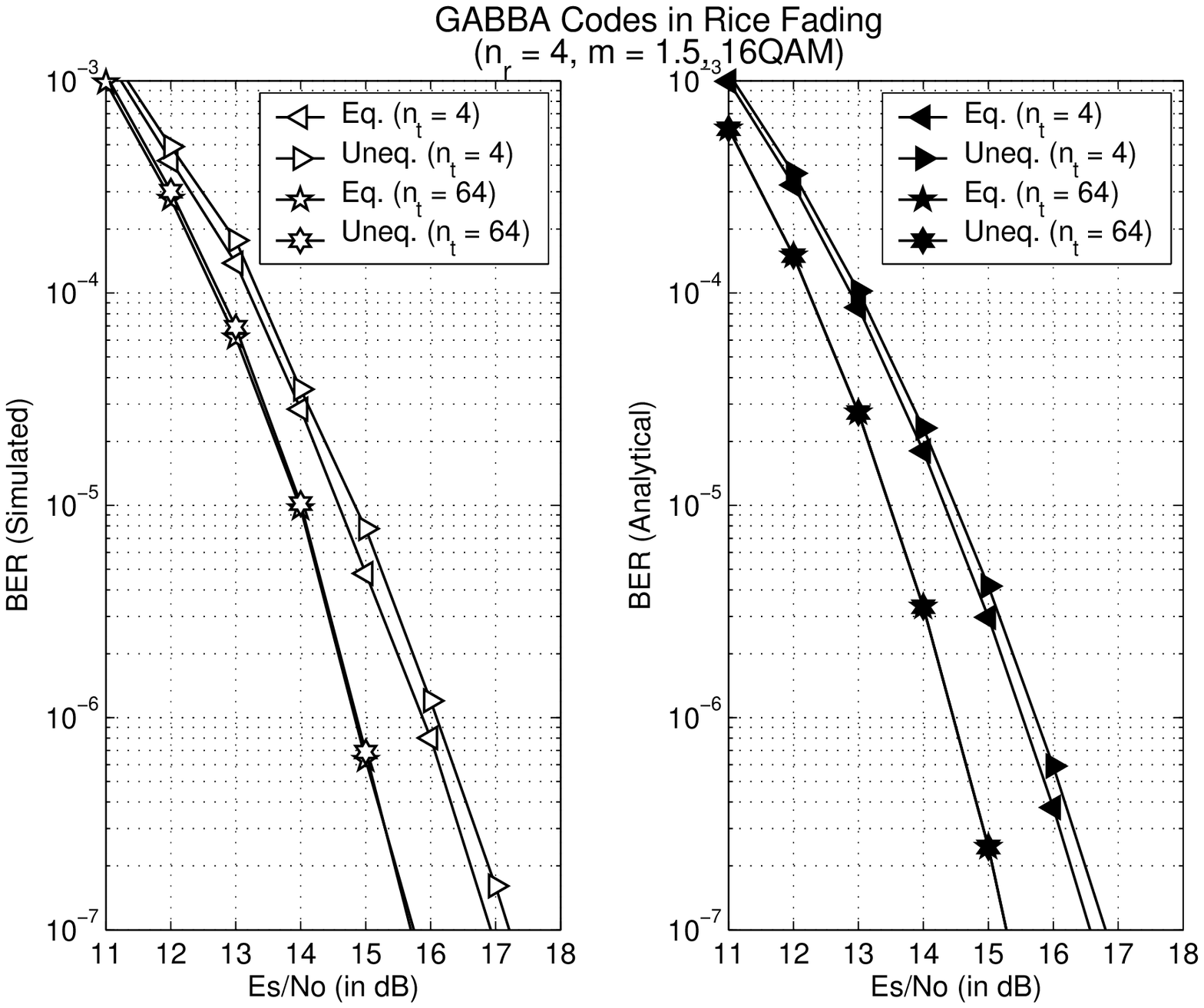}
\caption{Performance of GABBA codes of different sizes over $16$QAM
symbols in Rice block-fading channels with $4$ receive antennas.
Simulated results (white markers) were obtained using the orthogonal
GABBA decoder described in section
\ref{OrthogonalDecodabilityOfGABBACodes}, while analytical results
(black markers) were computed using equation
\eqref{BER_Fading_QAM_MGF}, with $\rho=1$ and $\eta=1$. Both
simulated and analytical results are given for channels with equal
and unequal (linear) power distribution profiles across the transmit
channels (see section
\ref{PerformanceUnequalBlock-fadingChannelsSameStatistics}).}
\label{GABBANtxVar_Nrx4_m15_16QAM}%
\end{figure}

In figure \ref{GABBANtxVar_Nrx4_m15_16QAM}, the simulated and
analytical performances of GABBA codes of sizes $4$ and $64$, over
$16$QAM-modulated symbols in an ensemble of Rice block-fading
channels with equal and unequal (linear) power distributions are
compared.
It can be seen that the impact of an unequal power distribution on
the ergodic performance of GABBA-encoded systems loses relevance
when large codes are used.
As a side product of the comparison, it is once again verified that
orthogonally decoded GABBA codes (simulated results) perform very
close to the analytical bound (in this case given by equation
\eqref{BER_Fading_QAM_MGF} in combination with equation
\eqref{ElementaryMGF_IntegralRhoEta}, with $N = K$, $\rho=1$ and
$\gamma_n = \dfrac{2k}{K}$).

\subsubsection{Block-fading Channels of Different Statistics}
\label{PerformanceBlock-fadingChannelsDifferentStatistics} \quad %

\begin{figure}[h!]
\includegraphics[width=\columnwidth]{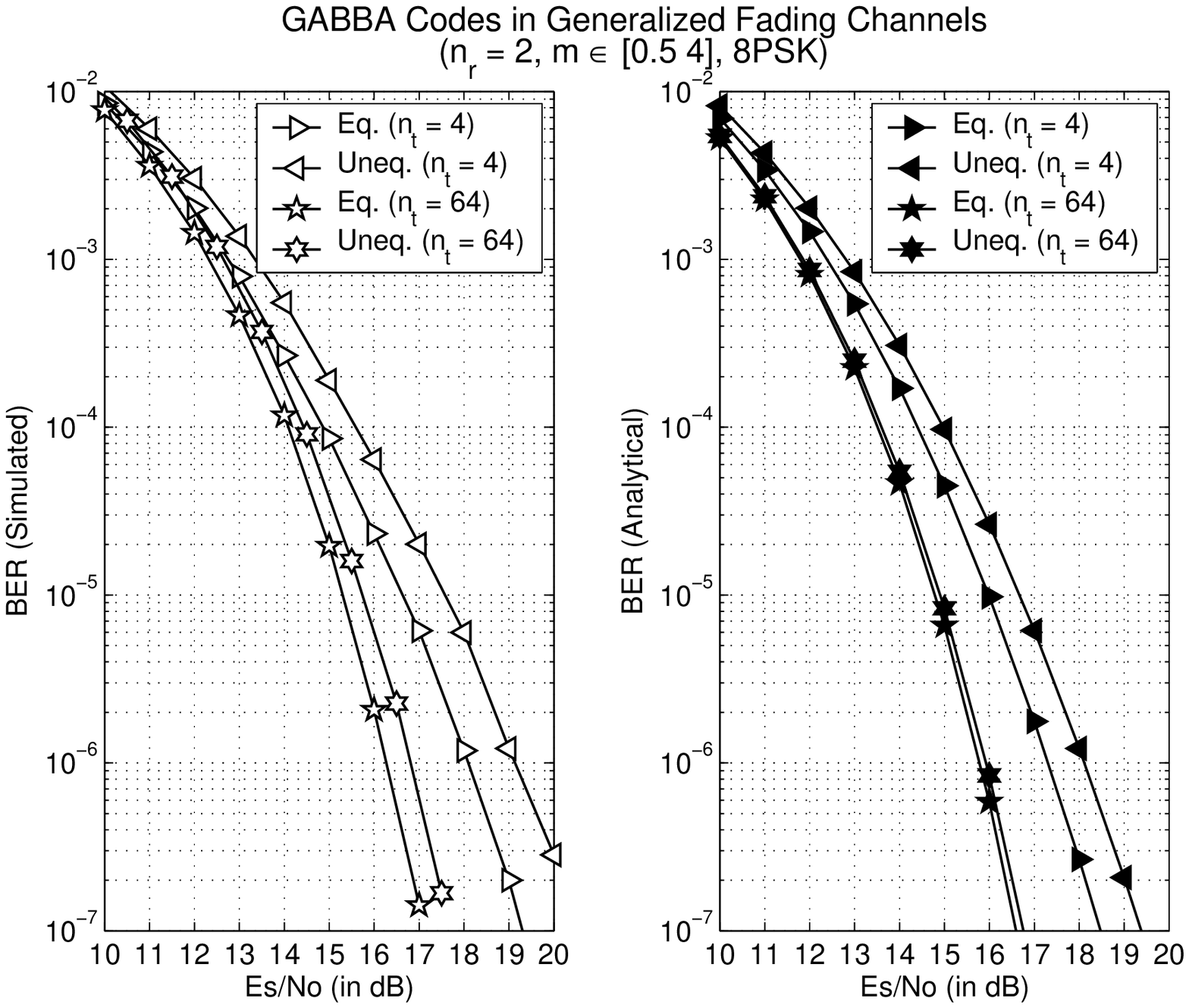}
\caption{Performance of GABBA codes of different sizes over $8$PSK
symbols in generalized block-fading channels with $2$ receive
antennas. Simulated results (white markers) were obtained using the
orthogonal GABBA decoder described in section
\ref{OrthogonalDecodabilityOfGABBACodes}, while analytical results
(black markers) were computed using equation
\eqref{BER_Fading_QAM_MGF}, with $\rho=1$ and $\eta=1$. Both
simulated and analytical results are given for channels with linear
power distribution (see section
\ref{PerformanceUnequalBlock-fadingChannelsSameStatistics}) and
fading severity factors $m$ spanning the interval $[0.5, 4]$, as
well as equipower channels, and a mean fading severity factor
$m=2.5$.}
\label{GABBANtxVar_Nrx_2_mVar_8PSK_OmegaVar}%
\end{figure}

Next, the performances of GABBA codes in uncorrelated block-fading
channels with equal and unequal statistics are compared.
Results obtained both analytically and through computer simulations
are shown in figure \ref{GABBANtxVar_Nrx_2_mVar_8PSK_OmegaVar}.
The curves corresponding to the unequal fading channels were
obtained by assigning to the $k$-th branch a fading severity factor
$m_k$ and a power factor $\Omega_k$ such that $m_k$ spans the
interval $[0.5,4]$ uniformly, with $m_{k+1}>m_k$, while
$\sum\Omega_k = 1$, with $\Omega_{k+1} < \Omega_k$.
In other words, the channels experiencing more severe fading are
also the most powerful ones, such that no particular channel in the
ensemble is ``dominant''.
This distribution of parameters is selected so as to prevent a few
channels to dominate the others, ensuring that, instead, all
branches have significance in providing diversity.
This selection of parameters is associated to scenario faced by
peer-to-peer MIMO systems employing diversity-optimized multi-mode
adaptive antennas with orthogonal pattern adaptability\footnote{It
is assumed here that the beamforming strategy is such that narrow
beams are pointed towards stronger multipath components while wider
beams are formed towards portions of the space characterized by more
dispersive multipath clusters. A consequence of such approach would
be that the narrower beams would collect less energy (fewer
multipath components), but exhibit less severe fading, In contrast,
while the wider beams would collect more energy (more multipath
components), but also exhibit severer fading.}.

The comparison in figure \ref{GABBANtxVar_Nrx_2_mVar_8PSK_OmegaVar}
demonstrates how the use of a large GABBA code results in more
robustness against shadowing and other effects that may affect the
power distribution and fading statistics of the diversity branches
available in peer-to-peer MIMO systems with multi-mode antennas.

The unitary rate and full diversity achieved by GABBA codes ensure
that they outperform all (presently known) alternative STBCs of
linear construction and comparable complexity (orthogonally
decodability).
It is, therefore, of little use to show comparisons between GABBA
codes and other linear STBCs such as the Alamouti scheme
\cite{MyListOfPapers:Alamouti1998}, the OSTBC
\cite{MyListOfPapers:Tarokh1999-2, MyListOfPapers:Tarokh1999-3}, and
the maximal-rate minimum-delay (MRMD) codes
\cite{MyListOfPapers:Tirkkonen-2002}.
The comparison of these codes can, nevertheless, be easily performed
analytically utilizing equations \eqref{BER_Fading_PSK_MGF} and
\eqref{BER_Fading_QAM_MGF}, by simply selecting for each technique
the corresponding values of $\rho$ (for rate) and $\eta$ (for
diversity order).

\section{Shannon, and Modulation-Constrained Rate Capacities}
\label{GABBA_Rate_and_Capacity}

In this section, the performance of the proposed GABBA codes is
analyzed from a capacity point of view.

The Shannon capacities (over Gaussian constellations) of MIMO
systems in ergodic and deterministic fading channels were derived in
\cite{MyListOfPapers:Telatar1995, MyListOfPapers:Foschini1998}.
More recently, the modulation-constrained rate capacity (maximum
average throughput) of MIMO systems over $M$-ary PSK constellations
has also been derived \cite{MyListOfPapers:Wenyan2005}.
It is known, however, that the rate capacity of linear STBCs (which
do not provide multiplexation gain) is bound by the Shannon capacity
of the corresponding SIMO channel \cite{MyListOfPapers:Sandhu2000}.

%
%
It is of interest, therefore, to compare the average throughput
theoretically achievable by GABBA codes in systems utilizing
different modulation schemes over ensembles of $n_t$-by-$n_r$
antennas against the ergodic capacity of the equivalent $1$-by-$n_r$
SIMO channel.
Given the exact BER probability formulae derived in section
\ref{GABBA_BER Analytical}, a simple yet effective approach at hand
is to consider the rate capacity of GABBA codes with hard-decision
on the orthogonal symbol estimates.
In this case, one can simply considering the binary symmetric
channel (as seen from the input of the space-time block encoder up
to the output of the corresponding decoder, with the physical MIMO
fading channel included), with the transition probabilities given by
$\bar{P}_{b}$ and $1-\bar{P}_{b}$, respectively
\cite{MyListOfPapers:ThomasCover}.
This leads to
\begin{equation}
\label{STBCModulationDependentCapacity}%
C_{_{\!M}}(\gamma) = \rho \cdot M \cdot\bigg[1 + \bar{P}_{b}\cdot\log_2 \bar{P}_{b} + \left(1\!-\!\bar{P}_{b}\!\right)\cdot\log_2\!\left(1\!-\!\bar{P}_{b}\!\right)\bigg],%
\end{equation}
where $\bar{P}_{b}$ is computed with equations
\eqref{BER_Fading_PSK_MGF} or \eqref{BER_Fading_QAM_MGF}, at the
given signal-to-noise ratio, considering the corresponding number of
transmit and receive antennas $n_t$ and $n_r$, and with parameters
$\rho = 1$ and $\eta = 1$\footnote{For simplicity, all these
parameters are omitted from the notation of $\bar{P}_{b}$.}.

\begin{figure}[t]
\centering  %
\subfigure[Dominant transmit diversity.] %
{\label{GABBAThroughput_X_Capacities_nt16_nr4_m1_Ptx_1}
\includegraphics[width=0.7\columnwidth]{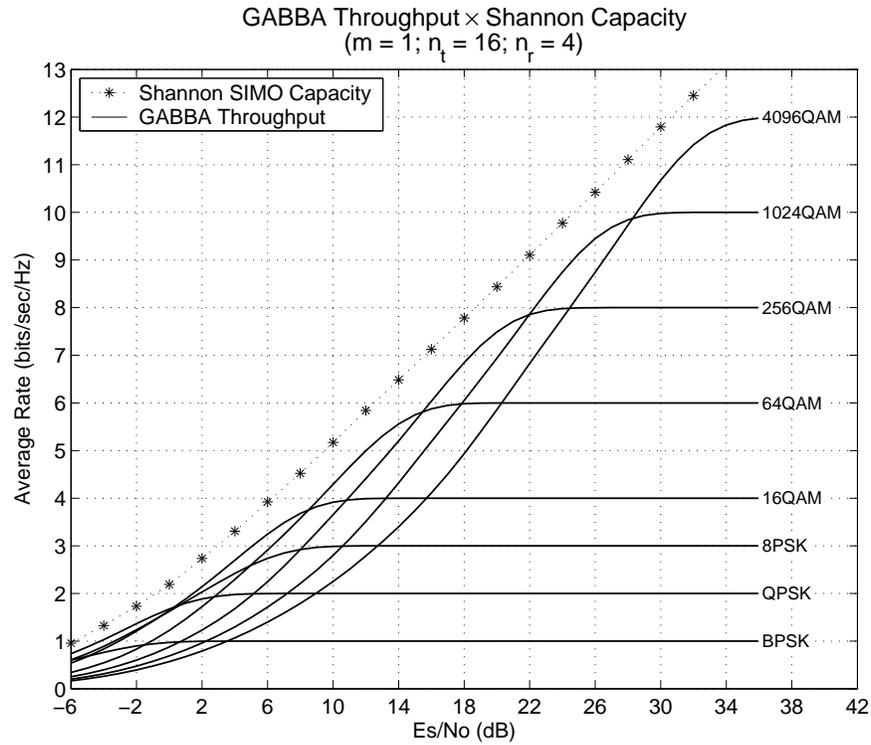}}\\%
%
\subfigure[Dominant receive diversity.] %
{\label{GABBAThroughput_X_Capacities_nt4_nr16_m1_Ptx_1}%
\includegraphics[width=0.7\columnwidth]{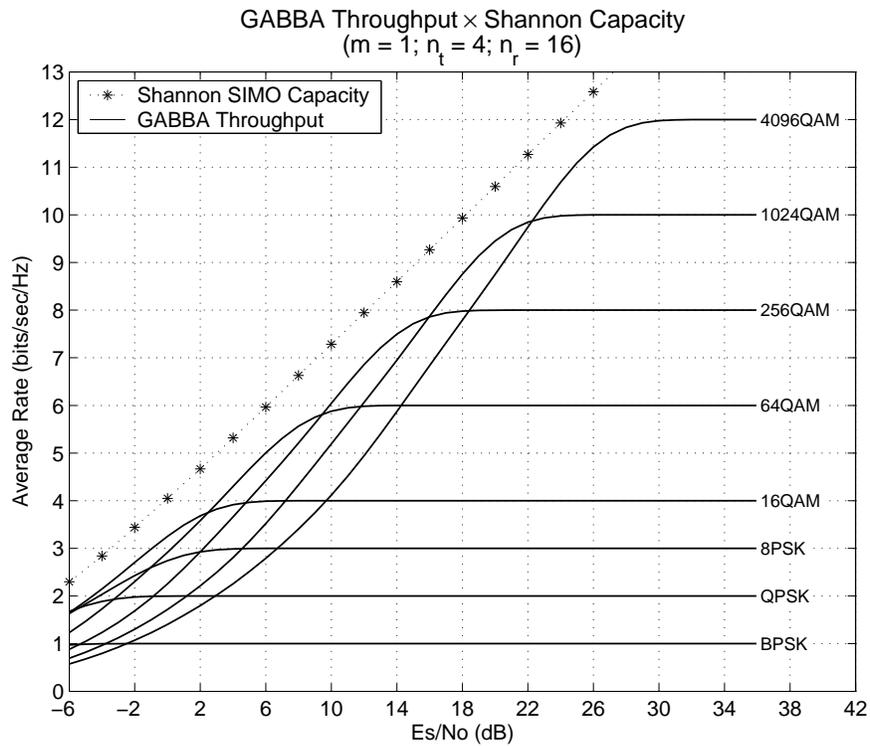}}
\caption{Modulation constrained achievable rate (throughput) of
GABBA codes of different sizes in Rayleigh fading channel and with
different modulation schemes against signal-to-noise ratio.}
\label{GABBAThroughput_X_Capacities_Ptx_1}
\end{figure}

Using equation \eqref{STBCModulationDependentCapacity}, the
hard-decision modulation-constrained rate capacities of GABBA codes
of different sizes, achieved with different modulation schemes, in
both dominant transmit ($n_t > n_r$) and receive ($n_r > n_t$)
diversity scenarios, are compared against the SIMO Shannon capacity
in figures \ref{GABBAThroughput_X_Capacities_nt16_nr4_m1_Ptx_1} and
\ref{GABBAThroughput_X_Capacities_nt4_nr16_m1_Ptx_1}, respectively.
It is found that GABBA-encoded systems with optimized modulation
comes to roughly $1$ bit/sec/Hz from the Shannon SISO capacity.
This is a rather remarkable performance when considered that the
Shannon capacity is related to an unfeasible Gaussian modulation and
soft-decision decoding.
In contrast, the modulation-dependent rates shown for the GABBA
scheme can be reached at significantly lower complexities, through
concatenation with efficient inner codes such LDPC or Turbo Codes
\cite{MyListOfPapers:ShuLinBook2004}, and employing hard decision at
the symbol-level.
In other words, it is learned from figures
\ref{GABBAThroughput_X_Capacities_nt16_nr4_m1_Ptx_1} and
\ref{GABBAThroughput_X_Capacities_nt4_nr16_m1_Ptx_1}, that the
considerably more complex soft-symbol decoding of GABBA codes
(concatenated with an inner code) would only improve the rate of the
system by $1$ bit/sec/Hz.

\begin{figure}[h!]
\includegraphics[width=\columnwidth]{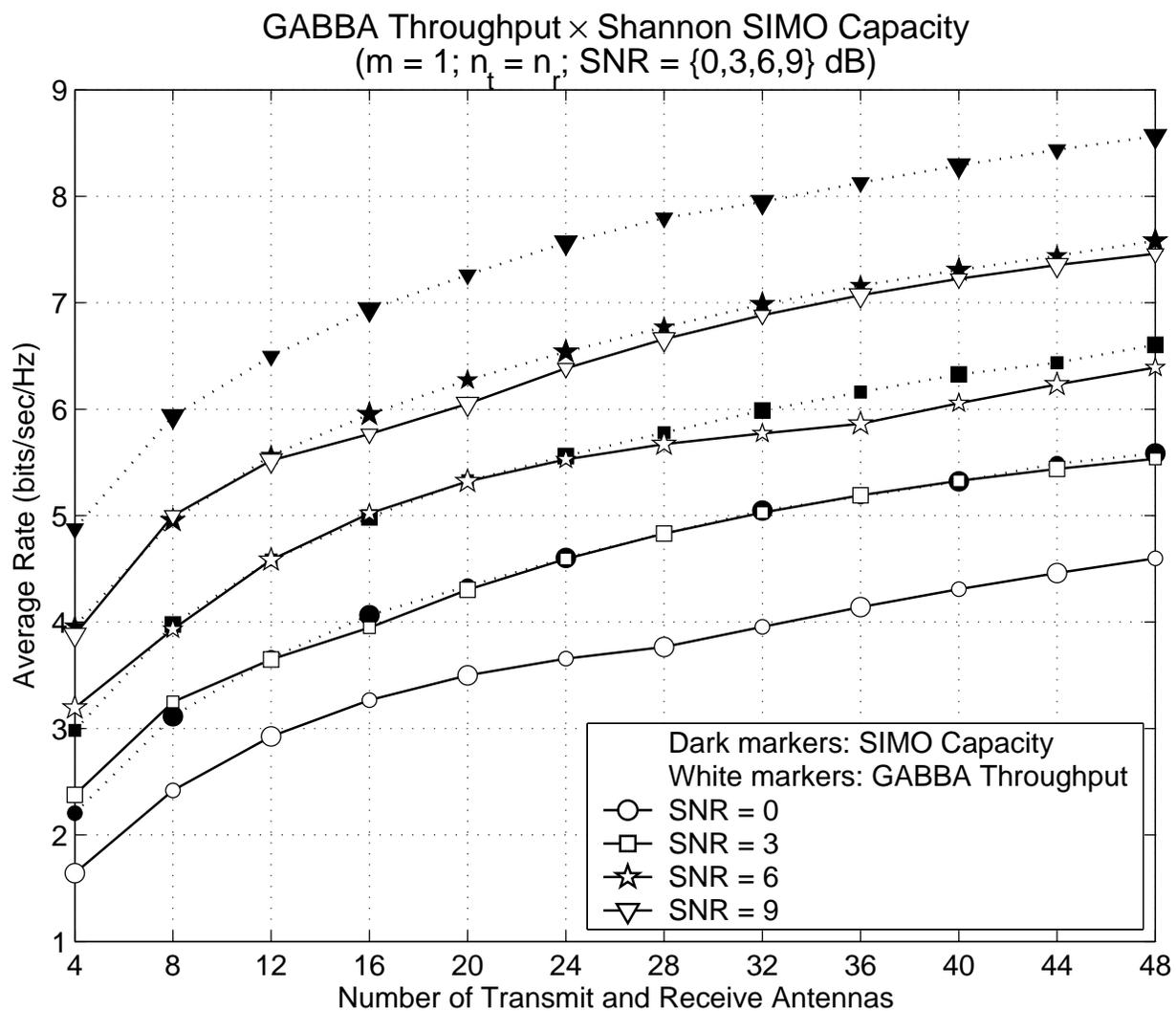}
\caption{Throughput of GABBA codes against number of transmit and
receive antennas ($n_t = n_r$) in Rayleigh fading channel with
optimized modulation.}
\label{GABBAThroughput_X_Capacities_SNR_Var_m1_Ntx_Nrx}%
\end{figure}

It is also noticeable from figures
\ref{GABBAThroughput_X_Capacities_nt16_nr4_m1_Ptx_1} and
\ref{GABBAThroughput_X_Capacities_nt4_nr16_m1_Ptx_1} that for lower
signal-to-noise ratios and relatively lower-order constellations (up
to $64$QAM), the GABBA codes considered ($n_t = 4$ and $n_t = 16$)
actually come to less than $1$ bit/sec/Hz from the Shannon bound.
This result is further illustrated in figure
\ref{GABBAThroughput_X_Capacities_SNR_Var_m1_Ntx_Nrx}, where the
achievable rates of balanced MIMO systems ($n_t = n_r$) employing
optimum modulation and GABBA codes of different sizes\footnote{The
curves are the envelopes of all possible regular modulations ranging
from BPSK to $4096$QAM, with hard-decision at the symbol level.} are
compared to the corresponding SIMO Shannon capacities.
It is interesting to notice that MIMO system with the GABBA scheme
and optimum modulation exhibits the same general behavior as the
``ideal'' Shannon bound when it comes to rate increase in proportion
to the number of antennas used in the system.

\section{Conclusions}

In this paper, a family of linear space-time block codes (STBCs) of
unitary rate and full diversity, systematic constructed over
arbitrary constellations for any number of transmit antennas are
introduced.
The codes were obtained from a generalization of the existing ABBA
STBCs, a.k.a \emph{quasi}-orthogonal STBCs (QO-STBCs) and,
therefore, are denominated generalized ABBA (GABBA) codes (see
section \ref{MotherCodeConstruction}).

It was also shown that the GABBA codes admit fully orthogonal
(symbol-by-symbol) decoding.
Two key ideas underlie the derivation of the orthogonal GABBA
decoder.
The first is a transformation applied to the system equation (see
section \ref{DecoderPreliminaries}), that exposes the relationship
between the code structure and the interference amongst symbols
embedded in the same encoding block, characterizing it in terms of
an equivalent encoded channel matrix (see section
\ref{StructureOfGABBAEncodedChannelMatrices}).
The second is the exploitation of partition orthogonality properties
present in structure of the GABBA codes, which are shown to be
transferred to the corresponding encoded channel matrices.
It was shown that, as a consequence of these techniques/properties,
the received signal vector can decomposed into lower-dimension
tuples, each dependent only on certain subsets of the transmitted
symbols, via simple linear combinations with coefficients derived
directly from the channel estimates.
Orthogonal decodability results from the nested application of such
linear combinations, with no matrix inversion or iterative signal
processing required.

The performance of the proposed GABBA codes were studied both
analytically and through computer simulations.
The exact BER probability of the codes with ML the decoder over
generalized (uncorrelated) fading channels was derived analytically,
under the assumption of perfect channel knowledge at the receiver
(see section \ref{GABBA_BER Analytical}).
The formulae were then compared to simulation results obtained with
the proposed orthogonal decoder.
The comparison reveals that the proposed GABBA solution, despite its
very low complexity, achieves nearly the same performance of the
bound corresponding to the ML-decoded system, especially in systems
with large numbers of antennas (see section
\ref{GABBA_BER_Simulated}).

A brief analysis of the average transmission rate achieved with
GABBA codes was also provided.
The analysis indicate that the transmission rates achievable with
the full(unitary)-rate full-diversity GABBA codes over optimized
constellations, with the corresponding orthogonal decoder and hard
symbol decision, have the potential to come $1$ bit/sec/Hz from the
theoretical bound given by the Shannol SIMO channel capacity.
It has been shown that higher rates can be achieved using
unitary-rate STBCs as building blocks in a layered structure.
Such approaches have already been demonstrated for the known ABBA
scheme and OSTBCs \cite[Chap. 9]{MyListOfPapers:TirkkonenBook2003},
\cite{MyListOfPapers:Sethuraman2003, MyListOfPapers:HassibiIT2003}.
It is foreseeable, therefore, that these techniques extend naturally
to the family of GABBA codes here introduced, with the added
flexibility, scalability and low complexity to its advantage.

In addition to the equations, proofs and figures found in the main
body of the paper, extensive material is provided in the appendices
to allow the verification of all claims, including explicit examples
and Matlab implementations of the encoder, the decoder and the BER
formulae.

\newpage

\newpage

\newpage

\newpage

\newpage

\appendices%
\clearpage
\newpage
\section{GABBA Encoder}
\label{Appendix_GABBAEncodingMatrix}%
\subsection{Example: GABBA Mother Encoding Matrix ($K = 32$)}
\label{ExampleGABBAEncodingMatrix} %
\vspace{-2em}
\begin{eqnarray*}
\label{EncodingMatrixC32}%
\renewcommand{\arraystretch}{0.5}
{\scriptstyle\mathbf{C}_{32}=}\!\!\left[\!\!%
{\tiny
\begin{array}{rrrrrrrrrrrrrrrrrrrrrrrrrrrrrrrr}
 &&&&&&&&&&&&&&&\\ [0.1ex]%
 \!\!\!\! s_{1}   \!\!\!\! & \!\!\!\!\!\!\! s_{2}   \!\!\!\! & \!\!\!\!\!\!\! s_{3}   \!\!\!\! & \!\!\!\!\!\!\! s_{4}   \!\!\!\! & \!\!\!\!\!\!\! s_{5}   \!\!\!\! & \!\!\!\!\!\!\! s_{6}   \!\!\!\! & \!\!\!\!\!\!\! s_{7}   \!\!\!\! & \!\!\!\!\!\!\! s_{8}   \!\!\!\! & \!\!\!\!\!\!\! s_{9}   \!\!\!\! & \!\!\!\!\!\!\! s_{10}  \!\!\!\! & \!\!\!\!\!\!\! s_{11}  \!\!\!\! & \!\!\!\!\!\!\! s_{12}  \!\!\!\! & \!\!\!\!\!\!\! s_{13}  \!\!\!\! & \!\!\!\!\!\!\! s_{14}  \!\!\!\! & \!\!\!\!\!\!\! s_{15}  \!\!\!\! & \!\!\!\!\!\!\! s_{16}  \!\!\!\! & \!\!\!\!\!\!\! s_{17}  \!\!\!\! & \!\!\!\!\!\!\! s_{18}  \!\!\!\! & \!\!\!\!\!\!\! s_{19}  \!\!\!\! & \!\!\!\!\!\!\! s_{20}  \!\!\!\! & \!\!\!\!\!\!\! s_{21}  \!\!\!\! & \!\!\!\!\!\!\! s_{22}  \!\!\!\! & \!\!\!\!\!\!\! s_{23}  \!\!\!\! & \!\!\!\!\!\!\! s_{24}  \!\!\!\! & \!\!\!\!\!\!\! s_{25}  \!\!\!\! & \!\!\!\!\!\!\! s_{26}  \!\!\!\! & \!\!\!\!\!\!\! s_{27}  \!\!\!\! & \!\!\!\!\!\!\! s_{28}  \!\!\!\! & \!\!\!\!\!\!\! s_{29}  \!\!\!\! & \!\!\!\!\!\!\! s_{30}  \!\!\!\! & \!\!\!\!\!\!\! s_{31}  \!\!\!\! & \!\!\!\!\!\!\! s_{32}  \\
 \!\!-\!\!s_{2}   \!\!\!\! & \!\!\!\!\!\!\! s_{1}   \!\!\!\! & \!\!\!\!\!-\!\!s_{4}   \!\!\!\! & \!\!\!\!\!\!\! s_{3}   \!\!\!\! & \!\!\!\!\!-\!\!s_{6}   \!\!\!\! & \!\!\!\!\!\!\! s_{5}   \!\!\!\! & \!\!\!\!\!-\!\!s_{8}   \!\!\!\! & \!\!\!\!\!\!\! s_{7}   \!\!\!\! & \!\!\!\!\!-\!\!s_{10}  \!\!\!\! & \!\!\!\!\!\!\! s_{9}   \!\!\!\! & \!\!\!\!\!-\!\!s_{12}  \!\!\!\! & \!\!\!\!\!\!\! s_{11}  \!\!\!\! & \!\!\!\!\!-\!\!s_{14}  \!\!\!\! & \!\!\!\!\!\!\! s_{13}  \!\!\!\! & \!\!\!\!\!-\!\!s_{16}  \!\!\!\! & \!\!\!\!\!\!\! s_{15}  \!\!\!\! & \!\!\!\!\!-\!\!s_{18}  \!\!\!\! & \!\!\!\!\!\!\! s_{17}  \!\!\!\! & \!\!\!\!\!-\!\!s_{20}  \!\!\!\! & \!\!\!\!\!\!\! s_{19}  \!\!\!\! & \!\!\!\!\!-\!\!s_{22}  \!\!\!\! & \!\!\!\!\!\!\! s_{21}  \!\!\!\! & \!\!\!\!\!-\!\!s_{24}  \!\!\!\! & \!\!\!\!\!\!\! s_{23}  \!\!\!\! & \!\!\!\!\!-\!\!s_{26}  \!\!\!\! & \!\!\!\!\!\!\! s_{25}  \!\!\!\! & \!\!\!\!\!-\!\!s_{28}  \!\!\!\! & \!\!\!\!\!\!\! s_{27}  \!\!\!\! & \!\!\!\!\!-\!\!s_{30}  \!\!\!\! & \!\!\!\!\!\!\! s_{29}  \!\!\!\! & \!\!\!\!\!-\!\!s_{32}  \!\!\!\! & \!\!\!\!\!\!\! s_{31}  \\
 \!\!-\!\!s_{3}   \!\!\!\! & \!\!\!\!\!-\!\!s_{4}   \!\!\!\! & \!\!\!\!\!\!\! s_{1}   \!\!\!\! & \!\!\!\!\!\!\! s_{2}   \!\!\!\! & \!\!\!\!\!-\!\!s_{7}   \!\!\!\! & \!\!\!\!\!-\!\!s_{8}   \!\!\!\! & \!\!\!\!\!\!\! s_{5}   \!\!\!\! & \!\!\!\!\!\!\! s_{6}   \!\!\!\! & \!\!\!\!\!-\!\!s_{11}  \!\!\!\! & \!\!\!\!\!-\!\!s_{12}  \!\!\!\! & \!\!\!\!\!\!\! s_{9}   \!\!\!\! & \!\!\!\!\!\!\! s_{10}  \!\!\!\! & \!\!\!\!\!-\!\!s_{15}  \!\!\!\! & \!\!\!\!\!-\!\!s_{16}  \!\!\!\! & \!\!\!\!\!\!\! s_{13}  \!\!\!\! & \!\!\!\!\!\!\! s_{14}  \!\!\!\! & \!\!\!\!\!-\!\!s_{19}  \!\!\!\! & \!\!\!\!\!-\!\!s_{20}  \!\!\!\! & \!\!\!\!\!\!\! s_{17}  \!\!\!\! & \!\!\!\!\!\!\! s_{18}  \!\!\!\! & \!\!\!\!\!-\!\!s_{23}  \!\!\!\! & \!\!\!\!\!-\!\!s_{24}  \!\!\!\! & \!\!\!\!\!\!\! s_{21}  \!\!\!\! & \!\!\!\!\!\!\! s_{22}  \!\!\!\! & \!\!\!\!\!-\!\!s_{27}  \!\!\!\! & \!\!\!\!\!-\!\!s_{28}  \!\!\!\! & \!\!\!\!\!\!\! s_{25}  \!\!\!\! & \!\!\!\!\!\!\! s_{26}  \!\!\!\! & \!\!\!\!\!-\!\!s_{31}  \!\!\!\! & \!\!\!\!\!-\!\!s_{32}  \!\!\!\! & \!\!\!\!\!\!\! s_{29}  \!\!\!\! & \!\!\!\!\!\!\! s_{30}  \\
 \!\!\!\! s_{4}   \!\!\!\! & \!\!\!\!\!-\!\!s_{3}   \!\!\!\! & \!\!\!\!\!-\!\!s_{2}   \!\!\!\! & \!\!\!\!\!\!\! s_{1}   \!\!\!\! & \!\!\!\!\!\!\! s_{8}   \!\!\!\! & \!\!\!\!\!-\!\!s_{7}   \!\!\!\! & \!\!\!\!\!-\!\!s_{6}   \!\!\!\! & \!\!\!\!\!\!\! s_{5}   \!\!\!\! & \!\!\!\!\!\!\! s_{12}  \!\!\!\! & \!\!\!\!\!-\!\!s_{11}  \!\!\!\! & \!\!\!\!\!-\!\!s_{10}  \!\!\!\! & \!\!\!\!\!\!\! s_{9}   \!\!\!\! & \!\!\!\!\!\!\! s_{16}  \!\!\!\! & \!\!\!\!\!-\!\!s_{15}  \!\!\!\! & \!\!\!\!\!-\!\!s_{14}  \!\!\!\! & \!\!\!\!\!\!\! s_{13}  \!\!\!\! & \!\!\!\!\!\!\! s_{20}  \!\!\!\! & \!\!\!\!\!-\!\!s_{19}  \!\!\!\! & \!\!\!\!\!-\!\!s_{18}  \!\!\!\! & \!\!\!\!\!\!\! s_{17}  \!\!\!\! & \!\!\!\!\!\!\! s_{24}  \!\!\!\! & \!\!\!\!\!-\!\!s_{23}  \!\!\!\! & \!\!\!\!\!-\!\!s_{22}  \!\!\!\! & \!\!\!\!\!\!\! s_{21}  \!\!\!\! & \!\!\!\!\!\!\! s_{28}  \!\!\!\! & \!\!\!\!\!-\!\!s_{27}  \!\!\!\! & \!\!\!\!\!-\!\!s_{26}  \!\!\!\! & \!\!\!\!\!\!\! s_{25}  \!\!\!\! & \!\!\!\!\!\!\! s_{32}  \!\!\!\! & \!\!\!\!\!-\!\!s_{31}  \!\!\!\! & \!\!\!\!\!-\!\!s_{30}  \!\!\!\! & \!\!\!\!\!\!\! s_{29}  \\
 \!\!-\!\!s_{5}   \!\!\!\! & \!\!\!\!\!-\!\!s_{6}   \!\!\!\! & \!\!\!\!\!-\!\!s_{7}   \!\!\!\! & \!\!\!\!\!-\!\!s_{8}   \!\!\!\! & \!\!\!\!\!\!\! s_{1}   \!\!\!\! & \!\!\!\!\!\!\! s_{2}   \!\!\!\! & \!\!\!\!\!\!\! s_{3}   \!\!\!\! & \!\!\!\!\!\!\! s_{4}   \!\!\!\! & \!\!\!\!\!-\!\!s_{13}  \!\!\!\! & \!\!\!\!\!-\!\!s_{14}  \!\!\!\! & \!\!\!\!\!-\!\!s_{15}  \!\!\!\! & \!\!\!\!\!-\!\!s_{16}  \!\!\!\! & \!\!\!\!\!\!\! s_{9}   \!\!\!\! & \!\!\!\!\!\!\! s_{10}  \!\!\!\! & \!\!\!\!\!\!\! s_{11}  \!\!\!\! & \!\!\!\!\!\!\! s_{12}  \!\!\!\! & \!\!\!\!\!-\!\!s_{21}  \!\!\!\! & \!\!\!\!\!-\!\!s_{22}  \!\!\!\! & \!\!\!\!\!-\!\!s_{23}  \!\!\!\! & \!\!\!\!\!-\!\!s_{24}  \!\!\!\! & \!\!\!\!\!\!\! s_{17}  \!\!\!\! & \!\!\!\!\!\!\! s_{18}  \!\!\!\! & \!\!\!\!\!\!\! s_{19}  \!\!\!\! & \!\!\!\!\!\!\! s_{20}  \!\!\!\! & \!\!\!\!\!-\!\!s_{29}  \!\!\!\! & \!\!\!\!\!-\!\!s_{30}  \!\!\!\! & \!\!\!\!\!-\!\!s_{31}  \!\!\!\! & \!\!\!\!\!-\!\!s_{32}  \!\!\!\! & \!\!\!\!\!\!\! s_{25}  \!\!\!\! & \!\!\!\!\!\!\! s_{26}  \!\!\!\! & \!\!\!\!\!\!\! s_{27}  \!\!\!\! & \!\!\!\!\!\!\! s_{28}  \\
 \!\!\!\! s_{6}   \!\!\!\! & \!\!\!\!\!-\!\!s_{5}   \!\!\!\! & \!\!\!\!\!\!\! s_{8}   \!\!\!\! & \!\!\!\!\!-\!\!s_{7}   \!\!\!\! & \!\!\!\!\!-\!\!s_{2}   \!\!\!\! & \!\!\!\!\!\!\! s_{1}   \!\!\!\! & \!\!\!\!\!-\!\!s_{4}   \!\!\!\! & \!\!\!\!\!\!\! s_{3}   \!\!\!\! & \!\!\!\!\!\!\! s_{14}  \!\!\!\! & \!\!\!\!\!-\!\!s_{13}  \!\!\!\! & \!\!\!\!\!\!\! s_{16}  \!\!\!\! & \!\!\!\!\!-\!\!s_{15}  \!\!\!\! & \!\!\!\!\!-\!\!s_{10}  \!\!\!\! & \!\!\!\!\!\!\! s_{9}   \!\!\!\! & \!\!\!\!\!-\!\!s_{12}  \!\!\!\! & \!\!\!\!\!\!\! s_{11}  \!\!\!\! & \!\!\!\!\!\!\! s_{22}  \!\!\!\! & \!\!\!\!\!-\!\!s_{21}  \!\!\!\! & \!\!\!\!\!\!\! s_{24}  \!\!\!\! & \!\!\!\!\!-\!\!s_{23}  \!\!\!\! & \!\!\!\!\!-\!\!s_{18}  \!\!\!\! & \!\!\!\!\!\!\! s_{17}  \!\!\!\! & \!\!\!\!\!-\!\!s_{20}  \!\!\!\! & \!\!\!\!\!\!\! s_{19}  \!\!\!\! & \!\!\!\!\!\!\! s_{30}  \!\!\!\! & \!\!\!\!\!-\!\!s_{29}  \!\!\!\! & \!\!\!\!\!\!\! s_{32}  \!\!\!\! & \!\!\!\!\!-\!\!s_{31}  \!\!\!\! & \!\!\!\!\!-\!\!s_{26}  \!\!\!\! & \!\!\!\!\!\!\! s_{25}  \!\!\!\! & \!\!\!\!\!-\!\!s_{28}  \!\!\!\! & \!\!\!\!\!\!\! s_{27}  \\
 \!\!\!\! s_{7}   \!\!\!\! & \!\!\!\!\!\!\! s_{8}   \!\!\!\! & \!\!\!\!\!-\!\!s_{5}   \!\!\!\! & \!\!\!\!\!-\!\!s_{6}   \!\!\!\! & \!\!\!\!\!-\!\!s_{3}   \!\!\!\! & \!\!\!\!\!-\!\!s_{4}   \!\!\!\! & \!\!\!\!\!\!\! s_{1}   \!\!\!\! & \!\!\!\!\!\!\! s_{2}   \!\!\!\! & \!\!\!\!\!\!\! s_{15}  \!\!\!\! & \!\!\!\!\!\!\! s_{16}  \!\!\!\! & \!\!\!\!\!-\!\!s_{13}  \!\!\!\! & \!\!\!\!\!-\!\!s_{14}  \!\!\!\! & \!\!\!\!\!-\!\!s_{11}  \!\!\!\! & \!\!\!\!\!-\!\!s_{12}  \!\!\!\! & \!\!\!\!\!\!\! s_{9}   \!\!\!\! & \!\!\!\!\!\!\! s_{10}  \!\!\!\! & \!\!\!\!\!\!\! s_{23}  \!\!\!\! & \!\!\!\!\!\!\! s_{24}  \!\!\!\! & \!\!\!\!\!-\!\!s_{21}  \!\!\!\! & \!\!\!\!\!-\!\!s_{22}  \!\!\!\! & \!\!\!\!\!-\!\!s_{19}  \!\!\!\! & \!\!\!\!\!-\!\!s_{20}  \!\!\!\! & \!\!\!\!\!\!\! s_{17}  \!\!\!\! & \!\!\!\!\!\!\! s_{18}  \!\!\!\! & \!\!\!\!\!\!\! s_{31}  \!\!\!\! & \!\!\!\!\!\!\! s_{32}  \!\!\!\! & \!\!\!\!\!-\!\!s_{29}  \!\!\!\! & \!\!\!\!\!-\!\!s_{30}  \!\!\!\! & \!\!\!\!\!-\!\!s_{27}  \!\!\!\! & \!\!\!\!\!-\!\!s_{28}  \!\!\!\! & \!\!\!\!\!\!\! s_{25}  \!\!\!\! & \!\!\!\!\!\!\! s_{26}  \\
 \!\!-\!\!s_{8}   \!\!\!\! & \!\!\!\!\!\!\! s_{7}   \!\!\!\! & \!\!\!\!\!\!\! s_{6}   \!\!\!\! & \!\!\!\!\!-\!\!s_{5}   \!\!\!\! & \!\!\!\!\!\!\! s_{4}   \!\!\!\! & \!\!\!\!\!-\!\!s_{3}   \!\!\!\! & \!\!\!\!\!-\!\!s_{2}   \!\!\!\! & \!\!\!\!\!\!\! s_{1}   \!\!\!\! & \!\!\!\!\!-\!\!s_{16}  \!\!\!\! & \!\!\!\!\!\!\! s_{15}  \!\!\!\! & \!\!\!\!\!\!\! s_{14}  \!\!\!\! & \!\!\!\!\!-\!\!s_{13}  \!\!\!\! & \!\!\!\!\!\!\! s_{12}  \!\!\!\! & \!\!\!\!\!-\!\!s_{11}  \!\!\!\! & \!\!\!\!\!-\!\!s_{10}  \!\!\!\! & \!\!\!\!\!\!\! s_{9}   \!\!\!\! & \!\!\!\!\!-\!\!s_{24}  \!\!\!\! & \!\!\!\!\!\!\! s_{23}  \!\!\!\! & \!\!\!\!\!\!\! s_{22}  \!\!\!\! & \!\!\!\!\!-\!\!s_{21}  \!\!\!\! & \!\!\!\!\!\!\! s_{20}  \!\!\!\! & \!\!\!\!\!-\!\!s_{19}  \!\!\!\! & \!\!\!\!\!-\!\!s_{18}  \!\!\!\! & \!\!\!\!\!\!\! s_{17}  \!\!\!\! & \!\!\!\!\!-\!\!s_{32}  \!\!\!\! & \!\!\!\!\!\!\! s_{31}  \!\!\!\! & \!\!\!\!\!\!\! s_{30}  \!\!\!\! & \!\!\!\!\!-\!\!s_{29}  \!\!\!\! & \!\!\!\!\!\!\! s_{28}  \!\!\!\! & \!\!\!\!\!-\!\!s_{27}  \!\!\!\! & \!\!\!\!\!-\!\!s_{26}  \!\!\!\! & \!\!\!\!\!\!\! s_{25}  \\
 \!\!-\!\!s_{9}   \!\!\!\! & \!\!\!\!\!-\!\!s_{10}  \!\!\!\! & \!\!\!\!\!-\!\!s_{11}  \!\!\!\! & \!\!\!\!\!-\!\!s_{12}  \!\!\!\! & \!\!\!\!\!-\!\!s_{13}  \!\!\!\! & \!\!\!\!\!-\!\!s_{14}  \!\!\!\! & \!\!\!\!\!-\!\!s_{15}  \!\!\!\! & \!\!\!\!\!-\!\!s_{16}  \!\!\!\! & \!\!\!\!\!\!\! s_{1}   \!\!\!\! & \!\!\!\!\!\!\! s_{2}   \!\!\!\! & \!\!\!\!\!\!\! s_{3}   \!\!\!\! & \!\!\!\!\!\!\! s_{4}   \!\!\!\! & \!\!\!\!\!\!\! s_{5}   \!\!\!\! & \!\!\!\!\!\!\! s_{6}   \!\!\!\! & \!\!\!\!\!\!\! s_{7}   \!\!\!\! & \!\!\!\!\!\!\! s_{8}   \!\!\!\! & \!\!\!\!\!-\!\!s_{25}  \!\!\!\! & \!\!\!\!\!-\!\!s_{26}  \!\!\!\! & \!\!\!\!\!-\!\!s_{27}  \!\!\!\! & \!\!\!\!\!-\!\!s_{28}  \!\!\!\! & \!\!\!\!\!-\!\!s_{29}  \!\!\!\! & \!\!\!\!\!-\!\!s_{30}  \!\!\!\! & \!\!\!\!\!-\!\!s_{31}  \!\!\!\! & \!\!\!\!\!-\!\!s_{32}  \!\!\!\! & \!\!\!\!\!\!\! s_{17}  \!\!\!\! & \!\!\!\!\!\!\! s_{18}  \!\!\!\! & \!\!\!\!\!\!\! s_{19}  \!\!\!\! & \!\!\!\!\!\!\! s_{20}  \!\!\!\! & \!\!\!\!\!\!\! s_{21}  \!\!\!\! & \!\!\!\!\!\!\! s_{22}  \!\!\!\! & \!\!\!\!\!\!\! s_{23}  \!\!\!\! & \!\!\!\!\!\!\! s_{24}  \\
 \!\!\!\! s_{10}  \!\!\!\! & \!\!\!\!\!-\!\!s_{9}   \!\!\!\! & \!\!\!\!\!\!\! s_{12}  \!\!\!\! & \!\!\!\!\!-\!\!s_{11}  \!\!\!\! & \!\!\!\!\!\!\! s_{14}  \!\!\!\! & \!\!\!\!\!-\!\!s_{13}  \!\!\!\! & \!\!\!\!\!\!\! s_{16}  \!\!\!\! & \!\!\!\!\!-\!\!s_{15}  \!\!\!\! & \!\!\!\!\!-\!\!s_{2}   \!\!\!\! & \!\!\!\!\!\!\! s_{1}   \!\!\!\! & \!\!\!\!\!-\!\!s_{4}   \!\!\!\! & \!\!\!\!\!\!\! s_{3}   \!\!\!\! & \!\!\!\!\!-\!\!s_{6}   \!\!\!\! & \!\!\!\!\!\!\! s_{5}   \!\!\!\! & \!\!\!\!\!-\!\!s_{8}   \!\!\!\! & \!\!\!\!\!\!\! s_{7}   \!\!\!\! & \!\!\!\!\!\!\! s_{26}  \!\!\!\! & \!\!\!\!\!-\!\!s_{25}  \!\!\!\! & \!\!\!\!\!\!\! s_{28}  \!\!\!\! & \!\!\!\!\!-\!\!s_{27}  \!\!\!\! & \!\!\!\!\!\!\! s_{30}  \!\!\!\! & \!\!\!\!\!-\!\!s_{29}  \!\!\!\! & \!\!\!\!\!\!\! s_{32}  \!\!\!\! & \!\!\!\!\!-\!\!s_{31}  \!\!\!\! & \!\!\!\!\!-\!\!s_{18}  \!\!\!\! & \!\!\!\!\!\!\! s_{17}  \!\!\!\! & \!\!\!\!\!-\!\!s_{20}  \!\!\!\! & \!\!\!\!\!\!\! s_{19}  \!\!\!\! & \!\!\!\!\!-\!\!s_{22}  \!\!\!\! & \!\!\!\!\!\!\! s_{21}  \!\!\!\! & \!\!\!\!\!-\!\!s_{24}  \!\!\!\! & \!\!\!\!\!\!\! s_{23}  \\
 \!\!\!\! s_{11}  \!\!\!\! & \!\!\!\!\!\!\! s_{12}  \!\!\!\! & \!\!\!\!\!-\!\!s_{9}   \!\!\!\! & \!\!\!\!\!-\!\!s_{10}  \!\!\!\! & \!\!\!\!\!\!\! s_{15}  \!\!\!\! & \!\!\!\!\!\!\! s_{16}  \!\!\!\! & \!\!\!\!\!-\!\!s_{13}  \!\!\!\! & \!\!\!\!\!-\!\!s_{14}  \!\!\!\! & \!\!\!\!\!-\!\!s_{3}   \!\!\!\! & \!\!\!\!\!-\!\!s_{4}   \!\!\!\! & \!\!\!\!\!\!\! s_{1}   \!\!\!\! & \!\!\!\!\!\!\! s_{2}   \!\!\!\! & \!\!\!\!\!-\!\!s_{7}   \!\!\!\! & \!\!\!\!\!-\!\!s_{8}   \!\!\!\! & \!\!\!\!\!\!\! s_{5}   \!\!\!\! & \!\!\!\!\!\!\! s_{6}   \!\!\!\! & \!\!\!\!\!\!\! s_{27}  \!\!\!\! & \!\!\!\!\!\!\! s_{28}  \!\!\!\! & \!\!\!\!\!-\!\!s_{25}  \!\!\!\! & \!\!\!\!\!-\!\!s_{26}  \!\!\!\! & \!\!\!\!\!\!\! s_{31}  \!\!\!\! & \!\!\!\!\!\!\! s_{32}  \!\!\!\! & \!\!\!\!\!-\!\!s_{29}  \!\!\!\! & \!\!\!\!\!-\!\!s_{30}  \!\!\!\! & \!\!\!\!\!-\!\!s_{19}  \!\!\!\! & \!\!\!\!\!-\!\!s_{20}  \!\!\!\! & \!\!\!\!\!\!\! s_{17}  \!\!\!\! & \!\!\!\!\!\!\! s_{18}  \!\!\!\! & \!\!\!\!\!-\!\!s_{23}  \!\!\!\! & \!\!\!\!\!-\!\!s_{24}  \!\!\!\! & \!\!\!\!\!\!\! s_{21}  \!\!\!\! & \!\!\!\!\!\!\! s_{22}  \\
 \!\!-\!\!s_{12}  \!\!\!\! & \!\!\!\!\!\!\! s_{11}  \!\!\!\! & \!\!\!\!\!\!\! s_{10}  \!\!\!\! & \!\!\!\!\!-\!\!s_{9}   \!\!\!\! & \!\!\!\!\!-\!\!s_{16}  \!\!\!\! & \!\!\!\!\!\!\! s_{15}  \!\!\!\! & \!\!\!\!\!\!\! s_{14}  \!\!\!\! & \!\!\!\!\!-\!\!s_{13}  \!\!\!\! & \!\!\!\!\!\!\! s_{4}   \!\!\!\! & \!\!\!\!\!-\!\!s_{3}   \!\!\!\! & \!\!\!\!\!-\!\!s_{2}   \!\!\!\! & \!\!\!\!\!\!\! s_{1}   \!\!\!\! & \!\!\!\!\!\!\! s_{8}   \!\!\!\! & \!\!\!\!\!-\!\!s_{7}   \!\!\!\! & \!\!\!\!\!-\!\!s_{6}   \!\!\!\! & \!\!\!\!\!\!\! s_{5}   \!\!\!\! & \!\!\!\!\!-\!\!s_{28}  \!\!\!\! & \!\!\!\!\!\!\! s_{27}  \!\!\!\! & \!\!\!\!\!\!\! s_{26}  \!\!\!\! & \!\!\!\!\!-\!\!s_{25}  \!\!\!\! & \!\!\!\!\!-\!\!s_{32}  \!\!\!\! & \!\!\!\!\!\!\! s_{31}  \!\!\!\! & \!\!\!\!\!\!\! s_{30}  \!\!\!\! & \!\!\!\!\!-\!\!s_{29}  \!\!\!\! & \!\!\!\!\!\!\! s_{20}  \!\!\!\! & \!\!\!\!\!-\!\!s_{19}  \!\!\!\! & \!\!\!\!\!-\!\!s_{18}  \!\!\!\! & \!\!\!\!\!\!\! s_{17}  \!\!\!\! & \!\!\!\!\!\!\! s_{24}  \!\!\!\! & \!\!\!\!\!-\!\!s_{23}  \!\!\!\! & \!\!\!\!\!-\!\!s_{22}  \!\!\!\! & \!\!\!\!\!\!\! s_{21}  \\
 \!\!\!\! s_{13}  \!\!\!\! & \!\!\!\!\!\!\! s_{14}  \!\!\!\! & \!\!\!\!\!\!\! s_{15}  \!\!\!\! & \!\!\!\!\!\!\! s_{16}  \!\!\!\! & \!\!\!\!\!-\!\!s_{9}   \!\!\!\! & \!\!\!\!\!-\!\!s_{10}  \!\!\!\! & \!\!\!\!\!-\!\!s_{11}  \!\!\!\! & \!\!\!\!\!-\!\!s_{12}  \!\!\!\! & \!\!\!\!\!-\!\!s_{5}   \!\!\!\! & \!\!\!\!\!-\!\!s_{6}   \!\!\!\! & \!\!\!\!\!-\!\!s_{7}   \!\!\!\! & \!\!\!\!\!-\!\!s_{8}   \!\!\!\! & \!\!\!\!\!\!\! s_{1}   \!\!\!\! & \!\!\!\!\!\!\! s_{2}   \!\!\!\! & \!\!\!\!\!\!\! s_{3}   \!\!\!\! & \!\!\!\!\!\!\! s_{4}   \!\!\!\! & \!\!\!\!\!\!\! s_{29}  \!\!\!\! & \!\!\!\!\!\!\! s_{30}  \!\!\!\! & \!\!\!\!\!\!\! s_{31}  \!\!\!\! & \!\!\!\!\!\!\! s_{32}  \!\!\!\! & \!\!\!\!\!-\!\!s_{25}  \!\!\!\! & \!\!\!\!\!-\!\!s_{26}  \!\!\!\! & \!\!\!\!\!-\!\!s_{27}  \!\!\!\! & \!\!\!\!\!-\!\!s_{28}  \!\!\!\! & \!\!\!\!\!-\!\!s_{21}  \!\!\!\! & \!\!\!\!\!-\!\!s_{22}  \!\!\!\! & \!\!\!\!\!-\!\!s_{23}  \!\!\!\! & \!\!\!\!\!-\!\!s_{24}  \!\!\!\! & \!\!\!\!\!\!\! s_{17}  \!\!\!\! & \!\!\!\!\!\!\! s_{18}  \!\!\!\! & \!\!\!\!\!\!\! s_{19}  \!\!\!\! & \!\!\!\!\!\!\! s_{20}  \\
 \!\!-\!\!s_{14}  \!\!\!\! & \!\!\!\!\!\!\! s_{13}  \!\!\!\! & \!\!\!\!\!-\!\!s_{16}  \!\!\!\! & \!\!\!\!\!\!\! s_{15}  \!\!\!\! & \!\!\!\!\!\!\! s_{10}  \!\!\!\! & \!\!\!\!\!-\!\!s_{9}   \!\!\!\! & \!\!\!\!\!\!\! s_{12}  \!\!\!\! & \!\!\!\!\!-\!\!s_{11}  \!\!\!\! & \!\!\!\!\!\!\! s_{6}   \!\!\!\! & \!\!\!\!\!-\!\!s_{5}   \!\!\!\! & \!\!\!\!\!\!\! s_{8}   \!\!\!\! & \!\!\!\!\!-\!\!s_{7}   \!\!\!\! & \!\!\!\!\!-\!\!s_{2}   \!\!\!\! & \!\!\!\!\!\!\! s_{1}   \!\!\!\! & \!\!\!\!\!-\!\!s_{4}   \!\!\!\! & \!\!\!\!\!\!\! s_{3}   \!\!\!\! & \!\!\!\!\!-\!\!s_{30}  \!\!\!\! & \!\!\!\!\!\!\! s_{29}  \!\!\!\! & \!\!\!\!\!-\!\!s_{32}  \!\!\!\! & \!\!\!\!\!\!\! s_{31}  \!\!\!\! & \!\!\!\!\!\!\! s_{26}  \!\!\!\! & \!\!\!\!\!-\!\!s_{25}  \!\!\!\! & \!\!\!\!\!\!\! s_{28}  \!\!\!\! & \!\!\!\!\!-\!\!s_{27}  \!\!\!\! & \!\!\!\!\!\!\! s_{22}  \!\!\!\! & \!\!\!\!\!-\!\!s_{21}  \!\!\!\! & \!\!\!\!\!\!\! s_{24}  \!\!\!\! & \!\!\!\!\!-\!\!s_{23}  \!\!\!\! & \!\!\!\!\!-\!\!s_{18}  \!\!\!\! & \!\!\!\!\!\!\! s_{17}  \!\!\!\! & \!\!\!\!\!-\!\!s_{20}  \!\!\!\! & \!\!\!\!\!\!\! s_{19}  \\
 \!\!-\!\!s_{15}  \!\!\!\! & \!\!\!\!\!-\!\!s_{16}  \!\!\!\! & \!\!\!\!\!\!\! s_{13}  \!\!\!\! & \!\!\!\!\!\!\! s_{14}  \!\!\!\! & \!\!\!\!\!\!\! s_{11}  \!\!\!\! & \!\!\!\!\!\!\! s_{12}  \!\!\!\! & \!\!\!\!\!-\!\!s_{9}   \!\!\!\! & \!\!\!\!\!-\!\!s_{10}  \!\!\!\! & \!\!\!\!\!\!\! s_{7}   \!\!\!\! & \!\!\!\!\!\!\! s_{8}   \!\!\!\! & \!\!\!\!\!-\!\!s_{5}   \!\!\!\! & \!\!\!\!\!-\!\!s_{6}   \!\!\!\! & \!\!\!\!\!-\!\!s_{3}   \!\!\!\! & \!\!\!\!\!-\!\!s_{4}   \!\!\!\! & \!\!\!\!\!\!\! s_{1}   \!\!\!\! & \!\!\!\!\!\!\! s_{2}   \!\!\!\! & \!\!\!\!\!-\!\!s_{31}  \!\!\!\! & \!\!\!\!\!-\!\!s_{32}  \!\!\!\! & \!\!\!\!\!\!\! s_{29}  \!\!\!\! & \!\!\!\!\!\!\! s_{30}  \!\!\!\! & \!\!\!\!\!\!\! s_{27}  \!\!\!\! & \!\!\!\!\!\!\! s_{28}  \!\!\!\! & \!\!\!\!\!-\!\!s_{25}  \!\!\!\! & \!\!\!\!\!-\!\!s_{26}  \!\!\!\! & \!\!\!\!\!\!\! s_{23}  \!\!\!\! & \!\!\!\!\!\!\! s_{24}  \!\!\!\! & \!\!\!\!\!-\!\!s_{21}  \!\!\!\! & \!\!\!\!\!-\!\!s_{22}  \!\!\!\! & \!\!\!\!\!-\!\!s_{19}  \!\!\!\! & \!\!\!\!\!-\!\!s_{20}  \!\!\!\! & \!\!\!\!\!\!\! s_{17}  \!\!\!\! & \!\!\!\!\!\!\! s_{18}  \\
 \!\!\!\! s_{16}  \!\!\!\! & \!\!\!\!\!-\!\!s_{15}  \!\!\!\! & \!\!\!\!\!-\!\!s_{14}  \!\!\!\! & \!\!\!\!\!\!\! s_{13}  \!\!\!\! & \!\!\!\!\!-\!\!s_{12}  \!\!\!\! & \!\!\!\!\!\!\! s_{11}  \!\!\!\! & \!\!\!\!\!\!\! s_{10}  \!\!\!\! & \!\!\!\!\!-\!\!s_{9}   \!\!\!\! & \!\!\!\!\!-\!\!s_{8}   \!\!\!\! & \!\!\!\!\!\!\! s_{7}   \!\!\!\! & \!\!\!\!\!\!\! s_{6}   \!\!\!\! & \!\!\!\!\!-\!\!s_{5}   \!\!\!\! & \!\!\!\!\!\!\! s_{4}   \!\!\!\! & \!\!\!\!\!-\!\!s_{3}   \!\!\!\! & \!\!\!\!\!-\!\!s_{2}   \!\!\!\! & \!\!\!\!\!\!\! s_{1}   \!\!\!\! & \!\!\!\!\!\!\! s_{32}  \!\!\!\! & \!\!\!\!\!-\!\!s_{31}  \!\!\!\! & \!\!\!\!\!-\!\!s_{30}  \!\!\!\! & \!\!\!\!\!\!\! s_{29}  \!\!\!\! & \!\!\!\!\!-\!\!s_{28}  \!\!\!\! & \!\!\!\!\!\!\! s_{27}  \!\!\!\! & \!\!\!\!\!\!\! s_{26}  \!\!\!\! & \!\!\!\!\!-\!\!s_{25}  \!\!\!\! & \!\!\!\!\!-\!\!s_{24}  \!\!\!\! & \!\!\!\!\!\!\! s_{23}  \!\!\!\! & \!\!\!\!\!\!\! s_{22}  \!\!\!\! & \!\!\!\!\!-\!\!s_{21}  \!\!\!\! & \!\!\!\!\!\!\! s_{20}  \!\!\!\! & \!\!\!\!\!-\!\!s_{19}  \!\!\!\! & \!\!\!\!\!-\!\!s_{18}  \!\!\!\! & \!\!\!\!\!\!\! s_{17}  \\
 \!\!-\!\!s_{17}^*\!\!\!\! & \!\!\!\!\!\!\! s_{18}^*\!\!\!\! & \!\!\!\!\!\!\! s_{19}^*\!\!\!\! & \!\!\!\!\!-\!\!s_{20}^*\!\!\!\! & \!\!\!\!\!\!\! s_{21}^*\!\!\!\! & \!\!\!\!\!-\!\!s_{22}^*\!\!\!\! & \!\!\!\!\!-\!\!s_{23}^*\!\!\!\! & \!\!\!\!\!\!\! s_{24}^*\!\!\!\! & \!\!\!\!\!\!\! s_{25}^*\!\!\!\! & \!\!\!\!\!-\!\!s_{26}^*\!\!\!\! & \!\!\!\!\!-\!\!s_{27}^*\!\!\!\! & \!\!\!\!\!\!\! s_{28}^*\!\!\!\! & \!\!\!\!\!-\!\!s_{29}^*\!\!\!\! & \!\!\!\!\!\!\! s_{30}^*\!\!\!\! & \!\!\!\!\!\!\! s_{31}^*\!\!\!\! & \!\!\!\!\!-\!\!s_{32}^*\!\!\!\! & \!\!\!\!\!\!\! s_{1}^*\!\!\!\!  & \!\!\!\!-\!\!s_{2}^*\!\!\!\!  & \!\!\!\!-\!\!s_{3}^*\!\!\!\!  & \!\!\!\!\!\! s_{4}^*\!\!\!\!  & \!\!\!\!-\!\!s_{5}^*\!\!\!\!  & \!\!\!\!\!\! s_{6}^*\!\!\!\!  & \!\!\!\!\!\! s_{7}^*\!\!\!\!  & \!\!\!\!-\!\!s_{8}^*\!\!\!\!  & \!\!\!\!-\!\!s_{9}^*\!\!\!\!  & \!\!\!\!\!\! s_{10}^*\!\!\!\! & \!\!\!\!\!\!\! s_{11}^*\!\!\!\! & \!\!\!\!\!-\!\!s_{12}^*\!\!\!\! & \!\!\!\!\!\!\! s_{13}^*\!\!\!\! & \!\!\!\!\!-\!\!s_{14}^*\!\!\!\! & \!\!\!\!\!-\!\!s_{15}^*\!\!\!\! & \!\!\!\!\!\!\! s_{16}^*\\
 \!\!-\!\!s_{18}^*\!\!\!\! & \!\!\!\!\!-\!\!s_{17}^*\!\!\!\! & \!\!\!\!\!\!\! s_{20}^*\!\!\!\! & \!\!\!\!\!\!\! s_{19}^*\!\!\!\! & \!\!\!\!\!\!\! s_{22}^*\!\!\!\! & \!\!\!\!\!\!\! s_{21}^*\!\!\!\! & \!\!\!\!\!-\!\!s_{24}^*\!\!\!\! & \!\!\!\!\!-\!\!s_{23}^*\!\!\!\! & \!\!\!\!\!\!\! s_{26}^*\!\!\!\! & \!\!\!\!\!\!\! s_{25}^*\!\!\!\! & \!\!\!\!\!-\!\!s_{28}^*\!\!\!\! & \!\!\!\!\!-\!\!s_{27}^*\!\!\!\! & \!\!\!\!\!-\!\!s_{30}^*\!\!\!\! & \!\!\!\!\!-\!\!s_{29}^*\!\!\!\! & \!\!\!\!\!\!\! s_{32}^*\!\!\!\! & \!\!\!\!\!\!\! s_{31}^*\!\!\!\! & \!\!\!\!\!\!\! s_{2}^*\!\!\!\!  & \!\!\!\!\!\! s_{1}^*\!\!\!\!  & \!\!\!\!-\!\!s_{4}^*\!\!\!\!  & \!\!\!\!-\!\!s_{3}^*\!\!\!\!  & \!\!\!\!-\!\!s_{6}^*\!\!\!\!  & \!\!\!\!-\!\!s_{5}^*\!\!\!\!  & \!\!\!\!\!\! s_{8}^*\!\!\!\!  & \!\!\!\!\!\! s_{7}^*\!\!\!\!  & \!\!\!\!-\!\!s_{10}^*\!\!\!\! & \!\!\!\!\!-\!\!s_{9}^*\!\!\!\!  & \!\!\!\!\!\! s_{12}^*\!\!\!\! & \!\!\!\!\!\!\! s_{11}^*\!\!\!\! & \!\!\!\!\!\!\! s_{14}^*\!\!\!\! & \!\!\!\!\!\!\! s_{13}^*\!\!\!\! & \!\!\!\!\!-\!\!s_{16}^*\!\!\!\! & \!\!\!\!\!-\!\!s_{15}^*\\
 \!\!-\!\!s_{19}^*\!\!\!\! & \!\!\!\!\!\!\! s_{20}^*\!\!\!\! & \!\!\!\!\!-\!\!s_{17}^*\!\!\!\! & \!\!\!\!\!\!\! s_{18}^*\!\!\!\! & \!\!\!\!\!\!\! s_{23}^*\!\!\!\! & \!\!\!\!\!-\!\!s_{24}^*\!\!\!\! & \!\!\!\!\!\!\! s_{21}^*\!\!\!\! & \!\!\!\!\!-\!\!s_{22}^*\!\!\!\! & \!\!\!\!\!\!\! s_{27}^*\!\!\!\! & \!\!\!\!\!-\!\!s_{28}^*\!\!\!\! & \!\!\!\!\!\!\! s_{25}^*\!\!\!\! & \!\!\!\!\!-\!\!s_{26}^*\!\!\!\! & \!\!\!\!\!-\!\!s_{31}^*\!\!\!\! & \!\!\!\!\!\!\! s_{32}^*\!\!\!\! & \!\!\!\!\!-\!\!s_{29}^*\!\!\!\! & \!\!\!\!\!\!\! s_{30}^*\!\!\!\! & \!\!\!\!\!\!\! s_{3}^*\!\!\!\!  & \!\!\!\!-\!\!s_{4}^*\!\!\!\!  & \!\!\!\!\!\! s_{1}^*\!\!\!\!  & \!\!\!\!-\!\!s_{2}^*\!\!\!\!  & \!\!\!\!-\!\!s_{7}^*\!\!\!\!  & \!\!\!\!\!\! s_{8}^*\!\!\!\!  & \!\!\!\!-\!\!s_{5}^*\!\!\!\!  & \!\!\!\!\!\! s_{6}^*\!\!\!\!  & \!\!\!\!-\!\!s_{11}^*\!\!\!\! & \!\!\!\!\!\!\! s_{12}^*\!\!\!\! & \!\!\!\!\!-\!\!s_{9}^*\!\!\!\!  & \!\!\!\!\!\! s_{10}^*\!\!\!\! & \!\!\!\!\!\!\! s_{15}^*\!\!\!\! & \!\!\!\!\!-\!\!s_{16}^*\!\!\!\! & \!\!\!\!\!\!\! s_{13}^*\!\!\!\! & \!\!\!\!\!-\!\!s_{14}^*\\
 \!\!-\!\!s_{20}^*\!\!\!\! & \!\!\!\!\!-\!\!s_{19}^*\!\!\!\! & \!\!\!\!\!-\!\!s_{18}^*\!\!\!\! & \!\!\!\!\!-\!\!s_{17}^*\!\!\!\! & \!\!\!\!\!\!\! s_{24}^*\!\!\!\! & \!\!\!\!\!\!\! s_{23}^*\!\!\!\! & \!\!\!\!\!\!\! s_{22}^*\!\!\!\! & \!\!\!\!\!\!\! s_{21}^*\!\!\!\! & \!\!\!\!\!\!\! s_{28}^*\!\!\!\! & \!\!\!\!\!\!\! s_{27}^*\!\!\!\! & \!\!\!\!\!\!\! s_{26}^*\!\!\!\! & \!\!\!\!\!\!\! s_{25}^*\!\!\!\! & \!\!\!\!\!-\!\!s_{32}^*\!\!\!\! & \!\!\!\!\!-\!\!s_{31}^*\!\!\!\! & \!\!\!\!\!-\!\!s_{30}^*\!\!\!\! & \!\!\!\!\!-\!\!s_{29}^*\!\!\!\! & \!\!\!\!\!\!\! s_{4}^*\!\!\!\!  & \!\!\!\!\!\! s_{3}^*\!\!\!\!  & \!\!\!\!\!\! s_{2}^*\!\!\!\!  & \!\!\!\!\!\! s_{1}^*\!\!\!\!  & \!\!\!\!-\!\!s_{8}^*\!\!\!\!  & \!\!\!\!-\!\!s_{7}^*\!\!\!\!  & \!\!\!\!-\!\!s_{6}^*\!\!\!\!  & \!\!\!\!-\!\!s_{5}^*\!\!\!\!  & \!\!\!\!-\!\!s_{12}^*\!\!\!\! & \!\!\!\!\!-\!\!s_{11}^*\!\!\!\! & \!\!\!\!\!-\!\!s_{10}^*\!\!\!\! & \!\!\!\!\!-\!\!s_{9}^*\!\!\!\!  & \!\!\!\!\!\! s_{16}^*\!\!\!\! & \!\!\!\!\!\!\! s_{15}^*\!\!\!\! & \!\!\!\!\!\!\! s_{14}^*\!\!\!\! & \!\!\!\!\!\!\! s_{13}^*\\
 \!\!-\!\!s_{21}^*\!\!\!\! & \!\!\!\!\!\!\! s_{22}^*\!\!\!\! & \!\!\!\!\!\!\! s_{23}^*\!\!\!\! & \!\!\!\!\!-\!\!s_{24}^*\!\!\!\! & \!\!\!\!\!-\!\!s_{17}^*\!\!\!\! & \!\!\!\!\!\!\! s_{18}^*\!\!\!\! & \!\!\!\!\!\!\! s_{19}^*\!\!\!\! & \!\!\!\!\!-\!\!s_{20}^*\!\!\!\! & \!\!\!\!\!\!\! s_{29}^*\!\!\!\! & \!\!\!\!\!-\!\!s_{30}^*\!\!\!\! & \!\!\!\!\!-\!\!s_{31}^*\!\!\!\! & \!\!\!\!\!\!\! s_{32}^*\!\!\!\! & \!\!\!\!\!\!\! s_{25}^*\!\!\!\! & \!\!\!\!\!-\!\!s_{26}^*\!\!\!\! & \!\!\!\!\!-\!\!s_{27}^*\!\!\!\! & \!\!\!\!\!\!\! s_{28}^*\!\!\!\! & \!\!\!\!\!\!\! s_{5}^*\!\!\!\!  & \!\!\!\!-\!\!s_{6}^*\!\!\!\!  & \!\!\!\!-\!\!s_{7}^*\!\!\!\!  & \!\!\!\!\!\! s_{8}^*\!\!\!\!  & \!\!\!\!\!\! s_{1}^*\!\!\!\!  & \!\!\!\!-\!\!s_{2}^*\!\!\!\!  & \!\!\!\!-\!\!s_{3}^*\!\!\!\!  & \!\!\!\!\!\! s_{4}^*\!\!\!\!  & \!\!\!\!-\!\!s_{13}^*\!\!\!\! & \!\!\!\!\!\!\! s_{14}^*\!\!\!\! & \!\!\!\!\!\!\! s_{15}^*\!\!\!\! & \!\!\!\!\!-\!\!s_{16}^*\!\!\!\! & \!\!\!\!\!-\!\!s_{9}^*\!\!\!\!  & \!\!\!\!\!\! s_{10}^*\!\!\!\! & \!\!\!\!\!\!\! s_{11}^*\!\!\!\! & \!\!\!\!\!-\!\!s_{12}^*\\
 \!\!-\!\!s_{22}^*\!\!\!\! & \!\!\!\!\!-\!\!s_{21}^*\!\!\!\! & \!\!\!\!\!\!\! s_{24}^*\!\!\!\! & \!\!\!\!\!\!\! s_{23}^*\!\!\!\! & \!\!\!\!\!-\!\!s_{18}^*\!\!\!\! & \!\!\!\!\!-\!\!s_{17}^*\!\!\!\! & \!\!\!\!\!\!\! s_{20}^*\!\!\!\! & \!\!\!\!\!\!\! s_{19}^*\!\!\!\! & \!\!\!\!\!\!\! s_{30}^*\!\!\!\! & \!\!\!\!\!\!\! s_{29}^*\!\!\!\! & \!\!\!\!\!-\!\!s_{32}^*\!\!\!\! & \!\!\!\!\!-\!\!s_{31}^*\!\!\!\! & \!\!\!\!\!\!\! s_{26}^*\!\!\!\! & \!\!\!\!\!\!\! s_{25}^*\!\!\!\! & \!\!\!\!\!-\!\!s_{28}^*\!\!\!\! & \!\!\!\!\!-\!\!s_{27}^*\!\!\!\! & \!\!\!\!\!\!\! s_{6}^*\!\!\!\!  & \!\!\!\!\!\! s_{5}^*\!\!\!\!  & \!\!\!\!-\!\!s_{8}^*\!\!\!\!  & \!\!\!\!-\!\!s_{7}^*\!\!\!\!  & \!\!\!\!\!\! s_{2}^*\!\!\!\!  & \!\!\!\!\!\! s_{1}^*\!\!\!\!  & \!\!\!\!-\!\!s_{4}^*\!\!\!\!  & \!\!\!\!-\!\!s_{3}^*\!\!\!\!  & \!\!\!\!-\!\!s_{14}^*\!\!\!\! & \!\!\!\!\!-\!\!s_{13}^*\!\!\!\! & \!\!\!\!\!\!\! s_{16}^*\!\!\!\! & \!\!\!\!\!\!\! s_{15}^*\!\!\!\! & \!\!\!\!\!-\!\!s_{10}^*\!\!\!\! & \!\!\!\!\!-\!\!s_{9}^*\!\!\!\!  & \!\!\!\!\!\! s_{12}^*\!\!\!\! & \!\!\!\!\!\!\! s_{11}^*\\
 \!\!-\!\!s_{23}^*\!\!\!\! & \!\!\!\!\!\!\! s_{24}^*\!\!\!\! & \!\!\!\!\!-\!\!s_{21}^*\!\!\!\! & \!\!\!\!\!\!\! s_{22}^*\!\!\!\! & \!\!\!\!\!-\!\!s_{19}^*\!\!\!\! & \!\!\!\!\!\!\! s_{20}^*\!\!\!\! & \!\!\!\!\!-\!\!s_{17}^*\!\!\!\! & \!\!\!\!\!\!\! s_{18}^*\!\!\!\! & \!\!\!\!\!\!\! s_{31}^*\!\!\!\! & \!\!\!\!\!-\!\!s_{32}^*\!\!\!\! & \!\!\!\!\!\!\! s_{29}^*\!\!\!\! & \!\!\!\!\!-\!\!s_{30}^*\!\!\!\! & \!\!\!\!\!\!\! s_{27}^*\!\!\!\! & \!\!\!\!\!-\!\!s_{28}^*\!\!\!\! & \!\!\!\!\!\!\! s_{25}^*\!\!\!\! & \!\!\!\!\!-\!\!s_{26}^*\!\!\!\! & \!\!\!\!\!\!\! s_{7}^*\!\!\!\!  & \!\!\!\!-\!\!s_{8}^*\!\!\!\!  & \!\!\!\!\!\! s_{5}^*\!\!\!\!  & \!\!\!\!-\!\!s_{6}^*\!\!\!\!  & \!\!\!\!\!\! s_{3}^*\!\!\!\!  & \!\!\!\!-\!\!s_{4}^*\!\!\!\!  & \!\!\!\!\!\! s_{1}^*\!\!\!\!  & \!\!\!\!-\!\!s_{2}^*\!\!\!\!  & \!\!\!\!-\!\!s_{15}^*\!\!\!\! & \!\!\!\!\!\!\! s_{16}^*\!\!\!\! & \!\!\!\!\!-\!\!s_{13}^*\!\!\!\! & \!\!\!\!\!\!\! s_{14}^*\!\!\!\! & \!\!\!\!\!-\!\!s_{11}^*\!\!\!\! & \!\!\!\!\!\!\! s_{12}^*\!\!\!\! & \!\!\!\!\!-\!\!s_{9}^*\!\!\!\!  & \!\!\!\!\!\! s_{10}^*\\
 \!\!-\!\!s_{24}^*\!\!\!\! & \!\!\!\!\!-\!\!s_{23}^*\!\!\!\! & \!\!\!\!\!-\!\!s_{22}^*\!\!\!\! & \!\!\!\!\!-\!\!s_{21}^*\!\!\!\! & \!\!\!\!\!-\!\!s_{20}^*\!\!\!\! & \!\!\!\!\!-\!\!s_{19}^*\!\!\!\! & \!\!\!\!\!-\!\!s_{18}^*\!\!\!\! & \!\!\!\!\!-\!\!s_{17}^*\!\!\!\! & \!\!\!\!\!\!\! s_{32}^*\!\!\!\! & \!\!\!\!\!\!\! s_{31}^*\!\!\!\! & \!\!\!\!\!\!\! s_{30}^*\!\!\!\! & \!\!\!\!\!\!\! s_{29}^*\!\!\!\! & \!\!\!\!\!\!\! s_{28}^*\!\!\!\! & \!\!\!\!\!\!\! s_{27}^*\!\!\!\! & \!\!\!\!\!\!\! s_{26}^*\!\!\!\! & \!\!\!\!\!\!\! s_{25}^*\!\!\!\! & \!\!\!\!\!\!\! s_{8}^*\!\!\!\!  & \!\!\!\!\!\! s_{7}^*\!\!\!\!  & \!\!\!\!\!\! s_{6}^*\!\!\!\!  & \!\!\!\!\!\! s_{5}^*\!\!\!\!  & \!\!\!\!\!\! s_{4}^*\!\!\!\!  & \!\!\!\!\!\! s_{3}^*\!\!\!\!  & \!\!\!\!\!\! s_{2}^*\!\!\!\!  & \!\!\!\!\!\! s_{1}^*\!\!\!\!  & \!\!\!\!-\!\!s_{16}^*\!\!\!\! & \!\!\!\!\!-\!\!s_{15}^*\!\!\!\! & \!\!\!\!\!-\!\!s_{14}^*\!\!\!\! & \!\!\!\!\!-\!\!s_{13}^*\!\!\!\! & \!\!\!\!\!-\!\!s_{12}^*\!\!\!\! & \!\!\!\!\!-\!\!s_{11}^*\!\!\!\! & \!\!\!\!\!-\!\!s_{10}^*\!\!\!\! & \!\!\!\!\!-\!\!s_{9}^* \\
 \!\!-\!\!s_{25}^*\!\!\!\! & \!\!\!\!\!\!\! s_{26}^*\!\!\!\! & \!\!\!\!\!\!\! s_{27}^*\!\!\!\! & \!\!\!\!\!-\!\!s_{28}^*\!\!\!\! & \!\!\!\!\!\!\! s_{29}^*\!\!\!\! & \!\!\!\!\!-\!\!s_{30}^*\!\!\!\! & \!\!\!\!\!-\!\!s_{31}^*\!\!\!\! & \!\!\!\!\!\!\! s_{32}^*\!\!\!\! & \!\!\!\!\!-\!\!s_{17}^*\!\!\!\! & \!\!\!\!\!\!\! s_{18}^*\!\!\!\! & \!\!\!\!\!\!\! s_{19}^*\!\!\!\! & \!\!\!\!\!-\!\!s_{20}^*\!\!\!\! & \!\!\!\!\!\!\! s_{21}^*\!\!\!\! & \!\!\!\!\!-\!\!s_{22}^*\!\!\!\! & \!\!\!\!\!-\!\!s_{23}^*\!\!\!\! & \!\!\!\!\!\!\! s_{24}^*\!\!\!\! & \!\!\!\!\!\!\! s_{9}^*\!\!\!\!  & \!\!\!\!-\!\!s_{10}^*\!\!\!\! & \!\!\!\!\!-\!\!s_{11}^*\!\!\!\! & \!\!\!\!\!\!\! s_{12}^*\!\!\!\! & \!\!\!\!\!-\!\!s_{13}^*\!\!\!\! & \!\!\!\!\!\!\! s_{14}^*\!\!\!\! & \!\!\!\!\!\!\! s_{15}^*\!\!\!\! & \!\!\!\!\!-\!\!s_{16}^*\!\!\!\! & \!\!\!\!\!\!\! s_{1}^*\!\!\!\!  & \!\!\!\!-\!\!s_{2}^*\!\!\!\!  & \!\!\!\!-\!\!s_{3}^*\!\!\!\!  & \!\!\!\!\!\! s_{4}^*\!\!\!\!  & \!\!\!\!-\!\!s_{5}^*\!\!\!\!  & \!\!\!\!\!\! s_{6}^*\!\!\!\!  & \!\!\!\!\!\! s_{7}^*\!\!\!\!  & \!\!\!\!-\!\!s_{8}^* \\
 \!\!-\!\!s_{26}^*\!\!\!\! & \!\!\!\!\!-\!\!s_{25}^*\!\!\!\! & \!\!\!\!\!\!\! s_{28}^*\!\!\!\! & \!\!\!\!\!\!\! s_{27}^*\!\!\!\! & \!\!\!\!\!\!\! s_{30}^*\!\!\!\! & \!\!\!\!\!\!\! s_{29}^*\!\!\!\! & \!\!\!\!\!-\!\!s_{32}^*\!\!\!\! & \!\!\!\!\!-\!\!s_{31}^*\!\!\!\! & \!\!\!\!\!-\!\!s_{18}^*\!\!\!\! & \!\!\!\!\!-\!\!s_{17}^*\!\!\!\! & \!\!\!\!\!\!\! s_{20}^*\!\!\!\! & \!\!\!\!\!\!\! s_{19}^*\!\!\!\! & \!\!\!\!\!\!\! s_{22}^*\!\!\!\! & \!\!\!\!\!\!\! s_{21}^*\!\!\!\! & \!\!\!\!\!-\!\!s_{24}^*\!\!\!\! & \!\!\!\!\!-\!\!s_{23}^*\!\!\!\! & \!\!\!\!\!\!\! s_{10}^*\!\!\!\! & \!\!\!\!\!\!\! s_{9}^*\!\!\!\!  & \!\!\!\!-\!\!s_{12}^*\!\!\!\! & \!\!\!\!\!-\!\!s_{11}^*\!\!\!\! & \!\!\!\!\!-\!\!s_{14}^*\!\!\!\! & \!\!\!\!\!-\!\!s_{13}^*\!\!\!\! & \!\!\!\!\!\!\! s_{16}^*\!\!\!\! & \!\!\!\!\!\!\! s_{15}^*\!\!\!\! & \!\!\!\!\!\!\! s_{2}^*\!\!\!\!  & \!\!\!\!\!\! s_{1}^*\!\!\!\!  & \!\!\!\!-\!\!s_{4}^*\!\!\!\!  & \!\!\!\!-\!\!s_{3}^*\!\!\!\!  & \!\!\!\!-\!\!s_{6}^*\!\!\!\!  & \!\!\!\!-\!\!s_{5}^*\!\!\!\!  & \!\!\!\!\!\! s_{8}^*\!\!\!\!  & \!\!\!\!\!\! s_{7}^* \\
 \!\!-\!\!s_{27}^*\!\!\!\! & \!\!\!\!\!\!\! s_{28}^*\!\!\!\! & \!\!\!\!\!-\!\!s_{25}^*\!\!\!\! & \!\!\!\!\!\!\! s_{26}^*\!\!\!\! & \!\!\!\!\!\!\! s_{31}^*\!\!\!\! & \!\!\!\!\!-\!\!s_{32}^*\!\!\!\! & \!\!\!\!\!\!\! s_{29}^*\!\!\!\! & \!\!\!\!\!-\!\!s_{30}^*\!\!\!\! & \!\!\!\!\!-\!\!s_{19}^*\!\!\!\! & \!\!\!\!\!\!\! s_{20}^*\!\!\!\! & \!\!\!\!\!-\!\!s_{17}^*\!\!\!\! & \!\!\!\!\!\!\! s_{18}^*\!\!\!\! & \!\!\!\!\!\!\! s_{23}^*\!\!\!\! & \!\!\!\!\!-\!\!s_{24}^*\!\!\!\! & \!\!\!\!\!\!\! s_{21}^*\!\!\!\! & \!\!\!\!\!-\!\!s_{22}^*\!\!\!\! & \!\!\!\!\!\!\! s_{11}^*\!\!\!\! & \!\!\!\!\!-\!\!s_{12}^*\!\!\!\! & \!\!\!\!\!\!\! s_{9}^*\!\!\!\!  & \!\!\!\!-\!\!s_{10}^*\!\!\!\! & \!\!\!\!\!-\!\!s_{15}^*\!\!\!\! & \!\!\!\!\!\!\! s_{16}^*\!\!\!\! & \!\!\!\!\!-\!\!s_{13}^*\!\!\!\! & \!\!\!\!\!\!\! s_{14}^*\!\!\!\! & \!\!\!\!\!\!\! s_{3}^*\!\!\!\!  & \!\!\!\!-\!\!s_{4}^*\!\!\!\!  & \!\!\!\!\!\! s_{1}^*\!\!\!\!  & \!\!\!\!-\!\!s_{2}^*\!\!\!\!  & \!\!\!\!-\!\!s_{7}^*\!\!\!\!  & \!\!\!\!\!\! s_{8}^*\!\!\!\!  & \!\!\!\!-\!\!s_{5}^*\!\!\!\!  & \!\!\!\!\!\! s_{6}^* \\
 \!\!-\!\!s_{28}^*\!\!\!\! & \!\!\!\!\!-\!\!s_{27}^*\!\!\!\! & \!\!\!\!\!-\!\!s_{26}^*\!\!\!\! & \!\!\!\!\!-\!\!s_{25}^*\!\!\!\! & \!\!\!\!\!\!\! s_{32}^*\!\!\!\! & \!\!\!\!\!\!\! s_{31}^*\!\!\!\! & \!\!\!\!\!\!\! s_{30}^*\!\!\!\! & \!\!\!\!\!\!\! s_{29}^*\!\!\!\! & \!\!\!\!\!-\!\!s_{20}^*\!\!\!\! & \!\!\!\!\!-\!\!s_{19}^*\!\!\!\! & \!\!\!\!\!-\!\!s_{18}^*\!\!\!\! & \!\!\!\!\!-\!\!s_{17}^*\!\!\!\! & \!\!\!\!\!\!\! s_{24}^*\!\!\!\! & \!\!\!\!\!\!\! s_{23}^*\!\!\!\! & \!\!\!\!\!\!\! s_{22}^*\!\!\!\! & \!\!\!\!\!\!\! s_{21}^*\!\!\!\! & \!\!\!\!\!\!\! s_{12}^*\!\!\!\! & \!\!\!\!\!\!\! s_{11}^*\!\!\!\! & \!\!\!\!\!\!\! s_{10}^*\!\!\!\! & \!\!\!\!\!\!\! s_{9}^*\!\!\!\!  & \!\!\!\!-\!\!s_{16}^*\!\!\!\! & \!\!\!\!\!-\!\!s_{15}^*\!\!\!\! & \!\!\!\!\!-\!\!s_{14}^*\!\!\!\! & \!\!\!\!\!-\!\!s_{13}^*\!\!\!\! & \!\!\!\!\!\!\! s_{4}^*\!\!\!\!  & \!\!\!\!\!\! s_{3}^*\!\!\!\!  & \!\!\!\!\!\! s_{2}^*\!\!\!\!  & \!\!\!\!\!\! s_{1}^*\!\!\!\!  & \!\!\!\!-\!\!s_{8}^*\!\!\!\!  & \!\!\!\!-\!\!s_{7}^*\!\!\!\!  & \!\!\!\!-\!\!s_{6}^*\!\!\!\!  & \!\!\!\!-\!\!s_{5}^* \\
 \!\!-\!\!s_{29}^*\!\!\!\! & \!\!\!\!\!\!\! s_{30}^*\!\!\!\! & \!\!\!\!\!\!\! s_{31}^*\!\!\!\! & \!\!\!\!\!-\!\!s_{32}^*\!\!\!\! & \!\!\!\!\!-\!\!s_{25}^*\!\!\!\! & \!\!\!\!\!\!\! s_{26}^*\!\!\!\! & \!\!\!\!\!\!\! s_{27}^*\!\!\!\! & \!\!\!\!\!-\!\!s_{28}^*\!\!\!\! & \!\!\!\!\!-\!\!s_{21}^*\!\!\!\! & \!\!\!\!\!\!\! s_{22}^*\!\!\!\! & \!\!\!\!\!\!\! s_{23}^*\!\!\!\! & \!\!\!\!\!-\!\!s_{24}^*\!\!\!\! & \!\!\!\!\!-\!\!s_{17}^*\!\!\!\! & \!\!\!\!\!\!\! s_{18}^*\!\!\!\! & \!\!\!\!\!\!\! s_{19}^*\!\!\!\! & \!\!\!\!\!-\!\!s_{20}^*\!\!\!\! & \!\!\!\!\!\!\! s_{13}^*\!\!\!\! & \!\!\!\!\!-\!\!s_{14}^*\!\!\!\! & \!\!\!\!\!-\!\!s_{15}^*\!\!\!\! & \!\!\!\!\!\!\! s_{16}^*\!\!\!\! & \!\!\!\!\!\!\! s_{9}^*\!\!\!\!  & \!\!\!\!-\!\!s_{10}^*\!\!\!\! & \!\!\!\!\!-\!\!s_{11}^*\!\!\!\! & \!\!\!\!\!\!\! s_{12}^*\!\!\!\! & \!\!\!\!\!\!\! s_{5}^*\!\!\!\!  & \!\!\!\!-\!\!s_{6}^*\!\!\!\!  & \!\!\!\!-\!\!s_{7}^*\!\!\!\!  & \!\!\!\!\!\! s_{8}^*\!\!\!\!  & \!\!\!\!\!\! s_{1}^*\!\!\!\!  & \!\!\!\!-\!\!s_{2}^*\!\!\!\!  & \!\!\!\!-\!\!s_{3}^*\!\!\!\!  & \!\!\!\!\!\! s_{4}^* \\
 \!\!-\!\!s_{30}^*\!\!\!\! & \!\!\!\!\!-\!\!s_{29}^*\!\!\!\! & \!\!\!\!\!\!\! s_{32}^*\!\!\!\! & \!\!\!\!\!\!\! s_{31}^*\!\!\!\! & \!\!\!\!\!-\!\!s_{26}^*\!\!\!\! & \!\!\!\!\!-\!\!s_{25}^*\!\!\!\! & \!\!\!\!\!\!\! s_{28}^*\!\!\!\! & \!\!\!\!\!\!\! s_{27}^*\!\!\!\! & \!\!\!\!\!-\!\!s_{22}^*\!\!\!\! & \!\!\!\!\!-\!\!s_{21}^*\!\!\!\! & \!\!\!\!\!\!\! s_{24}^*\!\!\!\! & \!\!\!\!\!\!\! s_{23}^*\!\!\!\! & \!\!\!\!\!-\!\!s_{18}^*\!\!\!\! & \!\!\!\!\!-\!\!s_{17}^*\!\!\!\! & \!\!\!\!\!\!\! s_{20}^*\!\!\!\! & \!\!\!\!\!\!\! s_{19}^*\!\!\!\! & \!\!\!\!\!\!\! s_{14}^*\!\!\!\! & \!\!\!\!\!\!\! s_{13}^*\!\!\!\! & \!\!\!\!\!-\!\!s_{16}^*\!\!\!\! & \!\!\!\!\!-\!\!s_{15}^*\!\!\!\! & \!\!\!\!\!\!\! s_{10}^*\!\!\!\! & \!\!\!\!\!\!\! s_{9}^*\!\!\!\!  & \!\!\!\!-\!\!s_{12}^*\!\!\!\! & \!\!\!\!\!-\!\!s_{11}^*\!\!\!\! & \!\!\!\!\!\!\! s_{6}^*\!\!\!\!  & \!\!\!\!\!\! s_{5}^*\!\!\!\!  & \!\!\!\!-\!\!s_{8}^*\!\!\!\!  & \!\!\!\!-\!\!s_{7}^*\!\!\!\!  & \!\!\!\!\!\! s_{2}^*\!\!\!\!  & \!\!\!\!\!\! s_{1}^*\!\!\!\!  & \!\!\!\!-\!\!s_{4}^*\!\!\!\!  & \!\!\!\!-\!\!s_{3}^* \\
 \!\!-\!\!s_{31}^*\!\!\!\! & \!\!\!\!\!\!\! s_{32}^*\!\!\!\! & \!\!\!\!\!-\!\!s_{29}^*\!\!\!\! & \!\!\!\!\!\!\! s_{30}^*\!\!\!\! & \!\!\!\!\!-\!\!s_{27}^*\!\!\!\! & \!\!\!\!\!\!\! s_{28}^*\!\!\!\! & \!\!\!\!\!-\!\!s_{25}^*\!\!\!\! & \!\!\!\!\!\!\! s_{26}^*\!\!\!\! & \!\!\!\!\!-\!\!s_{23}^*\!\!\!\! & \!\!\!\!\!\!\! s_{24}^*\!\!\!\! & \!\!\!\!\!-\!\!s_{21}^*\!\!\!\! & \!\!\!\!\!\!\! s_{22}^*\!\!\!\! & \!\!\!\!\!-\!\!s_{19}^*\!\!\!\! & \!\!\!\!\!\!\! s_{20}^*\!\!\!\! & \!\!\!\!\!-\!\!s_{17}^*\!\!\!\! & \!\!\!\!\!\!\! s_{18}^*\!\!\!\! & \!\!\!\!\!\!\! s_{15}^*\!\!\!\! & \!\!\!\!\!-\!\!s_{16}^*\!\!\!\! & \!\!\!\!\!\!\! s_{13}^*\!\!\!\! & \!\!\!\!\!-\!\!s_{14}^*\!\!\!\! & \!\!\!\!\!\!\! s_{11}^*\!\!\!\! & \!\!\!\!\!-\!\!s_{12}^*\!\!\!\! & \!\!\!\!\!\!\! s_{9}^*\!\!\!\!  & \!\!\!\!-\!\!s_{10}^*\!\!\!\! & \!\!\!\!\!\!\! s_{7}^*\!\!\!\!  & \!\!\!\!-\!\!s_{8}^*\!\!\!\!  & \!\!\!\!\!\! s_{5}^*\!\!\!\!  & \!\!\!\!-\!\!s_{6}^*\!\!\!\!  & \!\!\!\!\!\! s_{3}^*\!\!\!\!  & \!\!\!\!-\!\!s_{4}^*\!\!\!\!  & \!\!\!\!\!\! s_{1}^*\!\!\!\!  & \!\!\!\!-\!\!s_{2}^* \\
 \!\!-\!\!s_{32}^*\!\!\!\! & \!\!\!\!\!-\!\!s_{31}^*\!\!\!\! & \!\!\!\!\!-\!\!s_{30}^*\!\!\!\! & \!\!\!\!\!-\!\!s_{29}^*\!\!\!\! & \!\!\!\!\!-\!\!s_{28}^*\!\!\!\! & \!\!\!\!\!-\!\!s_{27}^*\!\!\!\! & \!\!\!\!\!-\!\!s_{26}^*\!\!\!\! & \!\!\!\!\!-\!\!s_{25}^*\!\!\!\! & \!\!\!\!\!-\!\!s_{24}^*\!\!\!\! & \!\!\!\!\!-\!\!s_{23}^*\!\!\!\! & \!\!\!\!\!-\!\!s_{22}^*\!\!\!\! & \!\!\!\!\!-\!\!s_{21}^*\!\!\!\! & \!\!\!\!\!-\!\!s_{20}^*\!\!\!\! & \!\!\!\!\!-\!\!s_{19}^*\!\!\!\! & \!\!\!\!\!-\!\!s_{18}^*\!\!\!\! & \!\!\!\!\!-\!\!s_{17}^*\!\!\!\! & \!\!\!\!\!\!\! s_{16}^*\!\!\!\! & \!\!\!\!\!\!\! s_{15}^*\!\!\!\! & \!\!\!\!\!\!\! s_{14}^*\!\!\!\! & \!\!\!\!\!\!\! s_{13}^*\!\!\!\! & \!\!\!\!\!\!\! s_{12}^*\!\!\!\! & \!\!\!\!\!\!\! s_{11}^*\!\!\!\! & \!\!\!\!\!\!\! s_{10}^*\!\!\!\! & \!\!\!\!\!\!\! s_{9}^*\!\!\!\!  & \!\!\!\!\!\! s_{8}^*\!\!\!\!  & \!\!\!\!\!\! s_{7}^*\!\!\!\!  & \!\!\!\!\!\! s_{6}^*\!\!\!\!  & \!\!\!\!\!\! s_{5}^*\!\!\!\!  & \!\!\!\!\!\! s_{4}^*\!\!\!\!  & \!\!\!\!\!\! s_{3}^*\!\!\!\!  & \!\!\!\!\!\! s_{2}^*\!\!\!\!  & \!\!\!\!\!\! s_{1}^*  \\ [2.5ex]%
\end{array}}
\!\!\!\right]%
\end{eqnarray*}

\subsection{Matlab Function: Generator of GABBA Mother Encoding Matrices}
\label{ProgramGABBAEncodingMatrix} %
\begin{verbatim}
function C = GABBAEncoder(s)
% Input:  vector s of numeric or symbolic entries corresponding to symbols
% Output: GABBA mother encoding matrix C
% Ex.: syms s1 s2 s3 s4 s5 s6 s7 s8 s9 s10 s11 s12 s13 s14 s15 s16
%      s = [s1 s2 s3 s4 s5 s6 s7 s8 s9 s10 s11 s12 s13 s14 s15 s16];
%      C = GABBAEncoder(s);
k = length(s); % Must be a power of 2
s = reshape(s,[1 1 k]);
while k > 2;
    for n = 1:k/2,
        C(:,:,n) = ABBA1(s(:,:,2*n-1),s(:,:,2*n));
    end
    s = C;
    clear C;
    k = k/2;
end
C = ABBA2(s(:,:,1),s(:,:,2));
% --- Auxiliary functions ---
function C = ABBA1(x,y); C = [[x y];[-y x]];
function C = ABBA2(x,y); C = [[x y];[-y' x']];
\end{verbatim}

\clearpage
\newpage
\section{GABBA Encoded Channel Matrix}
\label{Appendix_GABBAEncodedChannelMatrix} %
\subsection{Example: Minors of GABBA Encoded Channel Matrix ($K = 32$)}
\label{ExampleGABBAEncodedChannelMatrix} %
\vspace{-2em}
\begin{eqnarray}
\label{EncodedChannelMatrixH1_32}%
\renewcommand{\arraystretch}{0.5}
&\hspace{-1.5em}{\scriptstyle\mathbf{H}_{32_1}\!= }\!\!\left[\!\!%
{\tiny
\begin{array}{rrrrrrrrrrrrrrrrrrrrrrrrrrrrrrrr}
 &&&&&&&&&&&&&&&\\ [-3ex]%
 \!\!\!\! h_{1}   \!\!\!\! & \!\!\!\!\!\!\! h_{2}   \!\!\!\! & \!\!\!\!\!\!\! h_{3}   \!\!\!\! & \!\!\!\!\!\!\! h_{4}   \!\!\!\! & \!\!\!\!\!\!\! h_{5}   \!\!\!\! & \!\!\!\!\!\!\! h_{6}   \!\!\!\! & \!\!\!\!\!\!\! h_{7}   \!\!\!\! & \!\!\!\!\!\!\! h_{8}   \!\!\!\! & \!\!\!\!\!\!\! h_{9}   \!\!\!\! & \!\!\!\!\!\!\! h_{10}  \!\!\!\! & \!\!\!\!\!\!\! h_{11}  \!\!\!\! & \!\!\!\!\!\!\! h_{12}  \!\!\!\! & \!\!\!\!\!\!\! h_{13}  \!\!\!\! & \!\!\!\!\!\!\! h_{14}  \!\!\!\! & \!\!\!\!\!\!\! h_{15}  \!\!\!\! & \!\!\!\!\!\!\! h_{16}  \!\!\!\! & \!\!\! h_{17}   \!\!\!\! & \!\!\!\!\!\!\! h_{18}   \!\!\!\! & \!\!\!\!\!\!\! h_{19}   \!\!\!\! & \!\!\!\!\!\!\! h_{20}   \!\!\!\! & \!\!\!\!\!\!\! h_{21}   \!\!\!\! & \!\!\!\!\!\!\! h_{22}   \!\!\!\! & \!\!\!\!\!\!\! h_{23}   \!\!\!\! & \!\!\!\!\!\!\! h_{24}   \!\!\!\! & \!\!\!\!\!\!\! h_{25}   \!\!\!\! & \!\!\!\!\!\!\! h_{26}   \!\!\!\! & \!\!\!\!\!\!\! h_{27}   \!\!\!\! & \!\!\!\!\!\!\! h_{28}   \!\!\!\! & \!\!\!\!\!\!\! h_{29}   \!\!\!\! & \!\!\!\!\!\!\! h_{30}   \!\!\!\! & \!\!\!\!\!\!\! h_{31}   \!\!\!\! & \!\!\!\!\!\!\! h_{32} \\ %
 \!\!\!\! h_{2}   \!\!\!\! & \!\!\!\!\!-\!\!h_{1}   \!\!\!\! & \!\!\!\!\!\!\! h_{4}   \!\!\!\! & \!\!\!\!\!-\!\!h_{3}   \!\!\!\! & \!\!\!\!\!\!\! h_{6}   \!\!\!\! & \!\!\!\!\!-\!\!h_{5}   \!\!\!\! & \!\!\!\!\!\!\! h_{8}   \!\!\!\! & \!\!\!\!\!-\!\!h_{7}   \!\!\!\! & \!\!\!\!\!\!\! h_{10}  \!\!\!\! & \!\!\!\!\!-\!\!h_{9}   \!\!\!\! & \!\!\!\!\!\!\! h_{12}  \!\!\!\! & \!\!\!\!\!-\!\!h_{11}  \!\!\!\! & \!\!\!\!\!\!\! h_{14}  \!\!\!\! & \!\!\!\!\!-\!\!h_{13}  \!\!\!\! & \!\!\!\!\!\!\! h_{16}  \!\!\!\! & \!\!\!\!\!-\!\!h_{15}  \!\!\!\! & \!\!\! h_{18}   \!\!\!\! & \!\!\!\!\!-\!\!h_{17}   \!\!\!\! & \!\!\!\!\!\!\! h_{20}   \!\!\!\! & \!\!\!\!\!-\!\!h_{19}   \!\!\!\! & \!\!\!\!\!\!\! h_{22}   \!\!\!\! & \!\!\!\!\!-\!\!h_{21}   \!\!\!\! & \!\!\!\!\!\!\! h_{24}   \!\!\!\! & \!\!\!\!\!-\!\!h_{23}   \!\!\!\! & \!\!\!\!\!\!\! h_{26}   \!\!\!\! & \!\!\!\!\!-\!\!h_{25}   \!\!\!\! & \!\!\!\!\!\!\! h_{28}   \!\!\!\! & \!\!\!\!\!-\!\!h_{27}   \!\!\!\! & \!\!\!\!\!\!\! h_{30}   \!\!\!\! & \!\!\!\!\!-\!\!h_{29}   \!\!\!\! & \!\!\!\!\!\!\! h_{32}   \!\!\!\! & \!\!\!\!\!-\!\!h_{31} \\ %
 \!\!\!\! h_{3}   \!\!\!\! & \!\!\!\!\!\!\! h_{4}   \!\!\!\! & \!\!\!\!\!-\!\!h_{1}   \!\!\!\! & \!\!\!\!\!-\!\!h_{2}   \!\!\!\! & \!\!\!\!\!\!\! h_{7}   \!\!\!\! & \!\!\!\!\!\!\! h_{8}   \!\!\!\! & \!\!\!\!\!-\!\!h_{5}   \!\!\!\! & \!\!\!\!\!-\!\!h_{6}   \!\!\!\! & \!\!\!\!\!\!\! h_{11}  \!\!\!\! & \!\!\!\!\!\!\! h_{12}  \!\!\!\! & \!\!\!\!\!-\!\!h_{9}   \!\!\!\! & \!\!\!\!\!-\!\!h_{10}  \!\!\!\! & \!\!\!\!\!\!\! h_{15}  \!\!\!\! & \!\!\!\!\!\!\! h_{16}  \!\!\!\! & \!\!\!\!\!-\!\!h_{13}  \!\!\!\! & \!\!\!\!\!-\!\!h_{14}  \!\!\!\! & \!\!\! h_{19}   \!\!\!\! & \!\!\!\!\!\!\! h_{20}   \!\!\!\! & \!\!\!\!\!-\!\!h_{17}   \!\!\!\! & \!\!\!\!\!-\!\!h_{18}   \!\!\!\! & \!\!\!\!\!\!\! h_{23}   \!\!\!\! & \!\!\!\!\!\!\! h_{24}   \!\!\!\! & \!\!\!\!\!-\!\!h_{21}   \!\!\!\! & \!\!\!\!\!-\!\!h_{22}   \!\!\!\! & \!\!\!\!\!\!\! h_{27}   \!\!\!\! & \!\!\!\!\!\!\! h_{28}   \!\!\!\! & \!\!\!\!\!-\!\!h_{25}   \!\!\!\! & \!\!\!\!\!-\!\!h_{26}   \!\!\!\! & \!\!\!\!\!\!\! h_{31}   \!\!\!\! & \!\!\!\!\!\!\! h_{32}   \!\!\!\! & \!\!\!\!\!-\!\!h_{29}   \!\!\!\! & \!\!\!\!\!-\!\!h_{30} \\ %
 \!\!\!\! h_{4}   \!\!\!\! & \!\!\!\!\!-\!\!h_{3}   \!\!\!\! & \!\!\!\!\!-\!\!h_{2}   \!\!\!\! & \!\!\!\!\!\!\! h_{1}   \!\!\!\! & \!\!\!\!\!\!\! h_{8}   \!\!\!\! & \!\!\!\!\!-\!\!h_{7}   \!\!\!\! & \!\!\!\!\!-\!\!h_{6}   \!\!\!\! & \!\!\!\!\!\!\! h_{5}   \!\!\!\! & \!\!\!\!\!\!\! h_{12}  \!\!\!\! & \!\!\!\!\!-\!\!h_{11}  \!\!\!\! & \!\!\!\!\!-\!\!h_{10}  \!\!\!\! & \!\!\!\!\!\!\! h_{9}   \!\!\!\! & \!\!\!\!\!\!\! h_{16}  \!\!\!\! & \!\!\!\!\!-\!\!h_{15}  \!\!\!\! & \!\!\!\!\!-\!\!h_{14}  \!\!\!\! & \!\!\!\!\!\!\! h_{13}  \!\!\!\! & \!\!\! h_{20}   \!\!\!\! & \!\!\!\!\!-\!\!h_{19}   \!\!\!\! & \!\!\!\!\!-\!\!h_{18}   \!\!\!\! & \!\!\!\!\!\!\! h_{17}   \!\!\!\! & \!\!\!\!\!\!\! h_{24}   \!\!\!\! & \!\!\!\!\!-\!\!h_{23}   \!\!\!\! & \!\!\!\!\!-\!\!h_{22}   \!\!\!\! & \!\!\!\!\!\!\! h_{21}   \!\!\!\! & \!\!\!\!\!\!\! h_{28}   \!\!\!\! & \!\!\!\!\!-\!\!h_{27}   \!\!\!\! & \!\!\!\!\!-\!\!h_{26}   \!\!\!\! & \!\!\!\!\!\!\! h_{25}   \!\!\!\! & \!\!\!\!\!\!\! h_{32}   \!\!\!\! & \!\!\!\!\!-\!\!h_{31}   \!\!\!\! & \!\!\!\!\!-\!\!h_{30}   \!\!\!\! & \!\!\!\!\!\!\! h_{29} \\ %
 \!\!\!\! h_{5}   \!\!\!\! & \!\!\!\!\!\!\! h_{6}   \!\!\!\! & \!\!\!\!\!\!\! h_{7}   \!\!\!\! & \!\!\!\!\!\!\! h_{8}   \!\!\!\! & \!\!\!\!\!-\!\!h_{1}   \!\!\!\! & \!\!\!\!\!-\!\!h_{2}   \!\!\!\! & \!\!\!\!\!-\!\!h_{3}   \!\!\!\! & \!\!\!\!\!-\!\!h_{4}   \!\!\!\! & \!\!\!\!\!\!\! h_{13}  \!\!\!\! & \!\!\!\!\!\!\! h_{14}  \!\!\!\! & \!\!\!\!\!\!\! h_{15}  \!\!\!\! & \!\!\!\!\!\!\! h_{16}  \!\!\!\! & \!\!\!\!\!-\!\!h_{9}   \!\!\!\! & \!\!\!\!\!-\!\!h_{10}  \!\!\!\! & \!\!\!\!\!-\!\!h_{11}  \!\!\!\! & \!\!\!\!\!-\!\!h_{12}  \!\!\!\! & \!\!\! h_{21}   \!\!\!\! & \!\!\!\!\!\!\! h_{22}   \!\!\!\! & \!\!\!\!\!\!\! h_{23}   \!\!\!\! & \!\!\!\!\!\!\! h_{24}   \!\!\!\! & \!\!\!\!\!-\!\!h_{17}   \!\!\!\! & \!\!\!\!\!-\!\!h_{18}   \!\!\!\! & \!\!\!\!\!-\!\!h_{19}   \!\!\!\! & \!\!\!\!\!-\!\!h_{20}   \!\!\!\! & \!\!\!\!\!\!\! h_{29}   \!\!\!\! & \!\!\!\!\!\!\! h_{30}   \!\!\!\! & \!\!\!\!\!\!\! h_{31}   \!\!\!\! & \!\!\!\!\!\!\! h_{32}   \!\!\!\! & \!\!\!\!\!-\!\!h_{25}   \!\!\!\! & \!\!\!\!\!-\!\!h_{26}   \!\!\!\! & \!\!\!\!\!-\!\!h_{27}   \!\!\!\! & \!\!\!\!\!-\!\!h_{28} \\ %
 \!\!\!\! h_{6}   \!\!\!\! & \!\!\!\!\!-\!\!h_{5}   \!\!\!\! & \!\!\!\!\!\!\! h_{8}   \!\!\!\! & \!\!\!\!\!-\!\!h_{7}   \!\!\!\! & \!\!\!\!\!-\!\!h_{2}   \!\!\!\! & \!\!\!\!\!\!\! h_{1}   \!\!\!\! & \!\!\!\!\!-\!\!h_{4}   \!\!\!\! & \!\!\!\!\!\!\! h_{3}   \!\!\!\! & \!\!\!\!\!\!\! h_{14}  \!\!\!\! & \!\!\!\!\!-\!\!h_{13}  \!\!\!\! & \!\!\!\!\!\!\! h_{16}  \!\!\!\! & \!\!\!\!\!-\!\!h_{15}  \!\!\!\! & \!\!\!\!\!-\!\!h_{10}  \!\!\!\! & \!\!\!\!\!\!\! h_{9}   \!\!\!\! & \!\!\!\!\!-\!\!h_{12}  \!\!\!\! & \!\!\!\!\!\!\! h_{11}  \!\!\!\! & \!\!\! h_{22}   \!\!\!\! & \!\!\!\!\!-\!\!h_{21}   \!\!\!\! & \!\!\!\!\!\!\! h_{24}   \!\!\!\! & \!\!\!\!\!-\!\!h_{23}   \!\!\!\! & \!\!\!\!\!-\!\!h_{18}   \!\!\!\! & \!\!\!\!\!\!\! h_{17}   \!\!\!\! & \!\!\!\!\!-\!\!h_{20}   \!\!\!\! & \!\!\!\!\!\!\! h_{19}   \!\!\!\! & \!\!\!\!\!\!\! h_{30}   \!\!\!\! & \!\!\!\!\!-\!\!h_{29}   \!\!\!\! & \!\!\!\!\!\!\! h_{32}   \!\!\!\! & \!\!\!\!\!-\!\!h_{31}   \!\!\!\! & \!\!\!\!\!-\!\!h_{26}   \!\!\!\! & \!\!\!\!\!\!\! h_{25}   \!\!\!\! & \!\!\!\!\!-\!\!h_{28}   \!\!\!\! & \!\!\!\!\!\!\! h_{27} \\ %
 \!\!\!\! h_{7}   \!\!\!\! & \!\!\!\!\!\!\! h_{8}   \!\!\!\! & \!\!\!\!\!-\!\!h_{5}   \!\!\!\! & \!\!\!\!\!-\!\!h_{6}   \!\!\!\! & \!\!\!\!\!-\!\!h_{3}   \!\!\!\! & \!\!\!\!\!-\!\!h_{4}   \!\!\!\! & \!\!\!\!\!\!\! h_{1}   \!\!\!\! & \!\!\!\!\!\!\! h_{2}   \!\!\!\! & \!\!\!\!\!\!\! h_{15}  \!\!\!\! & \!\!\!\!\!\!\! h_{16}  \!\!\!\! & \!\!\!\!\!-\!\!h_{13}  \!\!\!\! & \!\!\!\!\!-\!\!h_{14}  \!\!\!\! & \!\!\!\!\!-\!\!h_{11}  \!\!\!\! & \!\!\!\!\!-\!\!h_{12}  \!\!\!\! & \!\!\!\!\!\!\! h_{9}   \!\!\!\! & \!\!\!\!\!\!\! h_{10}  \!\!\!\! & \!\!\! h_{23}   \!\!\!\! & \!\!\!\!\!\!\! h_{24}   \!\!\!\! & \!\!\!\!\!-\!\!h_{21}   \!\!\!\! & \!\!\!\!\!-\!\!h_{22}   \!\!\!\! & \!\!\!\!\!-\!\!h_{19}   \!\!\!\! & \!\!\!\!\!-\!\!h_{20}   \!\!\!\! & \!\!\!\!\!\!\! h_{17}   \!\!\!\! & \!\!\!\!\!\!\! h_{18}   \!\!\!\! & \!\!\!\!\!\!\! h_{31}   \!\!\!\! & \!\!\!\!\!\!\! h_{32}   \!\!\!\! & \!\!\!\!\!-\!\!h_{29}   \!\!\!\! & \!\!\!\!\!-\!\!h_{30}   \!\!\!\! & \!\!\!\!\!-\!\!h_{27}   \!\!\!\! & \!\!\!\!\!-\!\!h_{28}   \!\!\!\! & \!\!\!\!\!\!\! h_{25}   \!\!\!\! & \!\!\!\!\!\!\! h_{26} \\ %
 \!\!\!\! h_{8}   \!\!\!\! & \!\!\!\!\!-\!\!h_{7}   \!\!\!\! & \!\!\!\!\!-\!\!h_{6}   \!\!\!\! & \!\!\!\!\!\!\! h_{5}   \!\!\!\! & \!\!\!\!\!-\!\!h_{4}   \!\!\!\! & \!\!\!\!\!\!\! h_{3}   \!\!\!\! & \!\!\!\!\!\!\! h_{2}   \!\!\!\! & \!\!\!\!\!-\!\!h_{1}   \!\!\!\! & \!\!\!\!\!\!\! h_{16}  \!\!\!\! & \!\!\!\!\!-\!\!h_{15}  \!\!\!\! & \!\!\!\!\!-\!\!h_{14}  \!\!\!\! & \!\!\!\!\!\!\! h_{13}  \!\!\!\! & \!\!\!\!\!-\!\!h_{12}  \!\!\!\! & \!\!\!\!\!\!\! h_{11}  \!\!\!\! & \!\!\!\!\!\!\! h_{10}  \!\!\!\! & \!\!\!\!\!-\!\!h_{9}   \!\!\!\! & \!\!\! h_{24}   \!\!\!\! & \!\!\!\!\!-\!\!h_{23}   \!\!\!\! & \!\!\!\!\!-\!\!h_{22}   \!\!\!\! & \!\!\!\!\!\!\! h_{21}   \!\!\!\! & \!\!\!\!\!-\!\!h_{20}   \!\!\!\! & \!\!\!\!\!\!\! h_{19}   \!\!\!\! & \!\!\!\!\!\!\! h_{18}   \!\!\!\! & \!\!\!\!\!-\!\!h_{17}   \!\!\!\! & \!\!\!\!\!\!\! h_{32}   \!\!\!\! & \!\!\!\!\!-\!\!h_{31}   \!\!\!\! & \!\!\!\!\!-\!\!h_{30}   \!\!\!\! & \!\!\!\!\!\!\! h_{29}   \!\!\!\! & \!\!\!\!\!-\!\!h_{28}   \!\!\!\! & \!\!\!\!\!\!\! h_{27}   \!\!\!\! & \!\!\!\!\!\!\! h_{26}   \!\!\!\! & \!\!\!\!\!-\!\!h_{25} \\ %
 \!\!\!\! h_{9}   \!\!\!\! & \!\!\!\!\!\!\! h_{10}  \!\!\!\! & \!\!\!\!\!\!\! h_{11}  \!\!\!\! & \!\!\!\!\!\!\! h_{12}  \!\!\!\! & \!\!\!\!\!\!\! h_{13}  \!\!\!\! & \!\!\!\!\!\!\! h_{14}  \!\!\!\! & \!\!\!\!\!\!\! h_{15}  \!\!\!\! & \!\!\!\!\!\!\! h_{16}  \!\!\!\! & \!\!\!\!\!-\!\!h_{1}   \!\!\!\! & \!\!\!\!\!-\!\!h_{2}   \!\!\!\! & \!\!\!\!\!-\!\!h_{3}   \!\!\!\! & \!\!\!\!\!-\!\!h_{4}   \!\!\!\! & \!\!\!\!\!-\!\!h_{5}   \!\!\!\! & \!\!\!\!\!-\!\!h_{6}   \!\!\!\! & \!\!\!\!\!-\!\!h_{7}   \!\!\!\! & \!\!\!\!\!-\!\!h_{8}   \!\!\!\! & \!\!\! h_{25}   \!\!\!\! & \!\!\!\!\!\!\! h_{26}   \!\!\!\! & \!\!\!\!\!\!\! h_{27}   \!\!\!\! & \!\!\!\!\!\!\! h_{28}   \!\!\!\! & \!\!\!\!\!\!\! h_{29}   \!\!\!\! & \!\!\!\!\!\!\! h_{30}   \!\!\!\! & \!\!\!\!\!\!\! h_{31}   \!\!\!\! & \!\!\!\!\!\!\! h_{32}   \!\!\!\! & \!\!\!\!\!-\!\!h_{17}   \!\!\!\! & \!\!\!\!\!-\!\!h_{18}   \!\!\!\! & \!\!\!\!\!-\!\!h_{19}   \!\!\!\! & \!\!\!\!\!-\!\!h_{20}   \!\!\!\! & \!\!\!\!\!-\!\!h_{21}   \!\!\!\! & \!\!\!\!\!-\!\!h_{22}   \!\!\!\! & \!\!\!\!\!-\!\!h_{23}   \!\!\!\! & \!\!\!\!\!-\!\!h_{24} \\ %
   \!\!\! h_{10}  \!\!\!\! & \!\!\!\!\!-\!\!h_{9}   \!\!\!\! & \!\!\!\!\!\!\! h_{12}  \!\!\!\! & \!\!\!\!\!-\!\!h_{11}  \!\!\!\! & \!\!\!\!\!\!\! h_{14}  \!\!\!\! & \!\!\!\!\!-\!\!h_{13}  \!\!\!\! & \!\!\!\!\!\!\! h_{16}  \!\!\!\! & \!\!\!\!\!-\!\!h_{15}  \!\!\!\! & \!\!\!\!\!-\!\!h_{2}   \!\!\!\! & \!\!\!\!\!\!\! h_{1}   \!\!\!\! & \!\!\!\!\!-\!\!h_{4}   \!\!\!\! & \!\!\!\!\!\!\! h_{3}   \!\!\!\! & \!\!\!\!\!-\!\!h_{6}   \!\!\!\! & \!\!\!\!\!\!\! h_{5}   \!\!\!\! & \!\!\!\!\!-\!\!h_{8}   \!\!\!\! & \!\!\!\!\!\!\! h_{7}   \!\!\!\! & \!\!\! h_{26}   \!\!\!\! & \!\!\!\!\!-\!\!h_{25}   \!\!\!\! & \!\!\!\!\!\!\! h_{28}   \!\!\!\! & \!\!\!\!\!-\!\!h_{27}   \!\!\!\! & \!\!\!\!\!\!\! h_{30}   \!\!\!\! & \!\!\!\!\!-\!\!h_{29}   \!\!\!\! & \!\!\!\!\!\!\! h_{32}   \!\!\!\! & \!\!\!\!\!-\!\!h_{31}   \!\!\!\! & \!\!\!\!\!-\!\!h_{18}   \!\!\!\! & \!\!\!\!\!\!\! h_{17}   \!\!\!\! & \!\!\!\!\!-\!\!h_{20}   \!\!\!\! & \!\!\!\!\!\!\! h_{19}   \!\!\!\! & \!\!\!\!\!-\!\!h_{22}   \!\!\!\! & \!\!\!\!\!\!\! h_{21}   \!\!\!\! & \!\!\!\!\!-\!\!h_{24}   \!\!\!\! & \!\!\!\!\!\!\! h_{23} \\ %
   \!\!\! h_{11}  \!\!\!\! & \!\!\!\!\!\!\! h_{12}  \!\!\!\! & \!\!\!\!\!-\!\!h_{9}   \!\!\!\! & \!\!\!\!\!-\!\!h_{10}  \!\!\!\! & \!\!\!\!\!\!\! h_{15}  \!\!\!\! & \!\!\!\!\!\!\! h_{16}  \!\!\!\! & \!\!\!\!\!-\!\!h_{13}  \!\!\!\! & \!\!\!\!\!-\!\!h_{14}  \!\!\!\! & \!\!\!\!\!-\!\!h_{3}   \!\!\!\! & \!\!\!\!\!-\!\!h_{4}   \!\!\!\! & \!\!\!\!\!\!\! h_{1}   \!\!\!\! & \!\!\!\!\!\!\! h_{2}   \!\!\!\! & \!\!\!\!\!-\!\!h_{7}   \!\!\!\! & \!\!\!\!\!-\!\!h_{8}   \!\!\!\! & \!\!\!\!\!\!\! h_{5}   \!\!\!\! & \!\!\!\!\!\!\! h_{6}   \!\!\!\! & \!\!\! h_{27}   \!\!\!\! & \!\!\!\!\!\!\! h_{28}   \!\!\!\! & \!\!\!\!\!-\!\!h_{25}   \!\!\!\! & \!\!\!\!\!-\!\!h_{26}   \!\!\!\! & \!\!\!\!\!\!\! h_{31}   \!\!\!\! & \!\!\!\!\!\!\! h_{32}   \!\!\!\! & \!\!\!\!\!-\!\!h_{29}   \!\!\!\! & \!\!\!\!\!-\!\!h_{30}   \!\!\!\! & \!\!\!\!\!-\!\!h_{19}   \!\!\!\! & \!\!\!\!\!-\!\!h_{20}   \!\!\!\! & \!\!\!\!\!\!\! h_{17}   \!\!\!\! & \!\!\!\!\!\!\! h_{18}   \!\!\!\! & \!\!\!\!\!-\!\!h_{23}   \!\!\!\! & \!\!\!\!\!-\!\!h_{24}   \!\!\!\! & \!\!\!\!\!\!\! h_{21}   \!\!\!\! & \!\!\!\!\!\!\! h_{22} \\ %
   \!\!\! h_{12}  \!\!\!\! & \!\!\!\!\!-\!\!h_{11}  \!\!\!\! & \!\!\!\!\!-\!\!h_{10}  \!\!\!\! & \!\!\!\!\!\!\! h_{9}   \!\!\!\! & \!\!\!\!\!\!\! h_{16}  \!\!\!\! & \!\!\!\!\!-\!\!h_{15}  \!\!\!\! & \!\!\!\!\!-\!\!h_{14}  \!\!\!\! & \!\!\!\!\!\!\! h_{13}  \!\!\!\! & \!\!\!\!\!-\!\!h_{4}   \!\!\!\! & \!\!\!\!\!\!\! h_{3}   \!\!\!\! & \!\!\!\!\!\!\! h_{2}   \!\!\!\! & \!\!\!\!\!-\!\!h_{1}   \!\!\!\! & \!\!\!\!\!-\!\!h_{8}   \!\!\!\! & \!\!\!\!\!\!\! h_{7}   \!\!\!\! & \!\!\!\!\!\!\! h_{6}   \!\!\!\! & \!\!\!\!\!-\!\!h_{5}   \!\!\!\! & \!\!\! h_{28}   \!\!\!\! & \!\!\!\!\!-\!\!h_{27}   \!\!\!\! & \!\!\!\!\!-\!\!h_{26}   \!\!\!\! & \!\!\!\!\!\!\! h_{25}   \!\!\!\! & \!\!\!\!\!\!\! h_{32}   \!\!\!\! & \!\!\!\!\!-\!\!h_{31}   \!\!\!\! & \!\!\!\!\!-\!\!h_{30}   \!\!\!\! & \!\!\!\!\!\!\! h_{29}   \!\!\!\! & \!\!\!\!\!-\!\!h_{20}   \!\!\!\! & \!\!\!\!\!\!\! h_{19}   \!\!\!\! & \!\!\!\!\!\!\! h_{18}   \!\!\!\! & \!\!\!\!\!-\!\!h_{17}   \!\!\!\! & \!\!\!\!\!-\!\!h_{24}   \!\!\!\! & \!\!\!\!\!\!\! h_{23}   \!\!\!\! & \!\!\!\!\!\!\! h_{22}   \!\!\!\! & \!\!\!\!\!-\!\!h_{21} \\ %
   \!\!\! h_{13}  \!\!\!\! & \!\!\!\!\!\!\! h_{14}  \!\!\!\! & \!\!\!\!\!\!\! h_{15}  \!\!\!\! & \!\!\!\!\!\!\! h_{16}  \!\!\!\! & \!\!\!\!\!-\!\!h_{9}   \!\!\!\! & \!\!\!\!\!-\!\!h_{10}  \!\!\!\! & \!\!\!\!\!-\!\!h_{11}  \!\!\!\! & \!\!\!\!\!-\!\!h_{12}  \!\!\!\! & \!\!\!\!\!-\!\!h_{5}   \!\!\!\! & \!\!\!\!\!-\!\!h_{6}   \!\!\!\! & \!\!\!\!\!-\!\!h_{7}   \!\!\!\! & \!\!\!\!\!-\!\!h_{8}   \!\!\!\! & \!\!\!\!\!\!\! h_{1}   \!\!\!\! & \!\!\!\!\!\!\! h_{2}   \!\!\!\! & \!\!\!\!\!\!\! h_{3}   \!\!\!\! & \!\!\!\!\!\!\! h_{4}   \!\!\!\! & \!\!\! h_{29}   \!\!\!\! & \!\!\!\!\!\!\! h_{30}   \!\!\!\! & \!\!\!\!\!\!\! h_{31}   \!\!\!\! & \!\!\!\!\!\!\! h_{32}   \!\!\!\! & \!\!\!\!\!-\!\!h_{25}   \!\!\!\! & \!\!\!\!\!-\!\!h_{26}   \!\!\!\! & \!\!\!\!\!-\!\!h_{27}   \!\!\!\! & \!\!\!\!\!-\!\!h_{28}   \!\!\!\! & \!\!\!\!\!-\!\!h_{21}   \!\!\!\! & \!\!\!\!\!-\!\!h_{22}   \!\!\!\! & \!\!\!\!\!-\!\!h_{23}   \!\!\!\! & \!\!\!\!\!-\!\!h_{24}   \!\!\!\! & \!\!\!\!\!\!\! h_{17}   \!\!\!\! & \!\!\!\!\!\!\! h_{18}   \!\!\!\! & \!\!\!\!\!\!\! h_{19}   \!\!\!\! & \!\!\!\!\!\!\! h_{20} \\ %
   \!\!\! h_{14}  \!\!\!\! & \!\!\!\!\!-\!\!h_{13}  \!\!\!\! & \!\!\!\!\!\!\! h_{16}  \!\!\!\! & \!\!\!\!\!-\!\!h_{15}  \!\!\!\! & \!\!\!\!\!-\!\!h_{10}  \!\!\!\! & \!\!\!\!\!\!\! h_{9}   \!\!\!\! & \!\!\!\!\!-\!\!h_{12}  \!\!\!\! & \!\!\!\!\!\!\! h_{11}  \!\!\!\! & \!\!\!\!\!-\!\!h_{6}   \!\!\!\! & \!\!\!\!\!\!\! h_{5}   \!\!\!\! & \!\!\!\!\!-\!\!h_{8}   \!\!\!\! & \!\!\!\!\!\!\! h_{7}   \!\!\!\! & \!\!\!\!\!\!\! h_{2}   \!\!\!\! & \!\!\!\!\!-\!\!h_{1}   \!\!\!\! & \!\!\!\!\!\!\! h_{4}   \!\!\!\! & \!\!\!\!\!-\!\!h_{3}   \!\!\!\! & \!\!\! h_{30}   \!\!\!\! & \!\!\!\!\!-\!\!h_{29}   \!\!\!\! & \!\!\!\!\!\!\! h_{32}   \!\!\!\! & \!\!\!\!\!-\!\!h_{31}   \!\!\!\! & \!\!\!\!\!-\!\!h_{26}   \!\!\!\! & \!\!\!\!\!\!\! h_{25}   \!\!\!\! & \!\!\!\!\!-\!\!h_{28}   \!\!\!\! & \!\!\!\!\!\!\! h_{27}   \!\!\!\! & \!\!\!\!\!-\!\!h_{22}   \!\!\!\! & \!\!\!\!\!\!\! h_{21}   \!\!\!\! & \!\!\!\!\!-\!\!h_{24}   \!\!\!\! & \!\!\!\!\!\!\! h_{23}   \!\!\!\! & \!\!\!\!\!\!\! h_{18}   \!\!\!\! & \!\!\!\!\!-\!\!h_{17}   \!\!\!\! & \!\!\!\!\!\!\! h_{20}   \!\!\!\! & \!\!\!\!\!-\!\!h_{19} \\ %
   \!\!\! h_{15}  \!\!\!\! & \!\!\!\!\!\!\! h_{16}  \!\!\!\! & \!\!\!\!\!-\!\!h_{13}  \!\!\!\! & \!\!\!\!\!-\!\!h_{14}  \!\!\!\! & \!\!\!\!\!-\!\!h_{11}  \!\!\!\! & \!\!\!\!\!-\!\!h_{12}  \!\!\!\! & \!\!\!\!\!\!\! h_{9}   \!\!\!\! & \!\!\!\!\!\!\! h_{10}  \!\!\!\! & \!\!\!\!\!-\!\!h_{7}   \!\!\!\! & \!\!\!\!\!-\!\!h_{8}   \!\!\!\! & \!\!\!\!\!\!\! h_{5}   \!\!\!\! & \!\!\!\!\!\!\! h_{6}   \!\!\!\! & \!\!\!\!\!\!\! h_{3}   \!\!\!\! & \!\!\!\!\!\!\! h_{4}   \!\!\!\! & \!\!\!\!\!-\!\!h_{1}   \!\!\!\! & \!\!\!\!\!-\!\!h_{2}   \!\!\!\! & \!\!\! h_{31}   \!\!\!\! & \!\!\!\!\!\!\! h_{32}   \!\!\!\! & \!\!\!\!\!-\!\!h_{29}   \!\!\!\! & \!\!\!\!\!-\!\!h_{30}   \!\!\!\! & \!\!\!\!\!-\!\!h_{27}   \!\!\!\! & \!\!\!\!\!-\!\!h_{28}   \!\!\!\! & \!\!\!\!\!\!\! h_{25}   \!\!\!\! & \!\!\!\!\!\!\! h_{26}   \!\!\!\! & \!\!\!\!\!-\!\!h_{23}   \!\!\!\! & \!\!\!\!\!-\!\!h_{24}   \!\!\!\! & \!\!\!\!\!\!\! h_{21}   \!\!\!\! & \!\!\!\!\!\!\! h_{22}   \!\!\!\! & \!\!\!\!\!\!\! h_{19}   \!\!\!\! & \!\!\!\!\!\!\! h_{20}   \!\!\!\! & \!\!\!\!\!-\!\!h_{17}   \!\!\!\! & \!\!\!\!\!-\!\!h_{18} \\ %
   \!\!\! h_{16}  \!\!\!\! & \!\!\!\!\!-\!\!h_{15}  \!\!\!\! & \!\!\!\!\!-\!\!h_{14}  \!\!\!\! & \!\!\!\!\!\!\! h_{13}  \!\!\!\! & \!\!\!\!\!-\!\!h_{12}  \!\!\!\! & \!\!\!\!\!\!\! h_{11}  \!\!\!\! & \!\!\!\!\!\!\! h_{10}  \!\!\!\! & \!\!\!\!\!-\!\!h_{9}   \!\!\!\! & \!\!\!\!\!-\!\!h_{8}   \!\!\!\! & \!\!\!\!\!\!\! h_{7}   \!\!\!\! & \!\!\!\!\!\!\! h_{6}   \!\!\!\! & \!\!\!\!\!-\!\!h_{5}   \!\!\!\! & \!\!\!\!\!\!\! h_{4}   \!\!\!\! & \!\!\!\!\!-\!\!h_{3}   \!\!\!\! & \!\!\!\!\!-\!\!h_{2}   \!\!\!\! & \!\!\!\!\!\!\! h_{1}   \!\!\!\! & \!\!\! h_{32}   \!\!\!\! & \!\!\!\!\!-\!\!h_{31}   \!\!\!\! & \!\!\!\!\!-\!\!h_{30}   \!\!\!\! & \!\!\!\!\!\!\! h_{29}   \!\!\!\! & \!\!\!\!\!-\!\!h_{28}   \!\!\!\! & \!\!\!\!\!\!\! h_{27}   \!\!\!\! & \!\!\!\!\!\!\! h_{26}   \!\!\!\! & \!\!\!\!\!-\!\!h_{25}   \!\!\!\! & \!\!\!\!\!-\!\!h_{24}   \!\!\!\! & \!\!\!\!\!\!\! h_{23}   \!\!\!\! & \!\!\!\!\!\!\! h_{22}   \!\!\!\! & \!\!\!\!\!-\!\!h_{21}   \!\!\!\! & \!\!\!\!\!\!\! h_{20}   \!\!\!\! & \!\!\!\!\!-\!\!h_{19}   \!\!\!\! & \!\!\!\!\!-\!\!h_{18}   \!\!\!\! & \!\!\!\!\!\!\! h_{17} \\ [1.5ex]%
\end{array}}
\!\!\!\!\right]&\nonumber\\ %
\label{EncodedChannelMatrixH2_32}%
&\hspace{-1.5em}{\scriptstyle\mathbf{H}_{32_2}\!= }\!\!\left[\!\!%
{\tiny
\begin{array}{rrrrrrrrrrrrrrrrrrrrrrrrrrrrrrrr}
 &&&&&&&&&&&&&&&\\ [-3ex]%
 \!\! h_{17}  \!\!\!\! & \!\!\!\!\!-\!\!h_{18}  \!\!\!\! & \!\!\!\!\!-\!\!h_{19}  \!\!\!\! & \!\!\!\!\!\!\! h_{20}  \!\!\!\! & \!\!\!\!\!-\!\!h_{21}  \!\!\!\! & \!\!\!\!\!\!\! h_{22}  \!\!\!\! & \!\!\!\!\!\!\! h_{23}  \!\!\!\! & \!\!\!\!\!-\!\!h_{24}  \!\!\!\! & \!\!\!\!\!-\!\!h_{25}  \!\!\!\! & \!\!\!\!\!\!\! h_{26}  \!\!\!\! & \!\!\!\!\!\!\! h_{27}  \!\!\!\! & \!\!\!\!\!-\!\!h_{28}  \!\!\!\! & \!\!\!\!\!\!\! h_{29}  \!\!\!\! & \!\!\!\!\!-\!\!h_{30}  \!\!\!\! & \!\!\!\!\!-\!\!h_{31}  \!\!\!\! & \!\!\!\!\!\!\! h_{32}  \!\!\!\! & \!\!\!\!\!-\!\!h_{1}   \!\!\!\! & \!\!\!\!\!\!\! h_{2}   \!\!\!\! & \!\!\!\!\!\!\! h_{3}   \!\!\!\! & \!\!\!\!\!-\!\!h_{4}   \!\!\!\! & \!\!\!\!\!\!\! h_{5}   \!\!\!\! & \!\!\!\!\!-\!\!h_{6}   \!\!\!\! & \!\!\!\!\!-\!\!h_{7}   \!\!\!\! & \!\!\!\!\!\!\! h_{8}   \!\!\!\! & \!\!\!\!\!\!\! h_{9}   \!\!\!\! & \!\!\!\!\!-\!\!h_{10}  \!\!\!\! & \!\!\!\!\!-\!\!h_{11}  \!\!\!\! & \!\!\!\!\!\!\! h_{12}  \!\!\!\! & \!\!\!\!\!-\!\!h_{13}  \!\!\!\! & \!\!\!\!\!\!\! h_{14}  \!\!\!\! & \!\!\!\!\!\!\! h_{15}  \!\!\!\! & \!\!\!\!\!-\!\!h_{16}  \\%
 \!\! h_{18}  \!\!\!\! & \!\!\!\!\!\!\! h_{17}  \!\!\!\! & \!\!\!\!\!-\!\!h_{20}  \!\!\!\! & \!\!\!\!\!-\!\!h_{19}  \!\!\!\! & \!\!\!\!\!-\!\!h_{22}  \!\!\!\! & \!\!\!\!\!-\!\!h_{21}  \!\!\!\! & \!\!\!\!\!\!\! h_{24}  \!\!\!\! & \!\!\!\!\!\!\! h_{23}  \!\!\!\! & \!\!\!\!\!-\!\!h_{26}  \!\!\!\! & \!\!\!\!\!-\!\!h_{25}  \!\!\!\! & \!\!\!\!\!\!\! h_{28}  \!\!\!\! & \!\!\!\!\!\!\! h_{27}  \!\!\!\! & \!\!\!\!\!\!\! h_{30}  \!\!\!\! & \!\!\!\!\!\!\! h_{29}  \!\!\!\! & \!\!\!\!\!-\!\!h_{32}  \!\!\!\! & \!\!\!\!\!-\!\!h_{31}  \!\!\!\! & \!\!\!\!\!-\!\!h_{2}   \!\!\!\! & \!\!\!\!\!-\!\!h_{1}   \!\!\!\! & \!\!\!\!\!\!\! h_{4}   \!\!\!\! & \!\!\!\!\!\!\! h_{3}   \!\!\!\! & \!\!\!\!\!\!\! h_{6}   \!\!\!\! & \!\!\!\!\!\!\! h_{5}   \!\!\!\! & \!\!\!\!\!-\!\!h_{8}   \!\!\!\! & \!\!\!\!\!-\!\!h_{7}   \!\!\!\! & \!\!\!\!\!\!\! h_{10}  \!\!\!\! & \!\!\!\!\!\!\! h_{9}   \!\!\!\! & \!\!\!\!\!-\!\!h_{12}  \!\!\!\! & \!\!\!\!\!-\!\!h_{11}  \!\!\!\! & \!\!\!\!\!-\!\!h_{14}  \!\!\!\! & \!\!\!\!\!-\!\!h_{13}  \!\!\!\! & \!\!\!\!\!\!\! h_{16}  \!\!\!\! & \!\!\!\!\!\!\! h_{15}  \\%
 \!\! h_{19}  \!\!\!\! & \!\!\!\!\!-\!\!h_{20}  \!\!\!\! & \!\!\!\!\!\!\! h_{17}  \!\!\!\! & \!\!\!\!\!-\!\!h_{18}  \!\!\!\! & \!\!\!\!\!-\!\!h_{23}  \!\!\!\! & \!\!\!\!\!\!\! h_{24}  \!\!\!\! & \!\!\!\!\!-\!\!h_{21}  \!\!\!\! & \!\!\!\!\!\!\! h_{22}  \!\!\!\! & \!\!\!\!\!-\!\!h_{27}  \!\!\!\! & \!\!\!\!\!\!\! h_{28}  \!\!\!\! & \!\!\!\!\!-\!\!h_{25}  \!\!\!\! & \!\!\!\!\!\!\! h_{26}  \!\!\!\! & \!\!\!\!\!\!\! h_{31}  \!\!\!\! & \!\!\!\!\!-\!\!h_{32}  \!\!\!\! & \!\!\!\!\!\!\! h_{29}  \!\!\!\! & \!\!\!\!\!-\!\!h_{30}  \!\!\!\! & \!\!\!\!\!-\!\!h_{3}   \!\!\!\! & \!\!\!\!\!\!\! h_{4}   \!\!\!\! & \!\!\!\!\!-\!\!h_{1}   \!\!\!\! & \!\!\!\!\!\!\! h_{2}   \!\!\!\! & \!\!\!\!\!\!\! h_{7}   \!\!\!\! & \!\!\!\!\!-\!\!h_{8}   \!\!\!\! & \!\!\!\!\!\!\! h_{5}   \!\!\!\! & \!\!\!\!\!-\!\!h_{6}   \!\!\!\! & \!\!\!\!\!\!\! h_{11}  \!\!\!\! & \!\!\!\!\!-\!\!h_{12}  \!\!\!\! & \!\!\!\!\!\!\! h_{9}   \!\!\!\! & \!\!\!\!\!-\!\!h_{10}  \!\!\!\! & \!\!\!\!\!-\!\!h_{15}  \!\!\!\! & \!\!\!\!\!\!\! h_{16}  \!\!\!\! & \!\!\!\!\!-\!\!h_{13}  \!\!\!\! & \!\!\!\!\!\!\! h_{14}  \\%
 \!\! h_{20}  \!\!\!\! & \!\!\!\!\!\!\! h_{19}  \!\!\!\! & \!\!\!\!\!\!\! h_{18}  \!\!\!\! & \!\!\!\!\!\!\! h_{17}  \!\!\!\! & \!\!\!\!\!-\!\!h_{24}  \!\!\!\! & \!\!\!\!\!-\!\!h_{23}  \!\!\!\! & \!\!\!\!\!-\!\!h_{22}  \!\!\!\! & \!\!\!\!\!-\!\!h_{21}  \!\!\!\! & \!\!\!\!\!-\!\!h_{28}  \!\!\!\! & \!\!\!\!\!-\!\!h_{27}  \!\!\!\! & \!\!\!\!\!-\!\!h_{26}  \!\!\!\! & \!\!\!\!\!-\!\!h_{25}  \!\!\!\! & \!\!\!\!\!\!\! h_{32}  \!\!\!\! & \!\!\!\!\!\!\! h_{31}  \!\!\!\! & \!\!\!\!\!\!\! h_{30}  \!\!\!\! & \!\!\!\!\!\!\! h_{29}  \!\!\!\! & \!\!\!\!\!-\!\!h_{4}   \!\!\!\! & \!\!\!\!\!-\!\!h_{3}   \!\!\!\! & \!\!\!\!\!-\!\!h_{2}   \!\!\!\! & \!\!\!\!\!-\!\!h_{1}   \!\!\!\! & \!\!\!\!\!\!\! h_{8}   \!\!\!\! & \!\!\!\!\!\!\! h_{7}   \!\!\!\! & \!\!\!\!\!\!\! h_{6}   \!\!\!\! & \!\!\!\!\!\!\! h_{5}   \!\!\!\! & \!\!\!\!\!\!\! h_{12}  \!\!\!\! & \!\!\!\!\!\!\! h_{11}  \!\!\!\! & \!\!\!\!\!\!\! h_{10}  \!\!\!\! & \!\!\!\!\!\!\! h_{9}   \!\!\!\! & \!\!\!\!\!-\!\!h_{16}  \!\!\!\! & \!\!\!\!\!-\!\!h_{15}  \!\!\!\! & \!\!\!\!\!-\!\!h_{14}  \!\!\!\! & \!\!\!\!\!-\!\!h_{13}  \\%
 \!\! h_{21}  \!\!\!\! & \!\!\!\!\!-\!\!h_{22}  \!\!\!\! & \!\!\!\!\!-\!\!h_{23}  \!\!\!\! & \!\!\!\!\!\!\! h_{24}  \!\!\!\! & \!\!\!\!\!\!\! h_{17}  \!\!\!\! & \!\!\!\!\!-\!\!h_{18}  \!\!\!\! & \!\!\!\!\!-\!\!h_{19}  \!\!\!\! & \!\!\!\!\!\!\! h_{20}  \!\!\!\! & \!\!\!\!\!-\!\!h_{29}  \!\!\!\! & \!\!\!\!\!\!\! h_{30}  \!\!\!\! & \!\!\!\!\!\!\! h_{31}  \!\!\!\! & \!\!\!\!\!-\!\!h_{32}  \!\!\!\! & \!\!\!\!\!-\!\!h_{25}  \!\!\!\! & \!\!\!\!\!\!\! h_{26}  \!\!\!\! & \!\!\!\!\!\!\! h_{27}  \!\!\!\! & \!\!\!\!\!-\!\!h_{28}  \!\!\!\! & \!\!\!\!\!-\!\!h_{5}   \!\!\!\! & \!\!\!\!\!\!\! h_{6}   \!\!\!\! & \!\!\!\!\!\!\! h_{7}   \!\!\!\! & \!\!\!\!\!-\!\!h_{8}   \!\!\!\! & \!\!\!\!\!-\!\!h_{1}   \!\!\!\! & \!\!\!\!\!\!\! h_{2}   \!\!\!\! & \!\!\!\!\!\!\! h_{3}   \!\!\!\! & \!\!\!\!\!-\!\!h_{4}   \!\!\!\! & \!\!\!\!\!\!\! h_{13}  \!\!\!\! & \!\!\!\!\!-\!\!h_{14}  \!\!\!\! & \!\!\!\!\!-\!\!h_{15}  \!\!\!\! & \!\!\!\!\!\!\! h_{16}  \!\!\!\! & \!\!\!\!\!\!\! h_{9}   \!\!\!\! & \!\!\!\!\!-\!\!h_{10}  \!\!\!\! & \!\!\!\!\!-\!\!h_{11}  \!\!\!\! & \!\!\!\!\!\!\! h_{12}  \\%
 \!\! h_{22}  \!\!\!\! & \!\!\!\!\!\!\! h_{21}  \!\!\!\! & \!\!\!\!\!-\!\!h_{24}  \!\!\!\! & \!\!\!\!\!-\!\!h_{23}  \!\!\!\! & \!\!\!\!\!\!\! h_{18}  \!\!\!\! & \!\!\!\!\!\!\! h_{17}  \!\!\!\! & \!\!\!\!\!-\!\!h_{20}  \!\!\!\! & \!\!\!\!\!-\!\!h_{19}  \!\!\!\! & \!\!\!\!\!-\!\!h_{30}  \!\!\!\! & \!\!\!\!\!-\!\!h_{29}  \!\!\!\! & \!\!\!\!\!\!\! h_{32}  \!\!\!\! & \!\!\!\!\!\!\! h_{31}  \!\!\!\! & \!\!\!\!\!-\!\!h_{26}  \!\!\!\! & \!\!\!\!\!-\!\!h_{25}  \!\!\!\! & \!\!\!\!\!\!\! h_{28}  \!\!\!\! & \!\!\!\!\!\!\! h_{27}  \!\!\!\! & \!\!\!\!\!-\!\!h_{6}   \!\!\!\! & \!\!\!\!\!-\!\!h_{5}   \!\!\!\! & \!\!\!\!\!\!\! h_{8}   \!\!\!\! & \!\!\!\!\!\!\! h_{7}   \!\!\!\! & \!\!\!\!\!-\!\!h_{2}   \!\!\!\! & \!\!\!\!\!-\!\!h_{1}   \!\!\!\! & \!\!\!\!\!\!\! h_{4}   \!\!\!\! & \!\!\!\!\!\!\! h_{3}   \!\!\!\! & \!\!\!\!\!\!\! h_{14}  \!\!\!\! & \!\!\!\!\!\!\! h_{13}  \!\!\!\! & \!\!\!\!\!-\!\!h_{16}  \!\!\!\! & \!\!\!\!\!-\!\!h_{15}  \!\!\!\! & \!\!\!\!\!\!\! h_{10}  \!\!\!\! & \!\!\!\!\!\!\! h_{9}   \!\!\!\! & \!\!\!\!\!-\!\!h_{12}  \!\!\!\! & \!\!\!\!\!-\!\!h_{11}  \\%
 \!\! h_{23}  \!\!\!\! & \!\!\!\!\!-\!\!h_{24}  \!\!\!\! & \!\!\!\!\!\!\! h_{21}  \!\!\!\! & \!\!\!\!\!-\!\!h_{22}  \!\!\!\! & \!\!\!\!\!\!\! h_{19}  \!\!\!\! & \!\!\!\!\!-\!\!h_{20}  \!\!\!\! & \!\!\!\!\!\!\! h_{17}  \!\!\!\! & \!\!\!\!\!-\!\!h_{18}  \!\!\!\! & \!\!\!\!\!-\!\!h_{31}  \!\!\!\! & \!\!\!\!\!\!\! h_{32}  \!\!\!\! & \!\!\!\!\!-\!\!h_{29}  \!\!\!\! & \!\!\!\!\!\!\! h_{30}  \!\!\!\! & \!\!\!\!\!-\!\!h_{27}  \!\!\!\! & \!\!\!\!\!\!\! h_{28}  \!\!\!\! & \!\!\!\!\!-\!\!h_{25}  \!\!\!\! & \!\!\!\!\!\!\! h_{26}  \!\!\!\! & \!\!\!\!\!-\!\!h_{7}   \!\!\!\! & \!\!\!\!\!\!\! h_{8}   \!\!\!\! & \!\!\!\!\!-\!\!h_{5}   \!\!\!\! & \!\!\!\!\!\!\! h_{6}   \!\!\!\! & \!\!\!\!\!-\!\!h_{3}   \!\!\!\! & \!\!\!\!\!\!\! h_{4}   \!\!\!\! & \!\!\!\!\!-\!\!h_{1}   \!\!\!\! & \!\!\!\!\!\!\! h_{2}   \!\!\!\! & \!\!\!\!\!\!\! h_{15}  \!\!\!\! & \!\!\!\!\!-\!\!h_{16}  \!\!\!\! & \!\!\!\!\!\!\! h_{13}  \!\!\!\! & \!\!\!\!\!-\!\!h_{14}  \!\!\!\! & \!\!\!\!\!\!\! h_{11}  \!\!\!\! & \!\!\!\!\!-\!\!h_{12}  \!\!\!\! & \!\!\!\!\!\!\! h_{9}   \!\!\!\! & \!\!\!\!\!-\!\!h_{10}  \\%
 \!\! h_{24}  \!\!\!\! & \!\!\!\!\!\!\! h_{23}  \!\!\!\! & \!\!\!\!\!\!\! h_{22}  \!\!\!\! & \!\!\!\!\!\!\! h_{21}  \!\!\!\! & \!\!\!\!\!\!\! h_{20}  \!\!\!\! & \!\!\!\!\!\!\! h_{19}  \!\!\!\! & \!\!\!\!\!\!\! h_{18}  \!\!\!\! & \!\!\!\!\!\!\! h_{17}  \!\!\!\! & \!\!\!\!\!-\!\!h_{32}  \!\!\!\! & \!\!\!\!\!-\!\!h_{31}  \!\!\!\! & \!\!\!\!\!-\!\!h_{30}  \!\!\!\! & \!\!\!\!\!-\!\!h_{29}  \!\!\!\! & \!\!\!\!\!-\!\!h_{28}  \!\!\!\! & \!\!\!\!\!-\!\!h_{27}  \!\!\!\! & \!\!\!\!\!-\!\!h_{26}  \!\!\!\! & \!\!\!\!\!-\!\!h_{25}  \!\!\!\! & \!\!\!\!\!-\!\!h_{8}   \!\!\!\! & \!\!\!\!\!-\!\!h_{7}   \!\!\!\! & \!\!\!\!\!-\!\!h_{6}   \!\!\!\! & \!\!\!\!\!-\!\!h_{5}   \!\!\!\! & \!\!\!\!\!-\!\!h_{4}   \!\!\!\! & \!\!\!\!\!-\!\!h_{3}   \!\!\!\! & \!\!\!\!\!-\!\!h_{2}   \!\!\!\! & \!\!\!\!\!-\!\!h_{1}   \!\!\!\! & \!\!\!\!\!\!\! h_{16}  \!\!\!\! & \!\!\!\!\!\!\! h_{15}  \!\!\!\! & \!\!\!\!\!\!\! h_{14}  \!\!\!\! & \!\!\!\!\!\!\! h_{13}  \!\!\!\! & \!\!\!\!\!\!\! h_{12}  \!\!\!\! & \!\!\!\!\!\!\! h_{11}  \!\!\!\! & \!\!\!\!\!\!\! h_{10}  \!\!\!\! & \!\!\!\!\!\!\! h_{9}   \\%
 \!\! h_{25}  \!\!\!\! & \!\!\!\!\!-\!\!h_{26}  \!\!\!\! & \!\!\!\!\!-\!\!h_{27}  \!\!\!\! & \!\!\!\!\!\!\! h_{28}  \!\!\!\! & \!\!\!\!\!-\!\!h_{29}  \!\!\!\! & \!\!\!\!\!\!\! h_{30}  \!\!\!\! & \!\!\!\!\!\!\! h_{31}  \!\!\!\! & \!\!\!\!\!-\!\!h_{32}  \!\!\!\! & \!\!\!\!\!\!\! h_{17}  \!\!\!\! & \!\!\!\!\!-\!\!h_{18}  \!\!\!\! & \!\!\!\!\!-\!\!h_{19}  \!\!\!\! & \!\!\!\!\!\!\! h_{20}  \!\!\!\! & \!\!\!\!\!-\!\!h_{21}  \!\!\!\! & \!\!\!\!\!\!\! h_{22}  \!\!\!\! & \!\!\!\!\!\!\! h_{23}  \!\!\!\! & \!\!\!\!\!-\!\!h_{24}  \!\!\!\! & \!\!\!\!\!-\!\!h_{9}   \!\!\!\! & \!\!\!\!\!\!\! h_{10}  \!\!\!\! & \!\!\!\!\!\!\! h_{11}  \!\!\!\! & \!\!\!\!\!-\!\!h_{12}  \!\!\!\! & \!\!\!\!\!\!\! h_{13}  \!\!\!\! & \!\!\!\!\!-\!\!h_{14}  \!\!\!\! & \!\!\!\!\!-\!\!h_{15}  \!\!\!\! & \!\!\!\!\!\!\! h_{16}  \!\!\!\! & \!\!\!\!\!-\!\!h_{1}   \!\!\!\! & \!\!\!\!\!\!\! h_{2}   \!\!\!\! & \!\!\!\!\!\!\! h_{3}   \!\!\!\! & \!\!\!\!\!-\!\!h_{4}   \!\!\!\! & \!\!\!\!\!\!\! h_{5}   \!\!\!\! & \!\!\!\!\!-\!\!h_{6}   \!\!\!\! & \!\!\!\!\!-\!\!h_{7}   \!\!\!\! & \!\!\!\!\!\!\! h_{8}   \\%
 \!\! h_{26}  \!\!\!\! & \!\!\!\!\!\!\! h_{25}  \!\!\!\! & \!\!\!\!\!-\!\!h_{28}  \!\!\!\! & \!\!\!\!\!-\!\!h_{27}  \!\!\!\! & \!\!\!\!\!-\!\!h_{30}  \!\!\!\! & \!\!\!\!\!-\!\!h_{29}  \!\!\!\! & \!\!\!\!\!\!\! h_{32}  \!\!\!\! & \!\!\!\!\!\!\! h_{31}  \!\!\!\! & \!\!\!\!\!\!\! h_{18}  \!\!\!\! & \!\!\!\!\!\!\! h_{17}  \!\!\!\! & \!\!\!\!\!-\!\!h_{20}  \!\!\!\! & \!\!\!\!\!-\!\!h_{19}  \!\!\!\! & \!\!\!\!\!-\!\!h_{22}  \!\!\!\! & \!\!\!\!\!-\!\!h_{21}  \!\!\!\! & \!\!\!\!\!\!\! h_{24}  \!\!\!\! & \!\!\!\!\!\!\! h_{23}  \!\!\!\! & \!\!\!\!\!-\!\!h_{10}  \!\!\!\! & \!\!\!\!\!-\!\!h_{9}   \!\!\!\! & \!\!\!\!\!\!\! h_{12}  \!\!\!\! & \!\!\!\!\!\!\! h_{11}  \!\!\!\! & \!\!\!\!\!\!\! h_{14}  \!\!\!\! & \!\!\!\!\!\!\! h_{13}  \!\!\!\! & \!\!\!\!\!-\!\!h_{16}  \!\!\!\! & \!\!\!\!\!-\!\!h_{15}  \!\!\!\! & \!\!\!\!\!-\!\!h_{2}   \!\!\!\! & \!\!\!\!\!-\!\!h_{1}   \!\!\!\! & \!\!\!\!\!\!\! h_{4}   \!\!\!\! & \!\!\!\!\!\!\! h_{3}   \!\!\!\! & \!\!\!\!\!\!\! h_{6}   \!\!\!\! & \!\!\!\!\!\!\! h_{5}   \!\!\!\! & \!\!\!\!\!-\!\!h_{8}   \!\!\!\! & \!\!\!\!\!-\!\!h_{7}   \\%
 \!\! h_{27}  \!\!\!\! & \!\!\!\!\!-\!\!h_{28}  \!\!\!\! & \!\!\!\!\!\!\! h_{25}  \!\!\!\! & \!\!\!\!\!-\!\!h_{26}  \!\!\!\! & \!\!\!\!\!-\!\!h_{31}  \!\!\!\! & \!\!\!\!\!\!\! h_{32}  \!\!\!\! & \!\!\!\!\!-\!\!h_{29}  \!\!\!\! & \!\!\!\!\!\!\! h_{30}  \!\!\!\! & \!\!\!\!\!\!\! h_{19}  \!\!\!\! & \!\!\!\!\!-\!\!h_{20}  \!\!\!\! & \!\!\!\!\!\!\! h_{17}  \!\!\!\! & \!\!\!\!\!-\!\!h_{18}  \!\!\!\! & \!\!\!\!\!-\!\!h_{23}  \!\!\!\! & \!\!\!\!\!\!\! h_{24}  \!\!\!\! & \!\!\!\!\!-\!\!h_{21}  \!\!\!\! & \!\!\!\!\!\!\! h_{22}  \!\!\!\! & \!\!\!\!\!-\!\!h_{11}  \!\!\!\! & \!\!\!\!\!\!\! h_{12}  \!\!\!\! & \!\!\!\!\!-\!\!h_{9}   \!\!\!\! & \!\!\!\!\!\!\! h_{10}  \!\!\!\! & \!\!\!\!\!\!\! h_{15}  \!\!\!\! & \!\!\!\!\!-\!\!h_{16}  \!\!\!\! & \!\!\!\!\!\!\! h_{13}  \!\!\!\! & \!\!\!\!\!-\!\!h_{14}  \!\!\!\! & \!\!\!\!\!-\!\!h_{3}   \!\!\!\! & \!\!\!\!\!\!\! h_{4}   \!\!\!\! & \!\!\!\!\!-\!\!h_{1}   \!\!\!\! & \!\!\!\!\!\!\! h_{2}   \!\!\!\! & \!\!\!\!\!\!\! h_{7}   \!\!\!\! & \!\!\!\!\!-\!\!h_{8}   \!\!\!\! & \!\!\!\!\!\!\! h_{5}   \!\!\!\! & \!\!\!\!\!-\!\!h_{6}   \\%
 \!\! h_{28}  \!\!\!\! & \!\!\!\!\!\!\! h_{27}  \!\!\!\! & \!\!\!\!\!\!\! h_{26}  \!\!\!\! & \!\!\!\!\!\!\! h_{25}  \!\!\!\! & \!\!\!\!\!-\!\!h_{32}  \!\!\!\! & \!\!\!\!\!-\!\!h_{31}  \!\!\!\! & \!\!\!\!\!-\!\!h_{30}  \!\!\!\! & \!\!\!\!\!-\!\!h_{29}  \!\!\!\! & \!\!\!\!\!\!\! h_{20}  \!\!\!\! & \!\!\!\!\!\!\! h_{19}  \!\!\!\! & \!\!\!\!\!\!\! h_{18}  \!\!\!\! & \!\!\!\!\!\!\! h_{17}  \!\!\!\! & \!\!\!\!\!-\!\!h_{24}  \!\!\!\! & \!\!\!\!\!-\!\!h_{23}  \!\!\!\! & \!\!\!\!\!-\!\!h_{22}  \!\!\!\! & \!\!\!\!\!-\!\!h_{21}  \!\!\!\! & \!\!\!\!\!-\!\!h_{12}  \!\!\!\! & \!\!\!\!\!-\!\!h_{11}  \!\!\!\! & \!\!\!\!\!-\!\!h_{10}  \!\!\!\! & \!\!\!\!\!-\!\!h_{9}   \!\!\!\! & \!\!\!\!\!\!\! h_{16}  \!\!\!\! & \!\!\!\!\!\!\! h_{15}  \!\!\!\! & \!\!\!\!\!\!\! h_{14}  \!\!\!\! & \!\!\!\!\!\!\! h_{13}  \!\!\!\! & \!\!\!\!\!-\!\!h_{4}   \!\!\!\! & \!\!\!\!\!-\!\!h_{3}   \!\!\!\! & \!\!\!\!\!-\!\!h_{2}   \!\!\!\! & \!\!\!\!\!-\!\!h_{1}   \!\!\!\! & \!\!\!\!\!\!\! h_{8}   \!\!\!\! & \!\!\!\!\!\!\! h_{7}   \!\!\!\! & \!\!\!\!\!\!\! h_{6}   \!\!\!\! & \!\!\!\!\!\!\! h_{5}   \\%
 \!\! h_{29}  \!\!\!\! & \!\!\!\!\!-\!\!h_{30}  \!\!\!\! & \!\!\!\!\!-\!\!h_{31}  \!\!\!\! & \!\!\!\!\!\!\! h_{32}  \!\!\!\! & \!\!\!\!\!\!\! h_{25}  \!\!\!\! & \!\!\!\!\!-\!\!h_{26}  \!\!\!\! & \!\!\!\!\!-\!\!h_{27}  \!\!\!\! & \!\!\!\!\!\!\! h_{28}  \!\!\!\! & \!\!\!\!\!\!\! h_{21}  \!\!\!\! & \!\!\!\!\!-\!\!h_{22}  \!\!\!\! & \!\!\!\!\!-\!\!h_{23}  \!\!\!\! & \!\!\!\!\!\!\! h_{24}  \!\!\!\! & \!\!\!\!\!\!\! h_{17}  \!\!\!\! & \!\!\!\!\!-\!\!h_{18}  \!\!\!\! & \!\!\!\!\!-\!\!h_{19}  \!\!\!\! & \!\!\!\!\!\!\! h_{20}  \!\!\!\! & \!\!\!\!\!-\!\!h_{13}  \!\!\!\! & \!\!\!\!\!\!\! h_{14}  \!\!\!\! & \!\!\!\!\!\!\! h_{15}  \!\!\!\! & \!\!\!\!\!-\!\!h_{16}  \!\!\!\! & \!\!\!\!\!-\!\!h_{9}   \!\!\!\! & \!\!\!\!\!\!\! h_{10}  \!\!\!\! & \!\!\!\!\!\!\! h_{11}  \!\!\!\! & \!\!\!\!\!-\!\!h_{12}  \!\!\!\! & \!\!\!\!\!-\!\!h_{5}   \!\!\!\! & \!\!\!\!\!\!\! h_{6}   \!\!\!\! & \!\!\!\!\!\!\! h_{7}   \!\!\!\! & \!\!\!\!\!-\!\!h_{8}   \!\!\!\! & \!\!\!\!\!-\!\!h_{1}   \!\!\!\! & \!\!\!\!\!\!\! h_{2}   \!\!\!\! & \!\!\!\!\!\!\! h_{3}   \!\!\!\! & \!\!\!\!\!-\!\!h_{4}   \\%
 \!\! h_{30}  \!\!\!\! & \!\!\!\!\!\!\! h_{29}  \!\!\!\! & \!\!\!\!\!-\!\!h_{32}  \!\!\!\! & \!\!\!\!\!-\!\!h_{31}  \!\!\!\! & \!\!\!\!\!\!\! h_{26}  \!\!\!\! & \!\!\!\!\!\!\! h_{25}  \!\!\!\! & \!\!\!\!\!-\!\!h_{28}  \!\!\!\! & \!\!\!\!\!-\!\!h_{27}  \!\!\!\! & \!\!\!\!\!\!\! h_{22}  \!\!\!\! & \!\!\!\!\!\!\! h_{21}  \!\!\!\! & \!\!\!\!\!-\!\!h_{24}  \!\!\!\! & \!\!\!\!\!-\!\!h_{23}  \!\!\!\! & \!\!\!\!\!\!\! h_{18}  \!\!\!\! & \!\!\!\!\!\!\! h_{17}  \!\!\!\! & \!\!\!\!\!-\!\!h_{20}  \!\!\!\! & \!\!\!\!\!-\!\!h_{19}  \!\!\!\! & \!\!\!\!\!-\!\!h_{14}  \!\!\!\! & \!\!\!\!\!-\!\!h_{13}  \!\!\!\! & \!\!\!\!\!\!\! h_{16}  \!\!\!\! & \!\!\!\!\!\!\! h_{15}  \!\!\!\! & \!\!\!\!\!-\!\!h_{10}  \!\!\!\! & \!\!\!\!\!-\!\!h_{9}   \!\!\!\! & \!\!\!\!\!\!\! h_{12}  \!\!\!\! & \!\!\!\!\!\!\! h_{11}  \!\!\!\! & \!\!\!\!\!-\!\!h_{6}   \!\!\!\! & \!\!\!\!\!-\!\!h_{5}   \!\!\!\! & \!\!\!\!\!\!\! h_{8}   \!\!\!\! & \!\!\!\!\!\!\! h_{7}   \!\!\!\! & \!\!\!\!\!-\!\!h_{2}   \!\!\!\! & \!\!\!\!\!-\!\!h_{1}   \!\!\!\! & \!\!\!\!\!\!\! h_{4}   \!\!\!\! & \!\!\!\!\!\!\! h_{3}   \\%
 \!\! h_{31}  \!\!\!\! & \!\!\!\!\!-\!\!h_{32}  \!\!\!\! & \!\!\!\!\!\!\! h_{29}  \!\!\!\! & \!\!\!\!\!-\!\!h_{30}  \!\!\!\! & \!\!\!\!\!\!\! h_{27}  \!\!\!\! & \!\!\!\!\!-\!\!h_{28}  \!\!\!\! & \!\!\!\!\!\!\! h_{25}  \!\!\!\! & \!\!\!\!\!-\!\!h_{26}  \!\!\!\! & \!\!\!\!\!\!\! h_{23}  \!\!\!\! & \!\!\!\!\!-\!\!h_{24}  \!\!\!\! & \!\!\!\!\!\!\! h_{21}  \!\!\!\! & \!\!\!\!\!-\!\!h_{22}  \!\!\!\! & \!\!\!\!\!\!\! h_{19}  \!\!\!\! & \!\!\!\!\!-\!\!h_{20}  \!\!\!\! & \!\!\!\!\!\!\! h_{17}  \!\!\!\! & \!\!\!\!\!-\!\!h_{18}  \!\!\!\! & \!\!\!\!\!-\!\!h_{15}  \!\!\!\! & \!\!\!\!\!\!\! h_{16}  \!\!\!\! & \!\!\!\!\!-\!\!h_{13}  \!\!\!\! & \!\!\!\!\!\!\! h_{14}  \!\!\!\! & \!\!\!\!\!-\!\!h_{11}  \!\!\!\! & \!\!\!\!\!\!\! h_{12}  \!\!\!\! & \!\!\!\!\!-\!\!h_{9}   \!\!\!\! & \!\!\!\!\!\!\! h_{10}  \!\!\!\! & \!\!\!\!\!-\!\!h_{7}   \!\!\!\! & \!\!\!\!\!\!\! h_{8}   \!\!\!\! & \!\!\!\!\!-\!\!h_{5}   \!\!\!\! & \!\!\!\!\!\!\! h_{6}   \!\!\!\! & \!\!\!\!\!-\!\!h_{3}   \!\!\!\! & \!\!\!\!\!\!\! h_{4}   \!\!\!\! & \!\!\!\!\!-\!\!h_{1}   \!\!\!\! & \!\!\!\!\!\!\! h_{2}   \\%
 \!\! h_{32}  \!\!\!\! & \!\!\!\!\!\!\! h_{31}  \!\!\!\! & \!\!\!\!\!\!\! h_{30}  \!\!\!\! & \!\!\!\!\!\!\! h_{29}  \!\!\!\! & \!\!\!\!\!\!\! h_{28}  \!\!\!\! & \!\!\!\!\!\!\! h_{27}  \!\!\!\! & \!\!\!\!\!\!\! h_{26}  \!\!\!\! & \!\!\!\!\!\!\! h_{25}  \!\!\!\! & \!\!\!\!\!\!\! h_{24}  \!\!\!\! & \!\!\!\!\!\!\! h_{23}  \!\!\!\! & \!\!\!\!\!\!\! h_{22}  \!\!\!\! & \!\!\!\!\!\!\! h_{21}  \!\!\!\! & \!\!\!\!\!\!\! h_{20}  \!\!\!\! & \!\!\!\!\!\!\! h_{19}  \!\!\!\! & \!\!\!\!\!\!\! h_{18}  \!\!\!\! & \!\!\!\!\!\!\! h_{17}  \!\!\!\! & \!\!\!\!\!-\!\!h_{16}  \!\!\!\! & \!\!\!\!\!-\!\!h_{15}  \!\!\!\! & \!\!\!\!\!-\!\!h_{14}  \!\!\!\! & \!\!\!\!\!-\!\!h_{13}  \!\!\!\! & \!\!\!\!\!-\!\!h_{12}  \!\!\!\! & \!\!\!\!\!-\!\!h_{11}  \!\!\!\! & \!\!\!\!\!-\!\!h_{10}  \!\!\!\! & \!\!\!\!\!-\!\!h_{9}   \!\!\!\! & \!\!\!\!\!-\!\!h_{8}   \!\!\!\! & \!\!\!\!\!-\!\!h_{7}   \!\!\!\! & \!\!\!\!\!-\!\!h_{6}   \!\!\!\! & \!\!\!\!\!-\!\!h_{5}   \!\!\!\! & \!\!\!\!\!-\!\!h_{4}   \!\!\!\! & \!\!\!\!\!-\!\!h_{3}   \!\!\!\! & \!\!\!\!\!-\!\!h_{2}   \!\!\!\! & \!\!\!\!\!-\!\!h_{1} \\ [1.5ex]%
\end{array}}
\!\!\!\!\right]& \nonumber\\[-5ex]\nonumber%
\end{eqnarray}

\subsection{Matlab Function: Generator of GABBA Encoded Channel Matrices}
\label{ProgramGABBAEncodedChannelMatrix} %
\begin{verbatim}
function [H, H1, H2] = GABBAEncodedChannelMatrix(h,K)
% Input:   vector of numeric or symbolic entries corresponding to channel estimates
% Output:  GABBA encoded channel matrix H
% Example: syms h1 h2 h3 h4 h5 h6 h7 h8 h9 h10 h11 h12 h13 h14 h15 h16
%          h = [h1 h2 h3 h4 h5 h6 h7 h8 h9 h10 h11 h12 h13 h14 h15 h16]
%          H = GABBA_EncodedChannelMatrix(h)
if nargin == 2,
    for k = 1:K % Must be a power of 2
        eval(['syms h' num2str(k)]);
    end
    h = [];
    for k = 1:K,
        eval(['h = [h; h' num2str(k) '];']);
    end
else
    K = length(h); % Must be a power of 2
end
H1 = reshape(h,[1 1 K]);
H2 = reshape([h(K/2+1:end) h(1:K/2)],[1 1 K]);
while K >= 2;
    for n = 1:K/2,
        C1(:,:,n) = ABBA1(H1(:,:,2*n-1),H1(:,:,2*n));
        C2(:,:,n) = ABBA2(H2(:,:,2*n-1),H2(:,:,2*n));
    end
    H1 = C1;
    H2 = C2;
    clear C1 C2;
    K = K/2;
end
K = length(h);
H = [H1(1:K/2,:) zeros(K/2,K); zeros(K/2,K) H2(1:K/2,:)];
% --- Auxiliary functions ---
function C = ABBA1(x,y); C = [[ x  y];[ y -x]];
function C = ABBA2(x,y); C = [[ x -y];[ y  x]];
\end{verbatim}

\section{Relabeling of GABBA Reduced Encoded Channel Matrices}
\label{Appendix_RelabelGABBAReducedChannelMatrix}%
\subsection{Matlab Function: Relabels GABBA Reduced Encoded Channel Matrices} %
\label{ProgramRelabelGABBAReducedChannelMatrix} %
\begin{verbatim}
function [Hhat1,Hhat2] = RelabelH(H1,H2,flag)
% Input: H1, H2 - minors of GABBA (reduced) encoded channel matrix
%        flag - 1 if minors are from GABBA encoded channel matrix
%               2 if minors are from GABBA reduced encoded channel matrix
% Output: relabeled outcome of (sum) product
%
K = length(H1);
switch flag,
    case 1,
        H  = conj((H1*H1' + H2*H2')/2);
        H1 = H(1:K/2,1:K/2);
        H2 = H(1+K/2:K,1+K/2:K);
    case 2,
        H = H1.'*H2;
        [p0, p1] = PermutationIndexes(1:K);
        H1 = H(p0,p0);
        H2 = H(p1,p1);
end
Hhat1 = zeros(K/2);
Hhat2 = zeros(K/2);
for k = 1:K/2,
    eval(['syms h' num2str(k)]);
    Hhat1 = Hhat1 + ...
    eval(['double(expand(H1-H1(1,' num2str(k) '))==0).*h' num2str(k) ';'])...
    -eval(['double(expand(H1+H1(1,' num2str(k) '))==0).*h' num2str(k) ';']);
    Hhat2 = Hhat2 + ...
    eval(['double(expand(H2-H1(1,' num2str(k) '))==0).*h' num2str(k) ';'])...
    -eval(['double(expand(H2+H1(1,' num2str(k) '))==0).*h' num2str(k) ';']);
end
\end{verbatim}

\subsection{Matlab Code: Generator of Relabeled GABBA Reduced Encoded Channel Matrices} %
\label{ProgramGenerateGABBAReducedChannelMatrix} %
\begin{verbatim}
function H = RelabeledGABBAReducedEncodedChannelMatrix(K)
N = K/2;
h = [];
for n = 1:N % Must be a power of 2
    eval(['syms h' num2str(n)]);
    eval(['h = [h; h' num2str(n) '];']);
end
H = reshape(h,[1 1 N]);
while N >= 2;
    for n = 1:N/2,
        C(:,:,n) = ABBA2(H(:,:,2*n-1),H(:,:,2*n));
    end
    H = C; clear C;
    N = N/2;
end
% --- Auxiliary functions ---
function C = ABBA2(x,y); C = [[ x -y];[ y  x]];
\end{verbatim}

\clearpage
\newpage
\section{Permutation Indexes}
\label{Appendix_GeneratePermutationIndexes} %
\subsection{Matlab Function: Generator of Permutation Indexes $\boldsymbol{p}_{0:N}$ and $\boldsymbol{p}_{1:N}$} %
\label{ProgramGeneratePermutationIndexes} %
\begin{verbatim}
function [p0, p1] = PermutationIndexes(k)
% Input:   vector of indexes k = 1,2,...K
% Output:  permutation indexes describing the real quasi-orthogonality
%          of GABBA equivalent encoded channel matrices
% Example: [p0, p1] = PermutationIndexes(1:32)
N = length(k);
p = k;
for n = 1:ceil(log2(N))-1,
    p = p + floor((k-1)/(2^n));
end
p = mod(p,2);
p0 = find(p); p1 = find(p-1);
\end{verbatim}

\section{Example: Orthogonal Decoding of the $4$-by-$4$ GABBA Code}
\subsection{Matlab Script: Explicit Calculations for Orthogonal Decoding of the $4$-by-$4$ GABBA Code} %
\label{ProgramOrthogonalDecodingOfOriginalABBACode} %
\begin{verbatim}
% Script: orthogonal soft-decoding of the 4-by-4 GABBA code
clear all
K = 4;
syms s1 s2 s3 s4     % Symbolic variables declaration
s = [s1 s2 s3 s4].'; % Algebraic symbol vector
C = GABBAEncoder(s); % Algebraic GABBA code (variation of original ABBA)
H = GABBAEncodedChannelMatrix([],K); % Algebraic GABBA encoded channel matrix
r = H*[s;conj(s)];  % Algebraic receive vector
% Linear combinations corresponding to equations 38 and 39
r_hat(:,1) = collect(collect(expand(H(:,1:2)'*r + (r'*H(:,5:6)).'),s1),s2);
r_hat(:,2) = collect(collect(expand(H(:,3:4)'*r + (r'*H(:,7:8)).'),s3),s4);
% Constructing the GABBA encoded channel matrix algebraically
H1 = H(1:K/2, 1:K);             % First minor
H2 = H((1+K/2):K, (1+K):2*K);   % Second minor
% Algebraic GABBA reduced encoded channel matrix
HH = simple(conj((H1*H1' + H2*H2')/2));
% At this point verify that:
HH*[s1;s2] == collect(collect(r_hat(:,1),s1),s2) % Results ones if true
HH*[s3;s4] == collect(collect(r_hat(:,2),s3),s4) % Results ones if true
% Real quasi-orthogonality of GABBA reduced encoded channel matrix
HHH = HH.'*HH; % Results in a diagonal matrix with identical entries
% The soft symbol estimates are the result of the linear combinations
% corresponding to equqations 53 through and 55, normalized by the entries
% of the diagonal matrix HHH, which are identical
s_hat = [];
s_hat = [s_hat; simple((HH.'*r_hat(:,1))/HHH(1))];
s_hat = [s_hat; simple((HH.'*r_hat(:,2))/HHH(1))]
\end{verbatim}

\section{Example: Orthogonal Decoding of the $4$-by-$4$ GABBA Code}
\subsection{Matlab Script: Explicit Calculations for the Inspection of Orthogonal Decodability of GABBA Codes} %
\label{ProgramOrthogonalDecodabilityOfGABBACodes} %
\begin{verbatim}
clear
% Input the size of GABBA mother encoding matrix
K = input('Enter value of K (must be a power of 2): '); fprintf('\n');
for n = 1:log2(K)-1,
    if n == 1,
        % Generate relabeled algebraic first-order reduced encoded matrix
        H1 = RelabeledGABBAReducedEncodedChannelMatrix(K);   H2 = H1;
    else
        % Generate relabeled algebraic higher-order reduced encoded matrix
        [H1,H2] = RelabelH(H1,H2,2);
    end
    H = H1.'*H2;
    [p1,p2] = PermutationIndexes(1:K/2^n);
    % Displaying the real-orthogonality of n-th order reduced matrix
    fprintf(['Off-diagonal Minors of the Real-Product of Permuted GABBA' ...
              ' Reduced Encoded Channel Matrix of Order ' num2str(n) '\n'])
    H(p1,p2) H(p2,p1)
    fprintf(['Type "return" and press ENTER to continue\n'])
    % Waiting for input to continue
    keyboard
end
\end{verbatim}

\clearpage
\newpage
\section{The GABBA Orthogonal Decoder}
\label{Appendix_GABBAOrthogonalDecoder}%
\subsection{Matlab Function: GABBA Orthogonal Decoder (Nested Implementation)} %
\label{ProgramGABBAOrthogonalDecoderNested} %
\begin{verbatim}
function s_hat = GABBADecoder(rr,hh)
% Inputs: rr = received signal vector(s) (one comlum per receive antenna)
%         hh = channel estimate vector(s) (one comlum per receive antenna)
% Output: s_hat = ordered vector of solf orthogonal symbol estimates
[K,nr] = size(rr); r_hat  = zeros(K/2,2); H = zeros(K/2,K/2,2);
for a = 1:nr,
    r = rr(:,a);  h = hh(:,a);
    % -------- First Stage ---------
    Hk = GABBAEncodedChannelMatrix(h);
    for n = 1:2,
        r_hat(:,n) = r_hat(:,n) + Hk(:,1 + (n-1)*K/2 : n*K/2)'*r ...
                   + (r'*Hk(:, 1 + K + (n-1)*K/2 : K + n*K/2)).';
    end
    H1 = Hk(1:K/2,1:K);  H2 = Hk((1+K/2):K,(1+K):2*K);
    H(:,:,1) = H(:,:,1) + conj((H1*H1' + H2*H2')/2);
end
H(:,:,2) = H(:,:,1);
clear r rr h hh H1 H2 Hk
% --- Normalizing Intermediate Quantities ---
NormH = norm(H(:,:,1)); H = H/NormH; r_hat = r_hat/NormH;
% ------- Higher Stages --------
Ps = [[1:K/2].' [(1:K/2) + K/2].'];
if K > 2,
    [p1,p2] = PermutationIndexes(1:K/2);
    for i = 1:(log2(K)-1),
        r = r_hat; r_hat = [];
        Ps_hat = Ps; Ps = [];
        for n = 1:2^i,
            Ps = [Ps Ps_hat(p1(1:K/(2^(i+1))),n) Ps_hat(p2(1:K/(2^(i+1))),n)];
            v = H(:,:,mod(n,2)+1).'*r(:,n);
            r_hat = [r_hat v(p1(1:K/(2^(i+1)))) v(p2(1:K/(2^(i+1))))];
        end
        H_hat = H(:,:,1).'*H(:,:,2); H = [];
        H(:,:,1) = H_hat(p1(1:K/(2^(i+1))),p1(1:K/(2^(i+1))));
        H(:,:,2) = H_hat(p2(1:K/(2^(i+1))),p2(1:K/(2^(i+1))));
        % --- Normalizing Intermediate Quantities ---
        NormH = norm(H(:,:,1)); H = H/NormH; r_hat = r_hat/NormH;
        % -------------------------------------------
    end
end
% ----------------------------
for n = 1:K,
    s_hat(Ps(n),1) = r_hat(n)/H(:,:,mod(n+1,2)+1);
end
\end{verbatim}

\section{Exact BER Performance of STBCs in Hoyt, Rice, Rayleigh and Nakagami Fading Channels}
\label{Program_BER_STBCGeneral} %
\begin{verbatim}
function BER = BER_STBC_GeneralFading(esno,Omod,ModType,m,Omega,nr,rho,eta,Naka)
% Inputs:
%    esno    = vector of constellation-energy-to-noise-power ratio figures (in dB)
%    Omod    = number of bits per symbol
%    ModType = 0 for QAM, 1 for PSK
%    m       = vector of "m" parameters (one entry per transmit antenna)
%    Omega   = vector of "Omega" parameters (one entry per transmit antenna)
%    nr      = number of receive antennas
%    rho     = rate of the STBC as defined in Definition 5
%    eta     = diversity order, as defined in footnote 24
%    Naka    = leave empty for physically binding, exact, Rice-Hoyt channel model;
%             or input a dummy value for results with the Nakagami approximation.
global g phi s gamma flag
flag = 0;
if nargin == 9,
    flag = 1;
end
nt = length(Omega);
if length(m) ~= length(Omega),
    m = m(1)*ones(length(Omega),1);
end
K = 5000; M = 2^Omod;
sigma2Vec = (10.^(-esno/10));
if ModType
    % PSK case
    for k = 1:M-1,
        d(k) = 2*abs(k/M - round(k/M));
        for i = 2:Omod,
            d(k) = d(k) + 2*abs(k/(2^i) - round(k/(2^i)));
        end
    end
    for n = 1:length(esno),
        P_minus = []; P_plus = [];
        for k = 1:M-1,
            % - First integral
            delta = (2*k-1)/M;  phi = linspace(1e-50,pi*(1-delta),K);
            g = (sin(delta*pi)^2)/rho; s = -g./(sin(phi).^2);
            for a = 1:nt,
                gamma = Omega(a)/(sigma2Vec(n));
                MGF_minus(a,:) = MGF_Explicit(m(a)).^(nr*eta);
            end
            P_minus(k) = phi(2)*trapz(prod(MGF_minus,1));
            % - Second integral
            delta = (2*k+1)/M; phi = linspace(1e-50,pi*(1-delta),K);
            g = (sin(delta*pi)^2)/rho; s = -g./(sin(phi).^2);
            for a = 1:nt,
                gamma = Omega(a)/(sigma2Vec(n));
                MGF_plus(a,:) = MGF_Explicit(m(a)).^(nr*eta);
            end
            P_plus(k) = phi(2)*trapz(prod(MGF_plus,1));
        end
        BER(n) = sum(d.*(P_minus - P_plus))/(2*pi*Omod);
    end
else
    % QAM case
    phi = linspace(1e-50,pi/2,K);
    for n = 1:length(esno),
        for k = 1:Omod/2,
            I = [];
            for i = 0:((1-1/2^k)*sqrt(M)-1),
                g = (3*(2*i+1)^2)/(2*(M-1)*rho);
                s = -g./(sin(phi).^2);
                for a = 1:nt,
                    gamma = Omega(a)/(sigma2Vec(n));
                    MGF(a,:) = MGF_Explicit(m(a)).^(nr*eta);
                end
                P(i+1,k) = ((-1)^floor((i*2^(k - 1))/sqrt(M)))* ...
                           (2^(k-1)-floor( (i*2^(k - 1))/sqrt(M) + 1/2))* ...
                           trapz(prod(MGF,1));
            end
        end
        BER(n) = 4*phi(2)*sum(sum(P,1),2)/(sqrt(M)*pi*Omod);
    end
end
% --- Moment Generating Functions ---
function MGF = MGF_Explicit(m)
global g phi s gamma flag
if flag,
    % Nakagami channel
    MGF = (1 - s*gamma/m).^(-m);
else
    switch m > 1,
        case 0 % Hoyt channel (reduces to Rayleigh for m = 1)
            q = sqrt((1 - 2*sqrt(m-m^2))/(2*m - 1));
            MGF = 1./sqrt(1 - 2*s*gamma + ((2*s*gamma*q).^2)/((1+q^2)^2));
        case 1 % Rice channel (reduces to Rayleigh for m = 1)
            k = sqrt(m^2 - m)/(m - sqrt(m^2 - m));
            MGF = ((1 + k)./(1 + k - s*gamma)).*exp(k*s*gamma./(1 + k - s*gamma));
    end
end
\end{verbatim}

\newpage
\section{Derivation of Equation \eqref{EulerIntegralSolution}}
\label{DerivationEulerIntegralSolution} %

Let
\begin{equation}
\label{ElementaryIntegral} %
I_k = \int\limits_{0}^{1}{x^k\cdot(1-x)^{K-k}dx}, \ \textup{ with } \ k,K \in \mathbb{N}. %
\end{equation}

By parts we obtain
\begin{equation}
\label{ElementaryIntegralByParts} %
I_k = x^k\cdot(1-x)^{K-k}\Big|_{0}^{1} + \frac{k}{K-k+1}\cdot\int\limits_{0}^{1}{x^{k-1}\cdot(1-x)^{K-k+1}dx} %
    = \frac{k}{K-k+1}I_{k-1}. %
\end{equation}

At $k=0$ we have,
\begin{equation}
\label{ElementaryIntegralKzero} %
I_0 =\int\limits_{0}^{1}{(1-x)^{K}dx} = \frac{1}{K+1}. %
\end{equation}

By induction we then find
\begin{equation}
\label{ElementaryIntegralSolved} %
I_k = \frac{k \cdot (k-1) \cdots 1}{(K-k+1) \cdot (K-k+2) \cdots K} \cdot \frac{1}{K+1} = \frac{(K-k)!k!}{(K+1)!}. %
\end{equation} \hfill \QEDopen

\end{document}